\documentclass[fleqn,usenatbib]{mnras}

\usepackage{newtxtext,newtxmath}

\usepackage[T1]{fontenc}

\DeclareRobustCommand{\VAN}[3]{#2}
\let\VANthebibliography\thebibliography
\def\thebibliography{\DeclareRobustCommand{\VAN}[3]{##3}\VANthebibliography}

\usepackage{graphicx}	
\usepackage{amsmath}	
\usepackage{multicol}        
\usepackage{bm}		
\usepackage{pdflscape}	
\usepackage{longtable}  
\usepackage{caption} 
\usepackage{subfigure} 
\usepackage[flushleft]{threeparttable}
\usepackage{multirow}   
\usepackage{xfrac}   
\usepackage{nicefrac} 
\usepackage{xcolor} 
\usepackage[colorinlistoftodos,prependcaption,textsize=footnotesize]{todonotes} 
\usepackage{outlines} 
\usepackage{ulem} 
\usepackage{cancel} 
\usepackage{svg} 

\newcommand{\dm}{\,pc\,cm$^{-3}$} 
\newcommand{\uJy}{\,$\upmu$Jy} 
\newcommand{\us}{\,$\upmu$s} 
\newcommand{\gray}{$\upgamma$-ray} 
\newcommand{\grays}{$\upgamma$-rays}

\newcommand{\catversion}{v2.5.1} 

\newcommand{\python}{{\sc python}}

\def\psrchive{\mbox{\textsc{psrchive}}}
\def\pulsarx{\mbox{\textsc{pulsar}X}}
\def\mosaic{\mbox{\textsc{mosaic}}}
\def\clfd{\mbox{\textsc{clfd}}}

\def\dspsr{\mbox{\textsc{dspsr}}}

\def\tempo{\mbox{\textsc{tempo}2}}

\def\pop{\mbox{P{\sc sr}P{\sc op}P{\sc y}}}

\def\psrfoldfil{\mbox{\texttt{psrfold\_fil2}}}
\def\pdmp{\mbox{\texttt{pdmp}}}
\def\pam{\mbox{\texttt{pam}}}
\def\pazi{\mbox{\texttt{pazi}}}
\def\pac{\mbox{\texttt{pac}}}
\def\psrsh{\mbox{\texttt{psrsh}}}

\def\evolve{\mbox{\texttt{evolve}}}

\def\psrA{\mbox{J1831$-$0941}}
\def\psrB{\mbox{J1818$-$1502}}

\title[TRAPUM search for Pulsars in Supernova Remnants II]{TRAPUM search for pulsars in supernova remnants and pulsar wind nebulae -- II. Survey analysis and population study}

\author[J. D. Turner et al.]{J. D. Turner$^{1}$\thanks{E-mail: james.turner-2@manchester.ac.uk},
B. W. Stappers$^{1}$, 
E. Barr$^{2}$,
M. Burgay$^{3}$, 
M. Colom i Bernadich$^{3,4}$,
V. Graber$^{5}$, \and 
M.~J.~Keith$^1$,
M. Kramer$^{2}$, 
L. Levin$^{1}$, 
Y. P. Men$^{2}$,
C. Pardo-Araujo$^6$, 
T. Thongmeearkom$^{1,7}$, \and
J. Tian$^1$,
P. V. Padmanabh$^{8,9}$, 
P. Weltevrede$^{1}$,
J. Behrend$^{5}$, 
W. Chen$^{5}$, 
E.~F. Keane$^{10}$,
A. Ridolfi$^{11}$ \\
$^{1}$ Jodrell Bank Centre for Astrophysics, Department of Physics and Astronomy, The University of Manchester, Manchester M13 9PL, UK\\
$^{2}$ Max Planck Institute for Radio Astronomy, Backend Development Group, Electronics Department, Auf dem Hügel 69, 53121 Bonn, Germany\\
$^{3}$ INAF - Osservatorio Astronomico di Cagliari, via della Scienza 5, 09047 Selargius (CA), Italy\\
$^{4}$Max-Planck-Institut für Radioastronomie, Fundamental Physics in Radio Astronomy Group, Auf dem Hügel 69, D-53121 Bonn, Germany \\
$^{5}$Department of Physics, Royal Holloway, University of London, Egham, TW20 0EX, UK\\
$^{6}$ Institute of Space Sciences (ICE-CSIC), Campus UAB, C/ de Can Magrans s/n, Cerdanyola del Vallès (Barcelona) 08193, Spain\\
$^{7}$ National Astronomical Research Institute of Thailand, Don Kaeo, Mae Rim, Chiang Mai 50180, Thailand\\
$^{8}$Max Planck Institute for Gravitational Physics (Albert Einstein Institute), D-30167 Hannover, Germany\\
$^{9}$Leibniz Universit{\"a}t Hannover, D-30167 Hannover, Germany\\
$^{10}$ School of Physics, Trinity College Dublin, College Green, Dublin 2, D02 PN40, Ireland\\
$^{11}$ Fakult\"at f\"ur Physik, Universit\"at Bielefeld, Postfach 100131, D-33501 Bielefeld, Germany\\
}

\date{Accepted XXX. Received YYY; in original form ZZZ}

\pubyear{2025}

\begin{document}
\label{firstpage}
\pagerange{\pageref{firstpage}--\pageref{lastpage}}
\maketitle

\begin{abstract}
We present the second and final set of TRAPUM searches for pulsars at 1284\,MHz inside supernova remnants and pulsar wind nebulae with the MeerKAT telescope. No new pulsars were detected for any of the 80 targets, which include some unidentified TeV sources that could be pulsar wind nebulae. The mean upper limit on the flux density of undetected pulsars is 52\uJy{}, which includes the average sensitivity loss across the coherent beam tiling pattern. This survey is the largest and most sensitive multi-target campaign of its kind. We explore the selection effects that precluded discoveries by testing the parameters of the survey iteratively against many simulated populations of young pulsars in supernova remnants. For the synthetic pulsars that were undetected, we find evidence that, after beaming effects are accounted for, about 45 per cent of pulsars are too faint, 30 per cent are too smeared by scattering, and a further 25 per cent have a modelled projected location which places them outside their supernova remnant. The simulations are repeated for the S1 subband of the MeerKAT S-band receivers, resulting in a $50-150$ per cent increase in the number of discoveries compared to L-band depending on the flux density limit achieved. Therefore, higher frequency searches that can also achieve improved flux density limits are the best hope for future targeted searches. We also report updated properties for the two previous discoveries, including a polarimetry study of PSR \psrA{} finding a rotation measure of 401$\pm$1\,rad\,m$^{-2}$.

\end{abstract}




\section{Introduction}\label{sec: intro}
Pulsars are rapidly spinning neutron stars (NSs). Their high magnetic field strengths generate beams of radio waves along the magnetic field axis, which are then seen as radio pulses when the beam crosses the line of sight to an observer. NSs are born when a massive star undergoes a core collapse supernova (CCSN) at the end of its life \citep{Baade1934a, Heger2003}. CCSNe expel approximately 1\,M$_{\sun}$ of ejecta which moves outwards and sweeps up any circumstellar matter \citep{Chevalier1982a}, including $\sim10\,\text{M}_{\sun}$ of matter lost by the star prior to the supernova. The expanding structure of material is visible as a supernova remnant (SNR), so the youngest pulsars are therefore typically located in the vicinity of SNRs. The discovery of more pulsar-SNR systems is important for constraining the Galactic NS birth rate \citep{Keane2008} and understanding their formation channels. Young pulsars are also laboratories for studying the energy budgets, X-ray and very high energy (VHE) \gray{} emission of the pulsar wind nebula (PWN) and its interaction with surrounding media \citep[e.g.][]{Kargaltsev2008, Shibata2016}, and also the interiors of NSs \citep[e.g.][]{Haskell2015, Antonopoulou2022a}.

The TRAPUM \citep[TRAnsients and PUlsars with MeerKAT;][]{Stappers2016b} project is a large survey for pulsars with the MeerKAT telescope. Since 2022, TRAPUM has been targeting dozens of SNRs, PWNe and also some unidentified sources from the H.E.S.S. TeV Catalogue \citep{Abdalla2018a} that could be PWNe. This work follows \citet[][henceforth Paper I]{Turner2024}, which can be deferred to for an introduction to the survey, what it might tell us about the properties of the pulsar-SNR populations, the energetics of PWNe and the Galactic birthrates of pulsars. Paper I also presented the first set of observations and two pulsars that had been discovered. PSR \psrA{} is a young pulsar we discovered in candidate SNR G22.045$-$0.028 \citep{Dokara2021}, which we concluded must be powering the X-ray counterpart PWN G22.0+0.0 \citep{Yamauchi2016}. We also discovered PSR \psrB{} in the foreground of G15.9+0.2, although the pulsar and SNR are probably unassociated. This new work describes all remaining searches since then. In \S\ref{sec: selection}, we provide information about how the target list was compiled, and the selection criteria used to add or reject sources from the list. In \S\ref{sec: survey-part2}, we present the final observations of the survey, describe the most recent follow-up of the two discoveries and provide a brief interpretation and discussion of the results. The main analysis and discussion follows in \S\ref{sec: pop}, with the description of an in-depth population synthesis that aims to understand the discovery rate of the survey. New results for the two discoveries are presented and discussed in \S\ref{sec: psr-update}. Finally, we summarise the survey and make recommendations for future searches in \S\ref{sec: pop-conc}.

\section{Composing the target list}
\label{sec: selection}
\subsection{Target selection criteria}
\label{subsec: criteria}
One motivation of the TRAPUM survey was to improve the current understanding of the population of young pulsars in supernova remnants by i) analysing the upper limits and discovery rate for a large sample and ii) discovering many new pulsars and carrying out follow-up studies of their systems. This was feasible as the long total observing time allocated to the survey could be used to search a sample large enough to make inferences about the population, thus facilitating i). Moreover, ii) should follow in turn if a large number of searches yielded many new discoveries. In the early phases of the survey, a list of targets was compiled from three sources; the 309 Galactic supernova remnants in the December 2022 version of the Green catalogue \citep[G22;][]{Green2022}, the sources discovered and catalogued by the H.E.S.S. survey \citep{Abdalla2018a} labelled as PWNe (14) or unidentified (25), as listed in TeVCat\footnote{\href{http://tevcat.uchicago.edu/}{http://tevcat.uchicago.edu/} v.3.400} \citep{Wakely2008}. A further 216 candidates were identified across results from the MAGPIS \citep{Helfand2006}, second epoch Molonglo Galactic Plane Survey \citep[MGPS-2;][]{Green2014}, THOR \citep{Anderson2017}, GLEAM \citep{Hurley-Walker2019b} and GLOSTAR \citep{Dokara2021} surveys. The combined target list thus totalled 564 sources. Equal allocation of the telescope time would have allowed just under 10 minutes per target, which would not have significantly improved sensitivity over the previous best searches carried out with the Murriyang telescope at Parkes by \citet{Crawford2002}, which achieved flux density limits of $60-180$\uJy{} at $1384$~MHz for 5 SNRs. We therefore applied some criteria to reduce the list by around half and enable 20-40 minute integration times and flux density limits of 30-50\uJy.

\begin{figure}
    \centering
    \includegraphics[width=0.98\columnwidth]{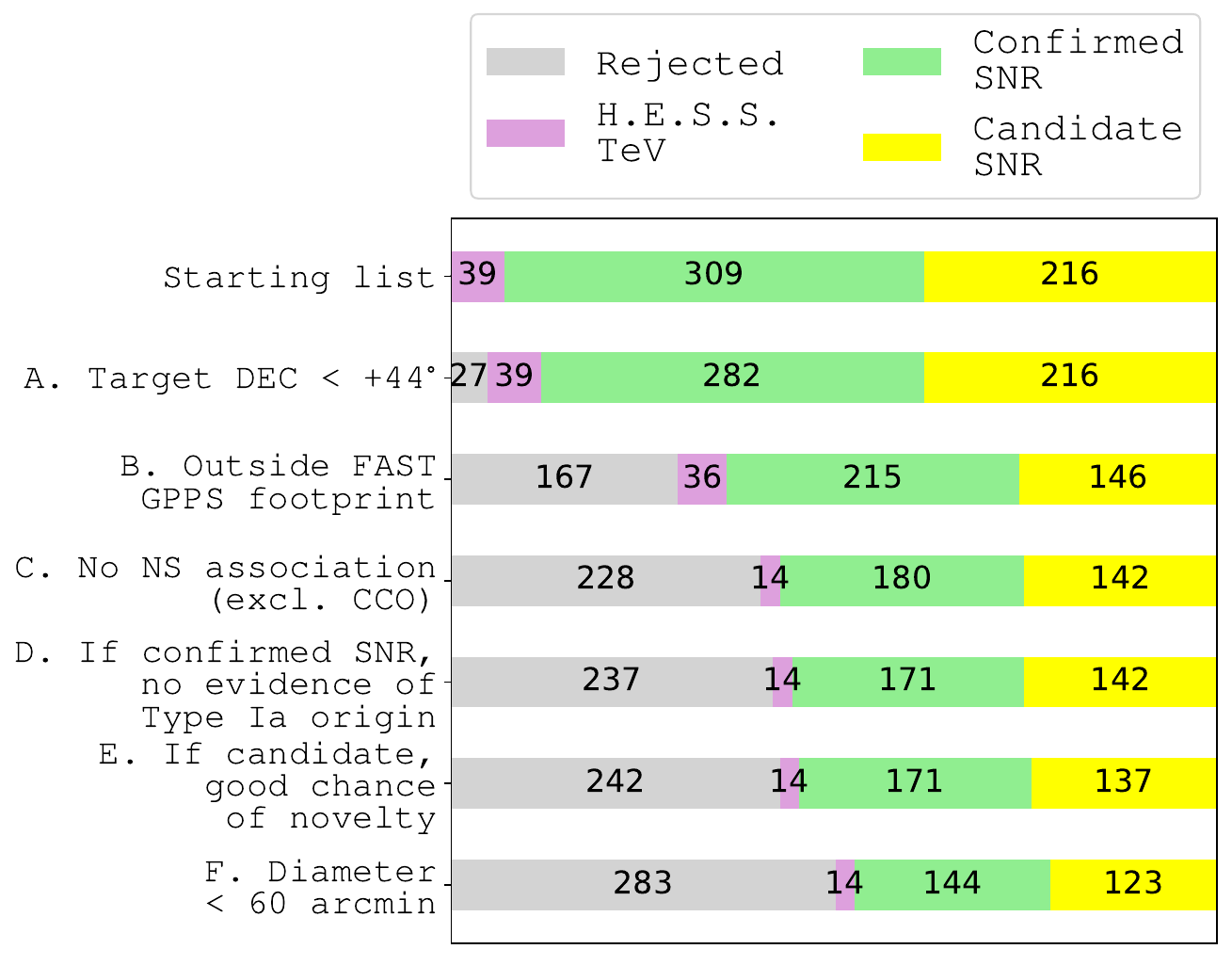}
    \caption[Breakdown of the target selection criteria]{Chart showing the sequential application of the criteria that were used on a starting list of 564 sources combined from the G22 and H.E.S.S. catalogues and SNR candidates in the literature. As can be seen, these criteria halved the number of sources, such that we could then select optimal targets on a case-by-case basis.}
    \label{fig:selcrit}
\end{figure}

\autoref{fig:selcrit} lists the criteria used and how they reduced the target list. Criterion A is of course necessary due to the location of MeerKAT, criterion B onwards were chosen to help maximise the number of discoveries. The final list of 281 targets was then further narrowed down by selecting candidates inferred to be filled centre or composite type, and by selecting against sources that already had competitive limits within the Max-Planck-Institut f\"{u}r Radioastronomie (MPIfR)--MeerKAT Galactic Plane Survey \citep[MMGPS;][]{Padmanabh2023} or from previous targeted searches. Multiple examples in particular concerned a large number of candidates from MAGPIS, the THOR survey and the GLOSTAR survey, which generally have small diameters and are located deep in the Galactic plane. MMGPS-S is highly competitive with our searches at L-band for such targets due to the sky temperature and predicted scattering along the lines of sight. It was therefore decided that small candidates overlapped by the MMGPS-S footprint would not be observed. The MeerKAT S-band receivers were also not yet available when we were preparing the target list. We chose not to switch to S-band at a later stage, instead deciding that the uniformity of a full survey at L-band would be more useful for later population-based analyses. 

Initially we did not apply criterion C to targets. We chose to observe the X-ray pulsar PSR \mbox{J1849$-$0001} associated with HESS J1849$-$000 to search for a radio signal at the known rotational periodicity. However, we subsequently decided to only observe known NSs if they were members of the class of central compact objects (CCOs). Conversely there were sources that did not match some of the criteria but that we still searched, because as important as observational uniformity was, there was a desire to maximise discoveries based on information available about the sources in the literature. For example, G28.8+1.5 has a very large angular diameter of $\sim$100~arcmin, but we found it to be worth targeting regions of interest within the shell.

\subsection{Data rate constraints}
\label{subsec: datarate}
The cap of 480 or 760 coherent beams (CBs) that the Filterbanking Beam Former User Supplied Equipment \citep[FBFUSE;][]{Barr2018} is capable of handling places a constraint on the size of the target that could be fully searched in a single pointing. This is the reason for criterion F in \autoref{fig:selcrit}. A majority of the targets are circular or elliptical, so the number of beams required to fully tile them scales with the square of the target's radius. There are several ways of increasing the size of the CBs to maximise coverage, though they all trade off sensitivity in some way:
\begin{itemize}
    \item By observing the target at a lower elevation, the projected baselines of the array in the direction pointing towards the source are smaller, thus the CBs are wider along that axis.
    \item The inner 44 telescopes of the MeerKAT array (the core) can be used instead of the full array. This reduces the longest baseline by a factor of $\sim$4, and so increases the CB coverage by a factor of about 8 after accounting for the relative weights of the baselines.
    \item The size of the beams scales with the inverse of the observing frequency, so using the MeerKAT Ultra High Frequency (UHF-band) receivers provides a larger CB coverage than at L-band. Whilst this could have been useful for nearby, more extended targets away from the plane where the sky temperature is lower, many of those targets already have associations.
\end{itemize}
\begin{figure}
    \centering
    \includegraphics[width=0.98\columnwidth]{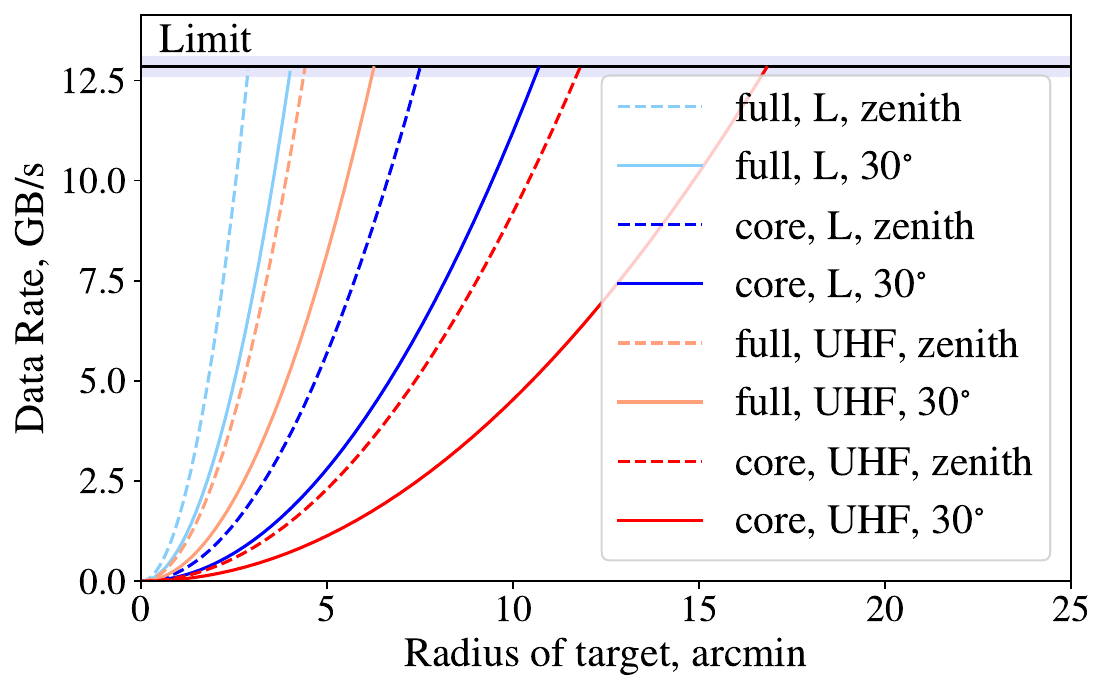}
    \caption[Data rate requirements against target size]{Plot of the data ingestion rates required to tile a circular target against its angular radius. A breakdown of different observing configurations that change the size of the CBs is shown: full array versus the core, L-band versus UHF-band and whether the target is at the zenith or at an elevation of 30\degr. A sampling time of 153\,$\upmu$s and 4096 channels are assumed. Thus the data limit is set by the cap of 480 on the number of CBs (see Paper I). The size of the CBs is calculated using the multibeam simulation package \mosaic{}\footnote{\href{https://github.com/wchenastro/Mosaic}{https://github.com/wchenastro/Mosaic} by WeiWei Chen} for an overlap level of 0.5.}
    \label{fig:datarates}
\end{figure}

\autoref{fig:datarates} shows the coverage supplied by using combinations of these practices. The legend is arranged approximately in order of declining sensitivity, though this strongly depends on the target's sky position as the degradation in sensitivity at UHF-band for locations on the Galactic plane is very high due to the sky temperature contribution and also strong scattering. \autoref{fig:datarates} demonstrates that the best balance between CB coverage and sensitivity, for targets larger than about 5\,arcmin in radius, is to use the core of the array at L-band. During the survey, we were able to push the CB coverage beyond these lines by organising blocks so that target elevations were as close as possible to MeerKAT's horizon limit of 15\degr.

\begin{figure}
    \centering
    \includegraphics[width=0.9\columnwidth]{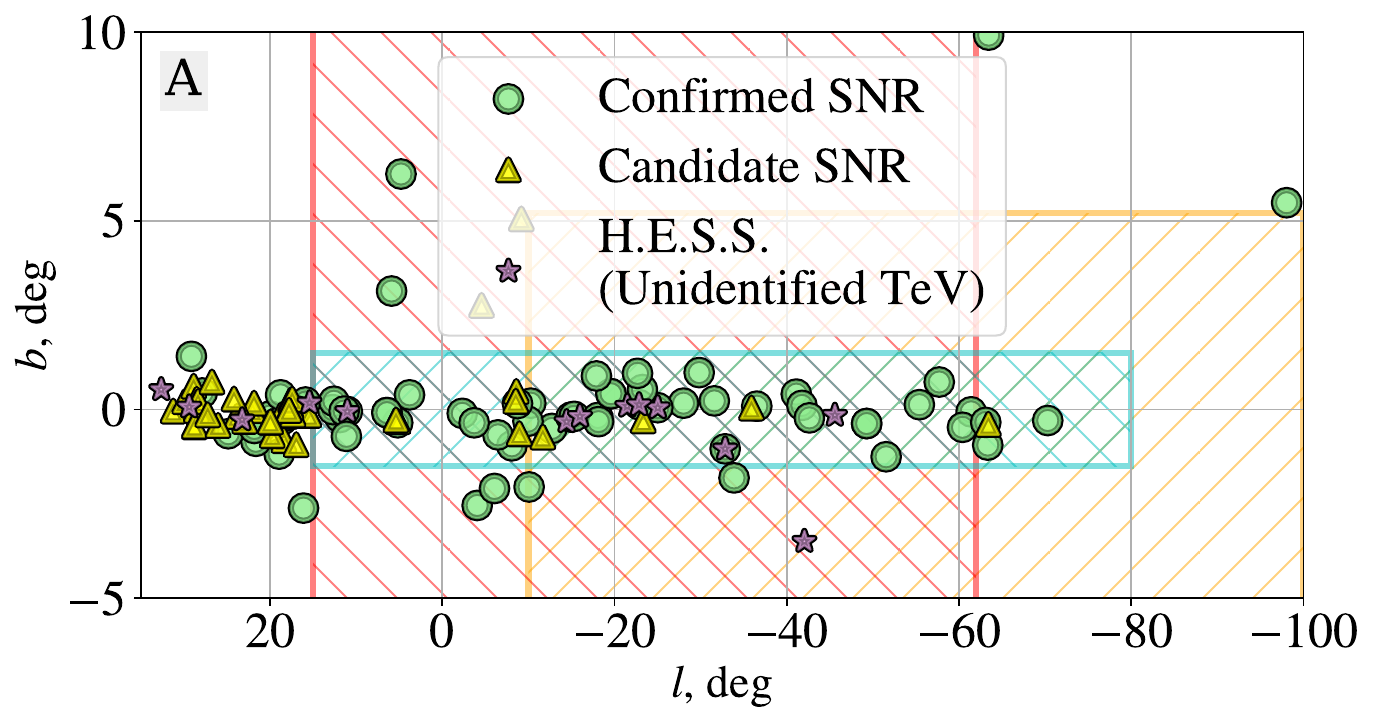}
    \includegraphics[width=0.9\columnwidth]{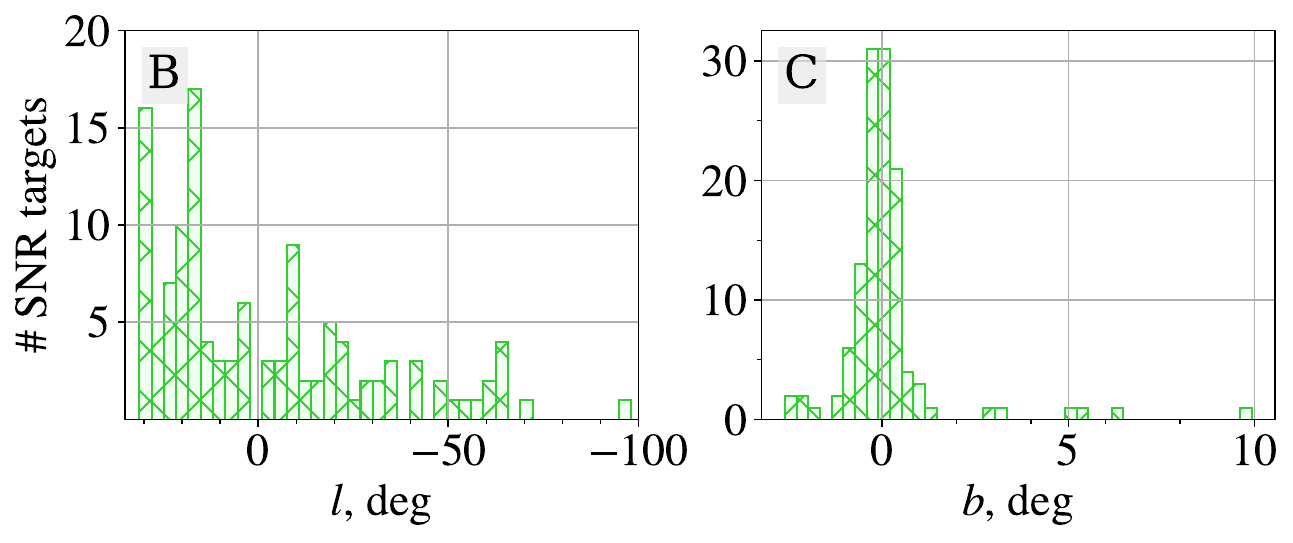}
    \includegraphics[width=0.9\columnwidth]{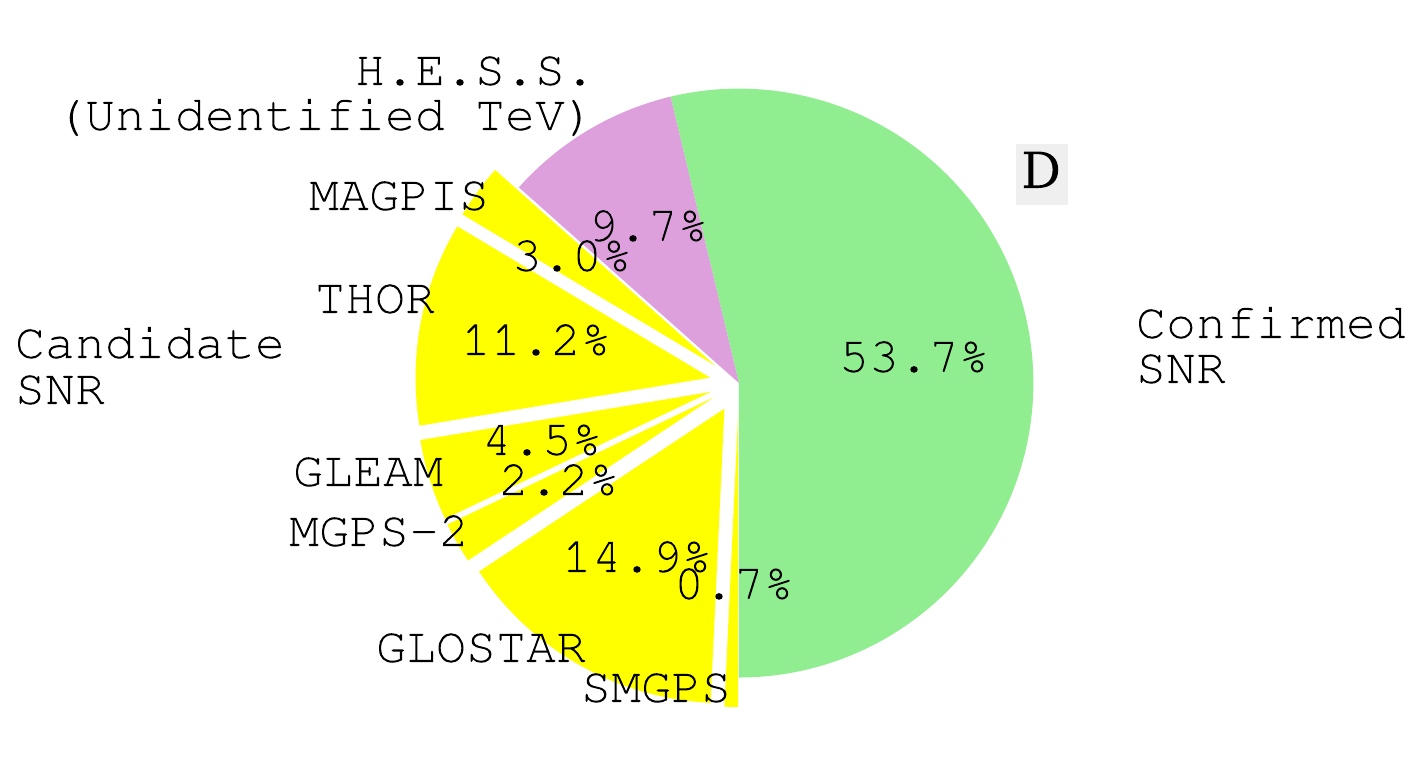}
    \caption[TRAPUM SNR targets: sky positions and types]{The positions and types for the 124 targets searched by this survey (CCOs are excluded). A map of the targets is shown in Galactic coordinates in panel A. The footprint of the MMGPS-UHF (red upper left-lower right hatching), MMGPS-L (orange lower left-upper right hatching) and MMGPS-S (cyan crossed hatching) surveys are overlaid. Panels B and C show histograms of the longitude and latitude of the targets, respectively. Panel D shows the share of targets by type. The candidates are shown divided between the surveys in which they were first identified.}
    \label{fig:targets}
\end{figure}

\section{Second portion of the TRAPUM SNR survey}
\label{sec: survey-part2}
\subsection{Searches since Paper I}
\label{subsec: newsearch}

Following the targets presented in Paper I, we have searched a further $80$ sources for pulsars. These observations were made under the project code SCI-20180923-MK-03. Information about the observations, properties and upper limits on the flux density and pseudo-luminosity for each target is listed in \autoref{tab:targets2}. This brings the full list of targets\footnote{One of the targets of the survey was the odd radio circle J0624$-$6948 (see Paper I), which has since been confirmed to be associated with the Large Magellanic Cloud after further study \citep{Sasaki2025}.} between Paper I and \autoref{tab:targets2} to 134. Ten of these targets are CCOs so a total of 124 sources were searched for new pulsars. In \autoref{fig:targets}, the positions of the targets are shown alongside their distributions in Galactic longitude, $l$, and latitude, $b$. As can be seen in panels A and B, there is a lower concentration of targets in the MMGPS survey regions, which reflects our tendency to avoid those footprints. Panel C demonstrates how concentrated our sample is on the Galactic plane (reflective of the distribution of SNRs; see \citealt{Green2025}), and panel D shows the fractional share of each target type.

\begin{landscape}
\begin{table}
\begin{threeparttable}
    \setlength\LTleft{0pt}
    \setlength\LTright{0pt}
\caption[Target list 2: target and observation properties for part II TRAPUM SNR/PWN/TeV survey]{80 targets searched for the second portion of the TRAPUM survey. Some sources have other names in the literature which are provided at the bottom of the table. As in Table 2 of Paper I, the type is defined as follows; `S' for shell-type G25 SNR, `C' for composite-type G25 SNR, `cand' for candidate SNR, `U' for unidentified TeV source. `?' denotes ambiguity in the literature. Distance references are as follows: [1] \citet{Ranasinghe2022},  [2] \citet{Wang2007}, [3] \citet{Sofue2023}, [4] \citet{Ranasinghe2018},  [5] \citet{Albert2006},  [6] \citet{Ranasinghe2018b},  [7] \citet{Prinz2013}, [8] \citet{Sun1999}, [9] \citet{Combi2005}, [10] \citet{Abdalla2018a}, [11] \citet{Kothes2007},  [12] \citet{Frail1994}, [13] \citet{Giacani2011} and  [14] \citet{Karpova2016}. The sensitivity limits are for the centres of CBs as their calculation does not account for the average degradation factor which is approximately 0.65 at an of overlap level of 0.5. A handful of sources had closely spaced CBs at overlap level of 0.75. \\ $^{a}$The associated object is a CCO or candidate CCO.}\label{tab:targets2}
\begin{tabular}{llllllllllll}
\hline
    \textbf{Source} & \textbf{Type} & \textbf{Associated objects} & \textbf{Size} & \textbf{Schedule Block ID} & \textbf{Observation Start UTC} & \textbf{$t_{\text{obs}}$} & \textbf{$N_{\text{d}}$} & \textbf{$S_{\text{min}}$} & \textbf{$D$} & \textbf{Dist. ref.} & \textbf{$L_{1284}$ limit} \\
     &  &  & (\arcmin) & & yyyy-mm-dd-hh:mm:ss & (s) & & (\uJy) & (kpc) & & (mJy kpc$^2$) \\
\hline
G4.8+6.2 & S & G4.5+6.2 \dotfill & 18 & 20230605$-$0014 & 2024$-$03$-$11$-$02:03:05 & 2390 & 44 & 28 &  &  & \\
G6.4$-$0.1 & C & \dotfill & 48 & 20230605$-$0014 & 2024$-$03$-$11$-$00:42:19 & 2384 & 44 & 44 & 1.8(3) & [1] & 0.14(5) \\
6.4500–0.5583 & cand & G6.5$-$0.4? \dotfill & 3.3 & 20230605$-$0014 & 2024$-$03$-$11$-$00:42:19 & 2384 & 44 & 44 & & & \\
G6.5$-$0.4 & S & 6.4500$-$0.5583? \dotfill & 18 & 20230605$-$0014 & 2024$-$03$-$11$-$00:42:19 & 2384 & 44 & 44 & 3.7(2) & [2] & 0.60(7) \\
G11.1$-$0.7 & S & \dotfill & 15 & 20240524$-$0003 & 2024$-$05$-$27$-$19:25:43 & 2381 & 60 & 27 &  &  & \\
G11.4$-$0.1 & S? & \dotfill & 8 & 20231212$-$0011 & 2024$-$05$-$01$-$06:33:39 & 2387 & 56 & 31 &  &  & \\
G16.2$-$2.7 & S & \dotfill & 17 & 20241128$-$0015 & 2024$-$12$-$02$-$14:17:59 & 2388 & 44 & 28 &  &  & \\
G016.956$-$0.933 & cand & \dotfill & 14.8 & 20241110$-$0005 & 2024$-$11$-$11$-$08:19:31 & 2398 & 44 & 34 &  &  & \\
G17.0$-$0.0 & S & \dotfill & 5 & 20241024$-$0011 & 2024$-$10$-$24$-$14:56:29 & 2383 & 60 & 29 &  &  & \\
G17.4$-$0.1 & S & \dotfill & 6 & 20240524$-$0003 & 2024$-$05$-$27$-$20:05:57 & 2395 & 60 & 28 &  &  & \\
G017.434+0.273 & cand & \dotfill & 4.2 & 20241024$-$0011 & 2024$-$10$-$24$-$15:36:46 & 2386 & 60 & 28 &  &  & \\
G17.80$-$0.02 & cand & \dotfill & 8.8 & 20241024$-$0011 & 2024$-$10$-$24$-$16:16:51 & 2385 & 60 & 28 &  &  & \\
G18.1$-$0.1 & S & \dotfill & 8 & 20230605$-$0014 & 2024$-$03$-$11$-$01:22:39 & 2400 & 44 & 40 & 6.07(13) & [3] & 1.5(1) \\
G018.393$-$0.816 & cand & \dotfill & 5.2 & 20231212$-$0011 & 2024$-$05$-$01$-$04:32:13 & 2394 & 56 & 29 &  &  & \\
G18.8+0.3 & S & \dotfill & 17 $\times$ 11 & 20241128$-$0015 & 2024$-$12$-$02$-$14:58:11 & 2389 & 44 & 37 & 13.8(4) & [4] & 7.2(4) \\
G18.9$-$1.1$^{a}$ & C? & CXOU J182913.1$-$125113 & 33 & 20231219$-$0009 & 2023$-$12$-$19$-$15:22:33 & 2383 & 44 & 34 & 3.1(7) & [2] & 0.3(1) \\
G19.75$-$0.69 & cand & \dotfill & 26.2 & 20241110$-$0005 & 2024$-$11$-$11$-$08:59:44 & 2380 & 44 & 37 &  &  & \\
G19.96$-$0.33 & cand & \dotfill & 11.8 & 20241128$-$0015 & 2024$-$12$-$02$-$15:38:23 & 2398 & 44 & 38 &  &  & \\
G21.0$-$0.4 & S & \dotfill & 9 $\times$ 7 & 20240524$-$0003 & 2024$-$05$-$27$-$20:46:13 & 2386 & 60 & 28 &  &  & \\
G21.6$-$0.8 & S & \dotfill & 13 & 20241117$-$0009 & 2024$-$11$-$19$-$08:06:51 & 2398 & 44 & 37 &  &  & \\
G21.8$-$0.6 & S & \dotfill & 20 & 20250308$-$0012 & 2025$-$03$-$13$-$00:43:20 & 2392 & 44 & 39 & 4.9(3) & [2] & 0.9(1) \\
G021.861+0.169 & cand & \dotfill & 5.4 & 20240722$-$0032 & 2024$-$07$-$25$-$23:21:58 & 2398 & 44 & 39 &  &  & \\
G022.951$-$0.311 & cand & \dotfill & 4.8 & 20240722$-$0032 & 2024$-$07$-$25$-$22:41:26 & 2399 & 44 & 43 &  &  & \\
HESS J1834$-$087 & U & \dotfill & 5.4 & 20241024$-$0011 & 2024$-$10$-$24$-$16:57:06 & 2397 & 60 & 32 & 4.1(3) & [2, 5] & 0.6(1) \\
G23.85$-$0.18 & cand & \dotfill & 5.4 & 20241024$-$0011 & 2024$-$10$-$24$-$17:37:19 & 2385 & 60 & 32 &  &  & \\
G024.193+0.284 & cand & \dotfill & 4.2 & 20240722$-$0032 & 2024$-$07$-$25$-$22:01:35 & 2399 & 44 & 43 &  &  & \\
G24.7$-$0.6 & S? & \dotfill & 15 & 20241117$-$0009 & 2024$-$11$-$19$-$08:47:05 & 2386 & 44 & 38 & 3.8(2) & [6] & 0.54(6) \\
G26.04$-$0.42 & cand & \dotfill & 27 & 20250308$-$0012 & 2025$-$03$-$13$-$01:23:34 & 2390 & 44 & 38 &  &  & \\
G26.75+0.73 & cand & \dotfill & 10.6 & 20250308$-$0012 & 2025$-$03$-$13$-$02:43:58 & 2389 & 44 & 35 &  &  & \\
G27.24$-$0.14 & cand & \dotfill & 12.2 & 20250308$-$0012 & 2025$-$03$-$13$-$02:03:48 & 2391 & 44 & 38 &  &  & \\
G28.21+0.02 & cand & \dotfill & 3.4 & 20240524$-$0003 & 2024$-$05$-$27$-$21:26:31 & 2378 & 60 & 30 &  &  & \\
G28.22$-$0.09 & cand & \dotfill & 6.4 & 20240524$-$0003 & 2024$-$05$-$27$-$21:26:31 & 2378 & 60 & 29 &  &  & \\
G28.3+0.2 & cand & \dotfill & 14 & 20241110$-$0005 & 2024$-$11$-$11$-$10:20:11 & 2390 & 44 & 40 &  &  & \\
G28.33+0.06 & cand & \dotfill & 5 & 20240524$-$0003 & 2024$-$05$-$27$-$21:26:31 & 2378 & 60 & 30 &  &  & \\
G028.524+0.268 & cand & \dotfill & 6.2 & 20241110$-$0005 & 2024$-$11$-$11$-$10:20:11 & 2388 & 44 & 39 &  &  & \\
G28.56+0.00 & cand & \dotfill & 1.5 & 20240524$-$0003 & 2024$-$05$-$27$-$21:26:31 & 2378 & 60 & 30 &  &  & \\
G28.78$-$0.44 & cand & \dotfill & 13.2 & 20241128$-$0015 & 2024$-$12$-$02$-$16:18:37 & 2397 & 44 & 38 &  &  & \\
\hline
\end{tabular}
\end{threeparttable}
\end{table}
\end{landscape}

\begin{landscape}
\begin{table}
\begin{threeparttable}
\contcaption{}
\label{tab:targetscont}
\begin{tabular}{llllllllllll}
\hline
    \textbf{Source} & \textbf{Type} & \textbf{Associated objects} & \textbf{Size} & \textbf{Schedule Block ID} & \textbf{Observation Start UTC} & \textbf{$t_{\text{obs}}$} & \textbf{$N_{\text{d}}$} & \textbf{$S_{\text{min}}$} & \textbf{$D$} & \textbf{Dist. ref.} & \textbf{$L_{1284}$ limit} \\
     &  &  & (\arcmin) & & yyyy-mm-dd-hh:mm:ss & (s) & & (\uJy) & (kpc) & & (mJy kpc$^2$) \\
\hline
G028.870+0.616 & cand & \dotfill & 4 & 20240722$-$0032 & 2024$-$07$-$25$-$21:21:55 & 2364 & 44 & 36 &  &  & \\
G028.877+0.241 & cand & \dotfill & 2.6 & 20231212$-$0011 & 2024$-$05$-$01$-$05:53:17 & 2387 & 56 & 31 &  &  & \\
G028.929+0.254 & cand & \dotfill & 4.4 & 20231212$-$0011 & 2024$-$05$-$01$-$05:53:17 & 2387 & 56 & 31 &  &  & \\
G28.92+0.26 & cand & \dotfill & 6.4 & 20231212$-$0011 & 2024$-$05$-$01$-$05:53:17 & 2387 & 56 & 31 &  &  & \\
G029.329+0.280 & cand & \dotfill & 5 & 20231212$-$0011 & 2024$-$05$-$01$-$05:53:17 & 2387 & 56 & 30 &  &  & \\
G29.92+0.21 & cand & \dotfill & 4.2 & 20241024$-$0011 & 2024$-$10$-$24$-$18:17:32 & 2393 & 60 & 30 &  &  & \\
G261.9+5.5 & S & \dotfill & 40 $\times$ 30 & 20250212$-$0011 & 2025$-$02$-$14$-$18:38:29 & 2343 & 44 & 40 &  &  & \\
G296.6$-$0.4 & cand & \dotfill & 14 $\times$ 10 & 20250212$-$0011 & 2025$-$02$-$14$-$19:18:31 & 2385 & 44 & 29 &  &  & \\
G296.7$-$0.9 & S & \dotfill & 15 $\times$ 8 & 20241117$-$0009 & 2024$-$11$-$19$-$05:25:20 & 2383 & 44 & 29 & 10(1) & [7] & 3(1) \\
G296.8$-$0.3$^{a}$ & S & 2XMMi J115836.1$-$623516 & 20 $\times$ 14 & 20250212$-$0011 & 2025$-$02$-$14$-$20:39:04 & 2396 & 44 & 29 & 9.6(6) & [1] & 2.6(3) \\
G299.6$-$0.5 & S & \dotfill & 13 & 20231219$-$0009 & 2023$-$12$-$19$-$12:01:18 & 2370 & 44 & 29 &  &  & \\
G302.3+0.7 & S & \dotfill & 17 & 20250212$-$0011 & 2025$-$02$-$14$-$19:58:47 & 2382 & 44 & 29 &  &  & \\
G304.6+0.1$^{(i)}$ & S & \dotfill & 8 & 20231212$-$0011 & 2024$-$05$-$01$-$03:09:15 & 2371 & 56 & 27 & 7.9(6) & [1] & 1.7(3) \\
G310.6$-$0.3$^{(ii)}$ & S & \dotfill & 8 & 20230605$-$0014 & 2024$-$03$-$10$-$22:01:17 & 2371 & 44 & 36 &  &  & \\
G310.8$-$0.4$^{(iii)}$ & S & \dotfill & 12 & 20230605$-$0014 & 2024$-$03$-$10$-$22:01:17 & 2371 & 44 & 36 &  &  & \\
G318.2+0.1 & S & HESS J1457$-$593? \dotfill & 40 $\times$ 35 & 20231219$-$0009 & 2023$-$12$-$19$-$12:41:16 & 2383 & 44 & 32 & 2.7(4) & [1] & 0.23(7) \\
G317.3$-$0.2 & S & \dotfill & 11 & 20231219$-$0009 & 2023$-$12$-$19$-$13:21:27 & 2391 & 44 & 34 &  &  & \\
HESS J1554$-$550$^{(iv)}$ & C & G327.1$-$1.1 \dotfill & 0 & 20240524$-$0003 & 2024$-$05$-$27$-$18:45:03 & 2392 & 60 & 25 & $\sim9$ & [8] & $\sim2$ \\
G337.3+1.0 & ? & \dotfill & 15 $\times$ 12 & 20241117$-$0009 & 2024$-$11$-$19$-$06:05:45 & 2393 & 44 & 35 &  &  & \\
G324.1+0.0 & cand & \dotfill & 11 $\times$ 7 & 20250212$-$0011 & 2025$-$02$-$14$-$21:19:42 & 2392 & 44 & 32 &  &  & \\
G326.3$-$1.8 & C & PWN G326.12$-$1.81 \dotfill & 38 & 20240524$-$0003 & 2024$-$05$-$27$-$18:04:50 & 2395 & 60 & 23 & 3.5(6) & [1] & 0.3(1) \\
HESS J1626$-$490 & U & G335.2+0.1 \dotfill & 6 & 20231219$-$0009 & 2023$-$12$-$19$-$14:01:47 & 2385 & 44 & 39 &  &  & \\
G336.7$-$0.3 & cand & \dotfill & 5 $\times$ 3 & 20231212$-$0011 & 2024$-$05$-$01$-$03:49:33 & 2390 & 56 & 36 &  &  & \\
G336.7+0.5 & S & \dotfill & 14 $\times$ 10 & 20230605$-$0014 & 2024$-$03$-$10$-$22:41:23 & 2394 & 44 & 40 &  &  & \\
HESS J1634$-$472 & U & G337.2+0.1 \dotfill & 6.6 & 20231219$-$0009 & 2023$-$12$-$19$-$14:42:01 & 2396 & 44 & 45 & $>13.5$ & [9, 10] & $>8.3$ \\
G340.6+0.3 & S & \dotfill & 6 & 20250130$-$0011 & 2025$-$02$-$05$-$01:36:29 & 2370 & 44 & 37 & $\sim15$ & [11] & $\sim8$ \\
G340.4+0.4 & S & \dotfill & 10 $\times$ 7 & 20250130$-$0011 & 2025$-$02$-$05$-$01:36:29 & 2370 & 44 & 37 &  &  & \\
G341.9$-$0.3 & S & \dotfill & 7 & 20241128$-$0015 & 2024$-$12$-$02$-$12:55:07 & 2374 & 44 & 38 & 15.8(6) & [1] & 9.4(7) \\
G342.0$-$0.2 & S & \dotfill & 12 $\times$ 9 & 20241128$-$0015 & 2024$-$12$-$02$-$12:55:07 & 2376 & 44 & 38 & 15.8(6) & [1] & 9.5(7) \\
G342.1+0.9 & S & \dotfill & 10 $\times$ 9 & 20250130$-$0011 & 2025$-$02$-$05$-$02:16:21 & 2391 & 44 & 34 & $\sim6.9$ & [12] & $\sim1.6$ \\
HESS J1702$-$420 & U & \dotfill & 3.6 & 20241117$-$0009 & 2024$-$11$-$19$-$06:46:01 & 2395 & 44 & 38 &  &  & \\
G344.7$-$0.1$^{a}$ & C? & CXOU J170357.8$-$414302 & 8 & 20240612$-$0011 & 2024$-$10$-$24$-$13:41:12 & 1577 & 60 & 34 & 6.3(1) & [13] & 1.33(4) \\
G345.1$-$0.2 & S & \dotfill & 6 $\times$ 5 & 20240612$-$0011 & 2024$-$10$-$24$-$13:41:12 & 1577 & 60 & 34 &  &  & \\
G348.32$-$0.73$^{\dagger}$ & cand & G348.324$-$00.735 \dotfill & 7 & 20240722$-$0032 & 2024$-$07$-$26$-$00:02:28 & 2389 & 44 & 37 &  &  & \\
G350.0$-$2.0$^{a}$ & S & 1RXS J172653.4$-$382157 \dotfill & 45 & 20230605$-$0014 & 2024$-$03$-$10$-$23:21:45 & 2388 & 44 & 31 & $\sim3$ & [14] & $\sim0.3$ \\
G351.0$-$0.6 & cand & \dotfill & 12 & 20241128$-$0015 & 2024$-$12$-$02$-$13:35:08 & 2393 & 44 & 35 &  &  & \\
G351.2+0.1 & C? & \dotfill & 7 & 20231212$-$0011 & 2024$-$05$-$01$-$05:12:43 & 2391 & 56 & 33 &  &  & \\
G351.4+0.2 & cand & \dotfill & 18 $\times$ 14 & 20241110$-$0005 & 2024$-$11$-$11$-$06:59:14 & 2359 & 44 & 42 &  &  & \\
G351.4+0.4 & cand & \dotfill & 9 & 20241110$-$0005 & 2024$-$11$-$11$-$06:59:14 & 2364 & 44 & 43 &  &  & \\
G353.9$-$2.0 & S & \dotfill & 13 & 20250308$-$0012 & 2025$-$03$-$13$-$00:03:13 & 2374 & 44 & 33 &  &  & \\
G355.4+2.7 & cand & \dotfill & 22 & 20241110$-$0005 & 2024$-$11$-$11$-$07:39:01 & 2388 & 44 & 31 &  &  & \\
G355.9$-$2.5 & S & \dotfill & 13 & 20241117$-$0009 & 2024$-$11$-$19$-$07:26:22 & 2393 & 44 & 30 &  &  & \\
G356.3$-$0.3 & S & \dotfill & 11 $\times$ 7 & 20230605$-$0014 & 2024$-$03$-$11$-$00:02:00 & 2398 & 44 & 38 &  &  & \\
\hline
\end{tabular}
\vspace{-1.0ex}
\footnotesize{\noindent$^{(i)}$Kes 17~$^{(ii)}$Kes 20B~$^{(iii)}$Kes 20A~$^{(iv)}$The Snail}
$^{\dagger}$The SNR candidate G348.32$-$0.73 identified by the SARAO MeerKAT Galactic Plane Survey \citep[SMGPS;][]{Goedhart2024, Anderson2025} was added to the list later on, as it was noted as having a central PWN-like feature.
\end{threeparttable}
\end{table}
\end{landscape}

Over half the target list consists of SNRs in the latest version of the online SNR catalogue\footnote{Green D. A., 2024, ‘A Catalogue of Galactic Supernova Remnants (2024 October version)’, Cavendish Laboratory, Cambridge, United Kingdom (available at \href{https://www.mrao.cam.ac.uk/surveys/snrs/}{https://www.mrao.cam.ac.uk/surveys/snrs/})} \citep[][; henceforth G25]{Green2025}. Eleven targets are unidentified TeV sources, which are presented in more detail in \S\ref{subsubsec: hess}. A plurality of the candidate SNRs are from the THOR and GLOSTAR surveys. These surveys reported the largest number of new candidates. However, they also generally reported smaller shells than the more extended GLEAM or MGPS-2 discoveries. This is linked to the bias mentioned in \S\ref{subsec: datarate} that less extended targets are easier to fully tile with CBs under our observing setup. These small-diameter candidate SNRs comprise a significant portion of the survey, especially in the region $15^{\circ}<l<30^{\circ}$. 

All targets in \autoref{tab:targets2} were observed with the L-band (856$-$1712\,MHz) receivers, in the 4k channel mode and with a sampling time of 306\,\us{}. As in Paper I, the flux limits are calculated using the radiometer equation and assuming broadening due to sampling and dispersion smearing only. The terms of the radiometer equation are calculated in the same way. $T_{\text{sky}}$ is calculated at 1284\,MHz using the GlobalSkyModel2016 \citep{Zheng2017} using {\sc pygdsm} \citep{Price2016}. The duty cycle, or fractional width, of young pulsars are measured to be mostly between 1--10 per cent. We choose a duty cycle of 10\,per cent, as this corresponds to wider profiles and thus represents the smallest provision in sensitivity provided by the width correction term. The spectral signal-to-noise ratio (S/N) threshold is 9 which, for young pulsar duty cycles, is on average degraded by the FFT efficiency factor of 0.7 \citep{Morello2020}. We also consider the degradation in sensitivity due to the multibeam tiling pattern where CBs overlap at their 0.5 level, and the corresponding average sensitivity across the CB is approximately 0.65 the value at the centre. This factor is not used when calculating the upper limits in \autoref{tab:targets2}, but is considered in all subsequent analyses.

\subsection{Observations of PSR \psrB{} and PSR \psrA{} since Paper I}
\label{subsec: newobs}
\subsubsection{Timing observations}\label{subsubsec: timing}
We have continued to observe both pulsars and calculate times of arrival (TOAs) in order to constrain their rotational properties. Paper I describes the pulsar timing procedure that is used to follow up the two discoveries. PSR \psrB{} did not have a measured rotational period derivative, $\dot{P}$, following the pseudo-logarithmic observing campaign we had carried out using the MeerKAT L-band receiver. Since then, we have observed PSR \psrB{} again\footnote{To observe this pulsar we used time allocated for our pulsar searches, project code SCI-20180923-MK-03.} for 2400\,s on 27 May 2024. This epoch was chosen as it is approximately one year after the discovery observation. Thus the effect of an erroneous position on the measurement of the period, $P$, due to the Doppler shift of Earth's orbit is minimised. This observation was at L-band and utilised the FBFUSE and Accelerated Pulsar Search User Supplied Equipment \citep[APSUSE;][]{Barr2017} back ends for data acquisition.

We have continued to observe PSR \psrA{} using the Murriyang telescope at the Parkes Observatory (see Paper I for a description of these observations and data reduction techniques). The observations are 1 hr in length and have been performed on an approximately monthly cadence under project code P1054 (Initial follow-up of pulsar discoveries from MeerKAT targeted searches) and also under some Director's Time with project code PX098. The Ultra Wide-bandwidth Low-frequency \citep[UWL;][]{Hobbs2020} receiver is used, which provides a continuous bandwidth of 704-4032\,MHz, and data are captured in fold mode. The fold mode coherently dedisperses the signal at its dispersion measure (DM) of 370.1\dm{} and folds at a resolution of 3328 frequency channels, 30\,s subintegrations and 1024 phase bins. Before each observation, we perform the switched calibration procedure by integrating at a position offset of about 2\,arcsec from the pulsar for 2\,min. These data are then used for polarisation calibration later on.

Furthermore, PSR \psrA{} is in the field of two other targets of this survey, G21.8$-$0.6\footnote{G21.8$-$0.6 was observed twice, firstly on 05 February 2025 and again on 13 March 2025 after a data acquisition failure during the first observation had catastrophically affected some CBs.} and G21.86+0.169, so we were also able to place a CB on it during these pointings. These observations also used the FBFUSE and APSUSE back ends, and provided additional TOAs at L-band.

\subsubsection{Fitting timing solutions}\label{subsubsec: timingfits}
In order to derive phase-connected timing solutions for the two discoveries, we compared the TOAs against the prediction of an ephemeris that contains, among other parameters, the rotational information and position of the pulsar. The evolution of the rotation of the pulsar is approximated as a polynomial of rotational frequency terms, $f=\frac{1}{P}, \dot{f}, \ddot{f}$, which are fitted using the pulsar timing package \tempo{} \citep{Hobbs2006, Edwards2006}. In Paper I, we were only able to place an upper limit on the period derivative of PSR \psrB{} using this method. However with the latest observations we were able to make a significant measurement, which is presented in \S\ref{sec: psr-update}. For PSR \psrA{}, we see pronounced timing noise across the data even after constraining $\dot{P}$. We use {\sc run\_enterprise} \citep{Keith2022}, a pulsar timing model analysis suite based on {\sc enterprise} \citep{Ellis2020}, to model red noise (quasi-random long baseline residual variations), white noise and a possible $\ddot{f}$ term~\citep{Keith2023}. Red noise is modelled as a power spectrum of modulation frequencies, $P(\nu)$:
\begin{equation}
    P(\nu) = \frac{A^2}{12\pi^2} \left(\frac{\nu}{1\,\text{yr}^{-1}}\right)^{-\gamma} \text{yr}^3,
\end{equation}
where $A$ is the amplitude and $\gamma$ is the spectral index of the red noise. We include uncorrelated white noise processes in the model by incorporating the conventional terms {\sc efac} and {\sc equad},\footnote{These labels match those used by the European Pulsar Timing Array \citep[e.g.][]{EPTA2023} among others.} which model white noise variability of individual TOAs \citep[see e.g.][]{Lentati2016}. The TOAs are all calculated for radio frequency-averaged data, therefore we do not include the chromatic white noise term, {\sc ecorr}. The evidences for two models, one omitting and one containing $\ddot{f}$ are compared using the {\sc dynesty} sampler before each model is refined using the {\tt emcee} Markov Chain Monte Carlo (MCMC) sampler within {\sc run\_enterprise}. The results of this analysis are provided in \S\ref{sec: psr-update}.

\subsubsection{Polarimetry of PSR \psrA{}}\label{subsubsec: pol}
Fold mode observations with the UWL receiver are recorded with full polarisation information for the Stokes parameters ($I$, $Q$, $U$, and $V$), so we use these observations for the polarimetry study. The folded data from each observation were individually cleaned and calibrated before being combined, all using the \psrchive\footnote{\href{https://psrchive.sourceforge.net/index.shtml}{https://psrchive.sourceforge.net/index.shtml}}~\citep{Hotan2004,VanStraten2011} pulsar data analysis package. The data were cleaned of radio frequency interference (RFI) at the full time and frequency resolution using \psrsh{} and two masks: i) a static channel mask that covers common RFI and ii) a mask produced by hand using \pazi{}. These data were eventually combined to produce a high S/N profile, so they had to be aligned in phase. To do this, we performed polynomial whitening by taking the TOAs that had already been calculated for timing purposes (see \S\ref{subsubsec: timing}) and fitted a high order of frequency terms in \tempo{} in order to forcefully whiten the distribution of timing residuals. The data were then adjusted by refolding with the resulting ephemeris using \pam{}, thus aligning the pulse profiles. The cleaned and refolded data were then downsampled to 1664 frequency channels and all subintegrations were combined with \pam{}. Then the flux was calibrated using the \pac{} command and the Parkes flux calibration files\footnote{The Parkes Observatory provides calibration files at \url{https://www.parkes.atnf.csiro.au/observing/Calibration\_and\_Data\_Processing\_Files.html}.} that were taken closest in time to our observation. We then calibrated the polarisation information using the switched calibration file, again using \pac{}. The fully calibrated data were then combined using the \texttt{psradd} command, producing an integrated profile with a S/N of 45.

To measure the polarisation properties we used the {\sc psrsalsa}\footnote{\href{https://github.com/weltevrede/psrsalsa}{https://github.com/weltevrede/psrsalsa}} suite of pulsar data analysis algorithms \citep{Weltevrede2016}. It was first necessary to correct for Faraday rotation due to interactions between the radio waves and free electrons in the magnetised interstellar medium (ISM). The plane of linear polarisation is rotated as the wave propagates. The overall strength of the effect is quantified by the rotation measure, RM, with units of rad\,m$^{-2}$. Using {\sc psrsalsa}/{\tt rmsynth}, we performed a brute-force $\rm{RM}$ search on the 128-phase-bin pulse profile between $-1000$ and $+1000$\,rad\,m$^{-2}$ with a step size of 1\,rad\,m$^{-2}$. A significant peak in the linearly polarised S/N was seen at an $\rm{RM}$ of 401$\pm$1\,rad\,m$^{-2}$. The folded data were de-rotated using this value by using \psrchive/\pam{}. We then used {\sc psrsalsa}/{\tt ppol} to extract measurements of the linear, $L$, and circular, $V$, polarisation as a fraction of the total intensity, $I$, and also the polarisation position angle (PA) across the pulse profile.

\subsection{Results and Discussion}
\label{subsec: results+disc}
\subsubsection{Search results}
\label{subsubsec: part2-results}
No new pulsars were detected during the searches of the targets in \autoref{tab:targets2}. If we exclude the 10 CCOs from the target count, then one associated pulsar, PSR \psrA, was found by this survey across 124 targets. We searched for pulsars in a total of 13 H.E.S.S. sources (excluding HESS J1849$-$000 which already has an associated radio-quiet pulsar). Three of these are known to be PWNe based on morphology, spectra, or spatial overlap with a multiwavelength counterpart. We discuss the observations and results for those sources in more detail in the next section. We have also performed an empirical sensitivity analysis of known young pulsars in the fields of our searches, which is presented in Appendix \ref{sec: disc-known}.

If we assume a beaming fraction of 20 per cent \citep[][equation 7 for an orthogonal rotator]{Tauris1998}, then we should expect 24 pulsars to have been detectable in theory. This conclusion, of course, assumes a random unbiased selection of SNRs, which is not the case for targeted searches in practice because SNRs that already have associated pulsars are not searched (see \S\ref{subsubsec: general-disc}). The mean upper limit across all targets is 34\uJy, rising to 52\uJy{} if we consider the CB degradation factor of 0.65. This survey is therefore the largest and deepest targeted survey for pulsars in SNRs and PWNe that has been conducted so far. The next largest survey was undertaken by \citet{Kaspi1996} with a sample of 40 remnants. Our survey is approximately 5 times more sensitive than those searches. The most sensitive prior survey of SNRs is that of \citet{Crawford2002}, who searched five remnants and set upper limits approximately 2 times worse than we were able to achieve with MeerKAT. Other targeted searches of specific sources have gone deeper than TRAPUM \citep[e.g.][]{Camilo2021, Liu2024a, Ahmad2025}, though these are usually targeting X-ray PWN features rather than SNR shells. To make these comparisons, those upper limits have been converted to L-band using a pulsar spectral index of $-$1.6 \citep{Jankowski2018}. The sky temperature is one of the dominant factors that limited our nominal sensitivity. The typical value was around 20\,K, which inflates the upper limits by approximately 75\,per cent (assuming the radiometer equation and values quoted in \S\ref{subsec: newsearch}).

\subsubsection{Upper limits on H.E.S.S. TeV sources}
\label{subsubsec: hess}
In Paper I, we highlighted that targeting high-energy components of young pulsar systems has been generally a more successful method for their discovery, especially compared to searches of radio SNRs. This echoes statements by \citet{Camilo2003}, who notes that X-ray sources that are spectrally and morphologically PWN-like should be a priority for deep searches. This practice was used by \citet{Liu2024a} who used FAST to take deep observations of two X-ray sources, one associated with SNR G22.7-0.2 and the other with SNR G74.9+1.2 (CTB 87). They found a very faint 50-ms pulsar, PSR J2016+3711, associated with G74.9+1.2 with a flux density of 15.5\uJy{} at 1250\,MHz. 

In young pulsar systems, the VHE \gray{} emission is produced by energetic particles that are freshly injected by the pulsar. Thus there exists a correlation between $\dot{E}$ and the TeV brightness \citep{Abdalla2018b}, leading us to expect to find highly energetic pulsars in these sources. Motivated by the previous discoveries made targeting high-energy components of pulsar systems, putative or known, we included TeV sources from the H.E.S.S. survey \citep{Abdalla2018a} in our sample. We mainly observed sources labelled as `unidentified' as the H.E.S.S. Collaboration often ensures there is a connection to a radio-emitting pulsar before labelling TeV sources as PWNe. Interestingly, our deep targeted searches of these thirteen TeV sources did not yield any new pulsar discoveries. This is not completely implausible based on beaming alone and the size of the sample. Assuming the pulsars have a beaming fraction of 0.2, there is a $0.8^{13}\approx5$\,per cent chance that all are beaming away. Nevertheless, we provide a discussion here, separate from the searches of SNRs, regarding these targets that explores other explanations.

In \autoref{fig:hessset}, we show the locations where we placed CBs for the 13 H.E.S.S. sources. In general, we tiled out the full angular extent of the source provided by the H.E.S.S. catalogue. Where a source was too extended, we targeted portions of the emission where studies have indicated the putative pulsar is most likely to be located. We did this for HESS J1554$-$550, HESS J1634$-$472, HESS J1708$-$410, HESS J1809$-$193 and HESS J1834$-$087. It should be noted that we search the incoherent beam (IB) data too which covered the entire TeV excess in all cases. The IB data are more affected by RFI \citep{Chen2021} and by the temperature of sources in the field. If we ignore these factors, the IB is approximately $\sqrt{N_{\text{d}}}$ less sensitive at boresight than the coherent beams, where $N_{\text{d}}$ is the number of MeerKAT dishes used. Any region within the IB can therefore be considered to have been searched down to a limit of around 200-500\uJy. We note that in some fields, new candidate supernova remnants have recently been identified by MeerKAT \citep{Anderson2025} and appear to overlap some of our 13  sources. For example, G314.338$-$00.204 overlaps the entire HESS J1427$-$608 excess, and G337.167+00.332 is located within HESS J1634$-$472. This indicates that there are other regions where the associated pulsar may be located, such as within the boundary of those candidate SNRs. Further to this, it is often the case that pulsars are offset by several parsecs from the bulk of the observed TeV excess \citep{Abdalla2018b, Tsirou2019}. We therefore conclude that for some of these H.E.S.S. sources, even those where a large angular extent was searched, it could be possible that some of the pulsar locations were missed. For the 11 TeV sources listed as unidentified, the deep upper limits we set do provide some constraints on the PWN scenario as an emission origin. We note it to be highly unlikely that these chosen targets are extragalactic in origin as this is not evidenced by their previous studies in the literature, and the sources are concentrated on or near to the Galactic plane.

For two of the TeV targets, we targeted compact X-ray sources thought to be associated with the VHE emission. The first is HESS J1554$-$550, a source that is spatially coincident with the SNR G327.1-1.1. This source is classified as a PWN (`The Snail') in the H.E.S.S. catalogue, as an exception to the rule that H.E.S.S. sources were not labelled as PWNe until a pulsed radio signal was found. The PWN scenario is instead evidenced by the multiwavelength counterparts seen at X-ray \citep{Sun1999, Temim2015} and radio \citep{Whiteoak1996} wavelengths. The regions where we searched for a pulsar can be seen in \autoref{fig:hessset}. The radio tail of the PWN was tiled at an overlap level of 0.5, and the head, which is the most likely location of the pulsar, was packed tightly with beams overlapping at their 0.75 level for an improved sensitivity coverage. The upper limit in the middle of the central CB is 24.5\uJy{}. The other source is HESS J1834$-$087, which is labelled as unidentified in the H.E.S.S. catalogue. \citet{Abramowski2015} studied the source in more detail and stated that a PWN scenario is favoured. They select the X-ray point source CXOU J183434.9$-$084443 \citep{Misanovic2011} as the most likely candidate. Rather than searching the extent of the TeV excess, we targeted this X-ray point source with CBs overlapping at the 0.75 level, as can be seen in \autoref{fig:hessset}. This overlap allowed us to go deeper on a smaller region, but means we could have missed the actual pulsar if CXOU J183434.9$-$084443 is not the source of the TeV emission. 

\subsubsection{Upper limits on CCOs}
\label{subsubsec: ccos-upperlims}
No radio signals were detected in searches of the two CCOs in \autoref{tab:targets2}, as was the case for the CCOs in Paper I. Upper limits on the radio emission from CCOs have become increasingly deep \citep{Mereghetti1996, Halpern2010b, Nayana2017}. Most recently, \citet{Lu2024} observed PSR J1852+0040, the CCO in Kes 79 (G33.6+0.1), with FAST and set a limit of 2.9\uJy{} at 1250\,MHz. While not as competitive as this value, our searches provide very deep upper limits of between 25-35\uJy{} for 10 of the existing 16 CCOs and candidates. The population of these objects is still fairly small, so the sample remains too small to ascertain whether or not some or all CCOs are truly radio-quiet. Unfavourable radio beam geometries could be one scenario that explains the lack of detections, as has been posited for non-detections of pulsed X-ray signals for some CCOs \citep[e.g.][]{Suleimanov2017, Doroshenko2018}. Moreover, the pseudo-luminosity limits set by this survey are 0.2-2.6\,mJy\,kpc$^2$ (or below 0.1\,mJy\,kpc$^2$ for PSR \mbox{J1210$-$5226} and 1WGA \mbox{J1713.4$-$3949}), which are still higher than the radio emission observed for approximately 200-900 pulsars \citep{Manchester2005}. Deeper limits than those set by this survey are needed to provide more evidence about the nature of CCOs one way or another.

\subsubsection{Interpretation of the low discovery rate}
\label{subsubsec: general-disc}
The discovery of PSR \psrA\,and PSR \psrB, the detections of known pulsars, and the general success of other TRAPUM searches \citep{Padmanabh2023, Carli2024a, Prayag2024} reassure us that the TRAPUM search pipeline performs as well as expected, thus we can be certain that the low number of detections is not down to an erroneous observing setup. We are therefore left in a similar situation as \citet{Kaspi1996}, \citet{Biggs1996}, \cite{Gorham1996} and \citet{Lorimer1998}, where we have a significantly lower discovery rate than would statistically be expected when taking beaming into account. \citet{Crawford2002}, \citet{Straal2019} and \citet{Sett2021} searched smaller samples but also saw no new associated pulsars. Those authors generally argue that low pulsar luminosities is the largest selection effect, with some consideration of smearing due to scattering and dispersion. In the next section, we extend this discussion by using the scale of the survey to make robust comparisons against a population-based analysis in order to try to understand the consistently low turnout across these surveys. 

We may, however, already be in possession of some clues. During the course of the survey, two noteworthy pulsars were discovered by other searches; PSR \mbox{J1638$-$4713} \citep{Lazarevic2024} with a period of 66 ms and a DM of 1553\dm{}, and PSR \mbox{J1631$-$4722} \citep{Ahmad2025}, an 118-ms pulsar with a DM of 873\dm{} that was detected at frequencies above 2\,GHz in one of the targets of this survey: SNR G336.7+0.5. As has tended to be the case in the past \citep{Camilo2003}, these searches were able to focus on a PWN component that had already been identified, which allowed the observers to plan more targeted and, by extension, more sensitive searches. A third pulsar, PSR \mbox{J1032$-$5804} was found by \citet{Wang2024a} in image-domain searches of the Variable and Slow Transients \citep[VAST;][]{Murphy2021} survey. This 79-ms pulsar has a DM of 819\dm{} and a scattering timescale, $\tau_{\text{sc}}$, of approximately $\nicefrac{1}{3}P$ at 3\,GHz. The TRAPUM survey at L-band is not capable of discovering these pulsars due to their scattering, as was proven to be the case for PSR \mbox{J1631$-$4722}.

In addition to this, the velocity of one of the pulsars, PSR \mbox{J1638$-$4713}, could be as large as 2000\,km\,s$^{-1}$ based on the significant displacement from its birth site. More pulsars are being found far from their still-visible shells. Work by \citet{Motta2023} showed that the Mini Mouse nebula and its pulsar are located over 2 shell radii from the supernova explosion site, implying a transverse birth velocity of $\sim$300\,km\,s$^{-1}$. As we do not place CBs outside the boundary of a SNR mainly due to data rate limits, we could have missed fast-moving pulsars. Furthermore as previously mentioned, \citet{Liu2024a} discovered PSR J2016+3711 using FAST in a targeted search of an X-ray source in SNR G74.9$+$1.2. The pulsar has a DM of 478\dm{} and a predicted distance of $\sim$6.1\,kpc, though it is not scattered. This source is outside the field of our survey. However, we would likely have not detected it with MeerKAT, as the flux density of 15\uJy{} is below our sensitivity limits. Given the properties of these new young pulsars, it is plausible that some pulsars have been missed by the TRAPUM survey if they are similarly scattered, fast-moving and have a low flux. These possibilities are all explored in more detail in the next section.

\begin{figure}
    \centering
    \includegraphics[width=0.98\columnwidth]{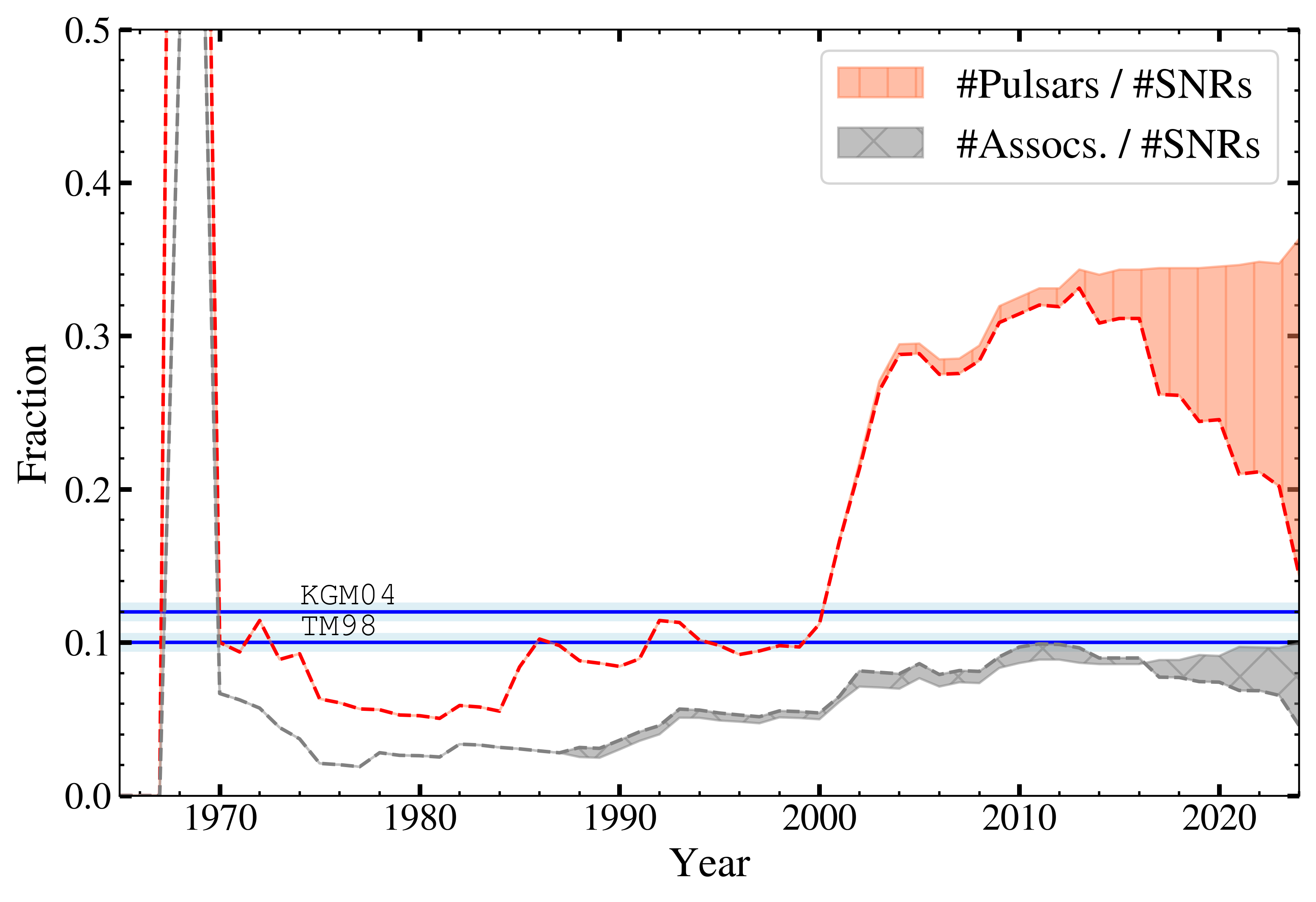}
    \caption[Fractional pulsar-SNR associations reported over time]{Fraction of the number of young ($\tau_{\text{c}}<100$\,kyr), radio-emitting pulsars against the total number of SNRs (red vertical hatchings) over time. Two beaming fractions are represented as horizontal lines: 0.12 from \citet{Kolonko2004} and 0.1 from \citet{Tauris1998}. A fraction below the line(s) indicates there are too many SNRs and not enough associations. The same fraction but explicitly calculated for those pulsars that have confirmed associations is also shown (grey crossed hatchings). The initial jump is due to the discovery of pulsars happening before SNRs began to be catalogued. Each filled region covers the difference between whether or not candidate SNRs are included in the total SNR count. The dashed edge corresponds to the inclusion of candidate SNRs. Depending on how many candidates are real, the true fractions lie within their shaded region.}
    \label{fig:snrvstime}
\end{figure}

It is worth re-examining, however, the point we made in Paper I regarding the number of young pulsars associated with SNRs. The current number of radio-emitting pulsars associated with SNRs is 38 (again, PSR \psrA{} is excluded from the count), and dividing this by the number of G25 SNRs gives a fraction of 0.123. This is equivalent to the beaming fraction predicted by \citet{Kolonko2004}, and is actually higher than the 0.1 predicted by \citet{Tauris1998}. One might reasonably ask if the reason we have not found pulsars is because they have already been found. This bias is cited by \citet{Kaspi1996} as a possible explanation for their low discovery rate. However, the inclusion of many candidate remnants in this survey changes these statistics. This is demonstrated in \autoref{fig:snrvstime}, which shows the fraction of pulsar-SNR associations over time. The evolution of this line generally follows the batches of discoveries reported by the large surveys, particularly the Parkes Multi-beam Pulsar Survey \citep{Manchester2001} between the years 2000 and 2005. As previously mentioned, most young pulsars in SNRs tend to be seen first by the large surveys, rather than targeted searches like ours. The rise in the number of pulsar-SNR associations is almost monotonic until around 2010 when it flattens out at 0.1, suggesting that the beaming limit had been reached. After 2010, we see the effect of many new candidate SNRs reported in the literature, and the fraction has roughly halved since then, with very few new associations being made. If we do not disregard the candidates, then there are currently half as many associations as there should be based on the beaming fraction, so we would expect to find pulsars in these candidates. Therefore, an initial expectation of the TRAPUM survey was that we would find many new pulsars associated with these previously untargeted locations and indeed, we found PSR \psrA{} in one of them - but only one. Also shown in \autoref{fig:snrvstime} is the nominal fraction of all young radio pulsars divided by the total number of SNRs. The interpretation of this line is that there are some pulsars missing an association, for example due to poor visibility or dissipation of the SNR.

\section{A pulsar population analysis approach}
\label{sec: pop}
The total number of radio-emitting pulsars in the Galaxy is estimated to be in the range of 10$^{5}$-10$^{6}$ \citep{Davies1973, Lorimer2006, FaucherGiguere2006}. We are only able to observe a fraction of them because the majority are beaming away or are too faint or smeared by the ISM for our instruments to be discovered. Pulsar population studies that (i) describe the complete population and (ii) produce the observed one \citep{Lyne1985, Narayan1990, Lorimer2004} are therefore an extremely useful tool for gauging the underlying properties of pulsars and the Galaxy. A common technique is to sample in a Monte Carlo fashion from a set of initial distributions, then apply an inference technique to compare the synthetic and real populations \citep[e.g.][]{FaucherGiguere2006, Gonthier2007, Cieslar2020, Ronchi2021, Graber2024, Sautron2024, Pardo-Araujo2025}. Tools exist that can generate such populations, such as the P{\sc srpop}\footnote{\href{http://psrpop.sourceforge.net/}{http://psrpop.sourceforge.net/}} package \citep{Lorimer2011b}. Population studies have limitations rooted in problems surrounding extrapolation from the known pulsar population, which is small and perhaps unrepresentative. This is particularly true for considering distances; a very small fraction of pulsars have reliable distances, so the associated systematic uncertainties leak into, for example, luminosity and velocity distributions. On a separate note, many processing hours may be required to handle a large number of pulsars, sometimes necessitating significant computational infrastructure.

Population analyses are, however, a powerful component in understanding and utilising the results of searches that are uniform in both sky coverage and integration time, such as Galactic plane surveys \citep[e.g.][]{FaucherGiguere2006, Lorimer2006, Levin2013, ColimiBernadich2023}, because biases should be minimised by a sustained sky and sensitivity coverage. Of course, ours is a targeted survey of specific locations, so is much more exposed to biases in how and where we searched. Nevertheless, there are two features of our survey that leave room for this approach. Firstly, it was mostly constrained to a very narrow section of the Galactic plane due to disfavouring or outright avoiding the footprints of other surveys. This means we have intensely sampled a particular region of the Galaxy. Secondly, a large and diverse set of targets, such as those outlined in \autoref{tab:targets2}, allows us to reasonably assume we have sampled the Galactic SNR and pulsar population within the surveyed sky area representatively. This does not contradict the first point in regards to the distribution of SNRs in the Galaxy; the vast majority of SNRs lie between $|b|\leq2\degr$ \citep[e.g.][]{Green2025}. We can therefore use the results of our survey to make inferences about the underlying population of SNRs and young pulsars. In this section, we begin with a description of the methodology of applying our survey to many synthetic populations, before going on to discuss the results and make recommendations for future study.

\subsection{Method}
\label{subsec: pop-method}
\subsubsection{Synthesising the population}
\label{subsubsec: pop-synth}
To ensure the number of pulsars reflects the real number that exist in the Milky Way, the population must be normalised using some constraint. It is common to produce pulsars at a uniform birth rate, which may be based on the observed CCSN rate \citep[e.g.][]{Dirson2022}. For this work, we use the results of the Parkes Multi-beam Pulsar Survey (henceforth PMPS) to produce a population within which an equivalent number of pulsars are seen by PMPS. PMPS is a very useful survey for this task, and has been used by others \citep[e.g.][]{Graber2024}, because the data have been intensely scrutinised by multiple follow-up analyses, so the sensitivity, parameter space and degradation of the survey are well understood. A total of 1160 pulsars have been detected in searches of PMPS data according to the ATNF pulsar catalogue \catversion{} \citep{Manchester2005}. If we exclude the 61 pulsars seen by rigorous reprocessing to search for highly accelerated binaries \citep{Knispel2013, Sengar2023}, we are left with 1099 pulsars to normalise to. The reason for excluding those 61 pulsars is that we did not search a wide enough acceleration range to detect most pulsars in binary systems. 

\begin{figure*}
    \centering
    \includegraphics[width=0.75\textwidth]{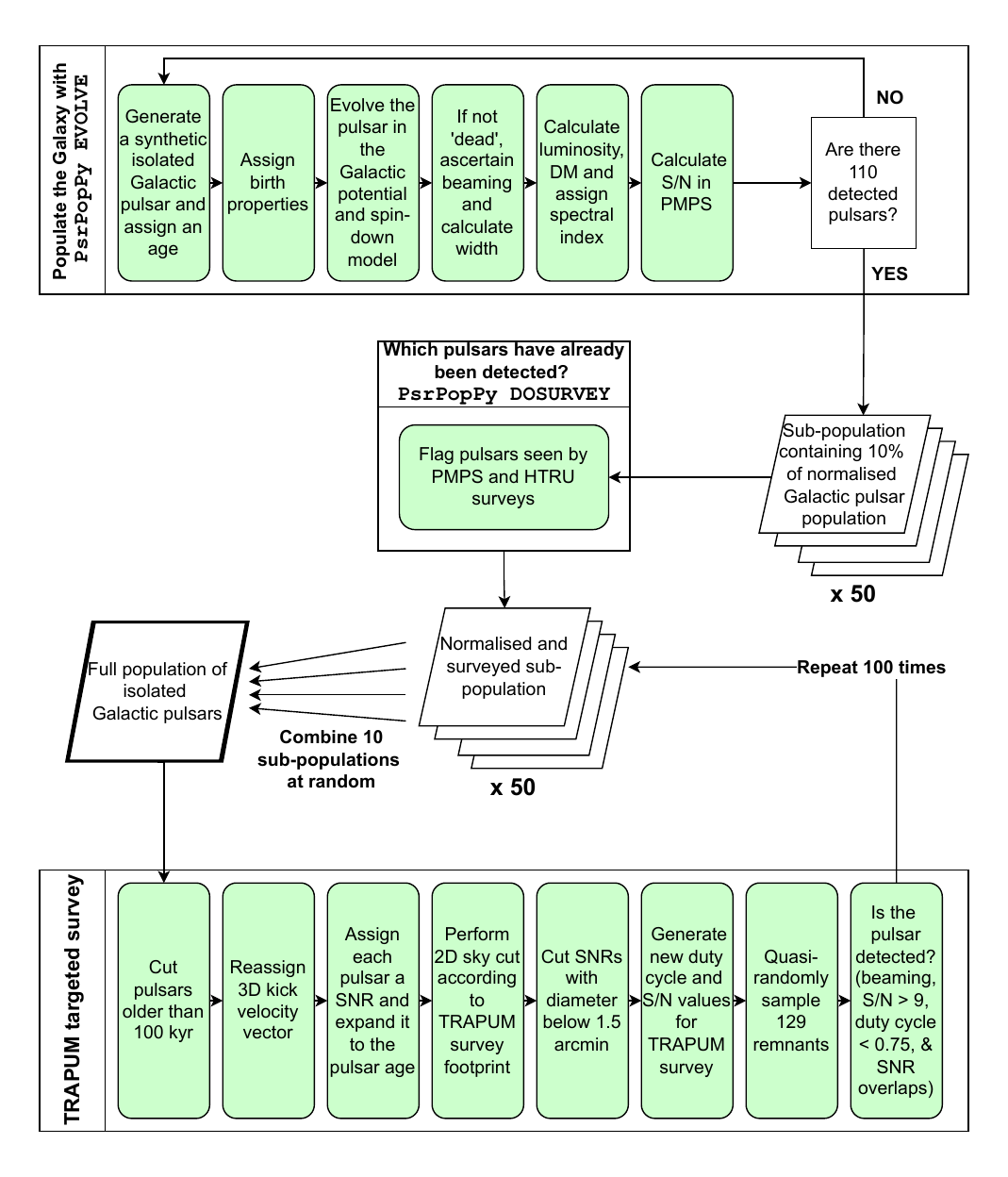}
    \caption[Population synthesis and analysis flowchart]{Flowchart showing the methodology for the population synthesis and analysis method. Each strip containing green shaded rounded boxes is a self-contained routine. This flowchart was created using \href{https://app.diagrams.net/}{draw.io}.}
    \label{fig:pop-flowchart}
\end{figure*}

\begin{table*}
\begin{center}
\caption[Model/distribution inputs to the population synthesis]{Models or distributions used as inputs for generating and evolving pulsars within \pop2/\evolve{}. For reference `default', the parameter is set out by \citet{Bates2014} and the default option in \pop2.}\label{tab:popmodels}
\begin{tabular}{llll}
\hline
Quantity & Model/Distribution & Description & Reference \\
\hline
\multicolumn{4}{c}{\textbf{Birth properties}} \\
\hline
Birth velocity, $\vec{v}_{\text{birth}}$ \dotfill & Normal & $\mu=0, \sigma=180$\,km\,s$^{-1}$ & Default \\
Birth position ($r, \theta$) \dotfill & Random & $r<15$\,kpc, $0\leq \theta <2\pi$ & This work \\
Birth position ($z$) \dotfill & Exponential & $\text{scale height}=50$\,pc & \citet{FaucherGiguere2006} \\
Birth period, $P_{0}$ \dotfill & Normal & $\mu=300, \sigma=150$\,ms & \citet{FaucherGiguere2006} \\
Braking index, $n$ \dotfill & Random & $2\leq n\leq 3$ & Default \\
Initial $B$-field, $B_{0}$ \dotfill & Log-normal & $\mu=12.65, \sigma=0.55$ & \citet{FaucherGiguere2006} \\
Magnetic inclination, $\chi$ \dotfill & Sinusoidal & $\chi=\text{cos}^{-1}(q), 0<q<1$ & \citet{Bates2014} \\
\hline
\multicolumn{4}{c}{\textbf{Evolution models}} \\
\hline
Age, $t$ \dotfill & Random & $0<t<1$\,Gyr & Default \\
Spin-down \dotfill & Function & \citet{Ridley2010}, equation 3, 6 & \citet{FaucherGiguere2006} \\
\hline
\multicolumn{4}{c}{\textbf{Other models}} \\
\hline
Width \dotfill & Log-normal & $\mu=0.05P^{0.9},\,\sigma=0.3$ & Default \\
Beaming fraction, $f_{\text{b}}$ \dotfill & Function & Equation \ref{eq: beaming} & \citet{Tauris1998} \\
Spectral index, $\alpha$ \dotfill & Normal & $\mu=-1.6, \sigma=0.35$ & \citet{Lorimer1995}; \citet{Smits2009} \\
Luminosity, L$_{\text{1400 MHz}}$ \dotfill & Log-normal in $P-\dot{P}$ & Equation \ref{eq: fk06lum} & \citet{FaucherGiguere2006} \\
Sky temperature, $T_{\text{sky}}$ \dotfill & Global sky model & GSM2016 & \citet{Price2016} \\
Electron density, $n_{\text{e}}$ \& DM \dotfill & {\sc ne}2001 & \dotfill & \citet{Cordes2004} \\
Scattering timescale, $\tau_{\text{sc}}$ $-$ & Parabolic in DM & $-6.46+0.154\text{log}_{10}\text{DM}+1.07(\text{log}_{10}\text{DM})^2$ & \citet{Bhat2004} \\
Scattering index & Constant & $-3.86$ & \citet{Bhat2004} \\
Scattering dither \dotfill & Constant & 0.8 & Default \\
\hline
\end{tabular}
\end{center}
\end{table*}

The full methodology is shown in the flowchart in \autoref{fig:pop-flowchart}, and each step is explained here. Our prescription for generating synthetic pulsars is the \evolve{} method in \pop2\footnote{\href{https://github.com/jamesdturner/PsrPopPy2\_trapumSNR}{https://github.com/jamesdturner/PsrPopPy2\_trapumSNR} by Devansh Agarwal, modified by James Turner}, a modified version of \pop\footnote{\href{https://github.com/samb8s/PsrPopPy}{https://github.com/samb8s/PsrPopPy} by Sam Bates} \citep{Bates2014}, which is itself a \python{} implementation of P{\sc srpop}. From these populations we cut all pulsars with an age above 100\,kyr which we set as the dissipation time of Galactic SNRs \citep{Frail1994, Vink2020}, because if the SNR is not visible then the associated pulsar would not be in our real survey sample. We require a knowledge of the real ages of the pulsars in order to model their SNR, hence the choice of \pop2/\evolve{} over the \texttt{snapshot} method. We modified the \evolve{} script to not reject pulsars that are beaming away, as these must remain in our sample to check which pulsars in SNRs are beaming away later on in the analysis. This introduced a problem as the already large number of pulsars that would need to be evolved through the Galactic potential differential equations \citep{Bates2014} increased ten-fold. Under the computational constraints\footnote{For reasons not fully confirmed, the run time does not scale linearly with the cumulative number of pulsars generated. So full populations could not be produced due to computational constraints. We explored ways to reduce the number of pulsars to be evolved by applying the age cut earlier, but this cannot produce a reliably normalised population.}, a solution was found. We produce a set of sub-populations normalised to 10 per cent of the number detected by PMPS (corresponding to 110 detections), and combine ten of the sub-populations at random to form a full population.

The models and distributions that are used by \evolve{} to generate and evolve pulsars are collectively summarised in \autoref{tab:popmodels}. The general method is set out by \citet{Bates2014}. Here, we reproduce this information in the context of the models that were chosen for this study. Firstly, an age, $t$ is chosen from a uniform distribution between 0 and 1\,Gyr. The birth spin period, $P_{0}$, magnetic field strength, $B_{0}$, Galactocentric position ($r$, $\theta$, $z$) and birth velocity vector, $\vec{v}_{\text{birth}}$ are sampled from their respective distributions and assigned to the pulsar. Note that the angle of $\vec{v}_{\text{birth}}$ is random and the distribution of the magnitude is the default in \pop2. In our case, the velocity is resampled later on after the age cut (see \S\ref{subsubsec: pop-survey}). A braking index, $n$, is sampled from a uniform distribution between $2\leq n\leq 3$. The magnetic inclination angle, $\chi$ is defined as the angle between the rotational axis and the magnetic axis\footnote{This angle is sometimes labelled at $\alpha$ in the literature, but $\chi$, is adopted instead as $\alpha$ appears throughout this study as the symbol for spectral index.}. We choose to follow \citet{Gullon2014} and \citet{Graber2024} and sample uniformly in $\chi$. No alignment timescale is specified, so $\chi$ is not evolved in time. The choice of $\chi$ can be important; young pulsars will have larger inclinations in a realistic scenario where $\chi$ decays over time \citep{Weltevrede2008, Young2010} and measurements of real $\chi$ values \citep[e.g.][]{Rankin1990, Posselt2021} show evidence for this. Furthermore, $P-\dot{P}$ values are dependent on $\chi$, which then go on to affect, for example, the luminosity of the evolved pulsar. 

The pulsar is then evolved to its age. A position is calculated using the Galactic potential differential equations in \citet{Kuijken1989}. The distance to Earth, $D$, and the DM are calculated. Then, following the spin-down framework of \citet{FaucherGiguere2006}, the period is calculated using equation 6 of \citet{Ridley2010}:
\begin{equation}
    P = \left[P_0^{n-1} +(n-1) \frac{8\pi R^6}{3Ic^3} B_0^2\,t\,\text{sin}^2(\chi) \right]^{\frac{1}{n-1}},
    \label{eq: fk06period}
\end{equation}
where $R$ and $I$ are the radius and moment of inertia of pulsars (conventionally $R=10$\,km, $I=10^{45}$\,g\,cm$^{2}$). From this, $\dot{P}$ is calculated using \citet{Ridley2010}, equation 3:
\begin{equation}
    \dot{P} = P^{2-n} \frac{8\pi R^6}{3Ic^3} B_0^{2}\,\text{sin}^2(\chi).
    \label{eq: fk06pdot}
\end{equation}
We note that \citet{Spitkovsky2006} showed that considering the spin-down prescription for a pulsar rotating with a plasma-filled magnetosphere instead of in a vacuum results in a substitution of $\text{sin}^2(\chi)$ for $1+\text{sin}^2(\chi)$. We do not expect our choice to have a significant effect on the properties of the young pulsar populations we are simulating, nevertheless, we briefly provide a comparison to the \citet{Spitkovsky2006} spin-down model on our results in \S\ref{subsubsec: disc-lum}. At this stage, the pulsar is assessed to check if it is radio-quiet (`dead') or not. We calculate the characteristic magnetic field strength using
\begin{equation}
    \frac{B}{\text{1\,G}} = 3.203\times10^{19}\sqrt{\frac{P\dot{P}}{\text{1\,s}}},
    \label{eq: bfield}
\end{equation}
and use the death line relation of \citet{Bhattacharya1992}, whereby the location of the pulsar on the $P-\dot{P}$ diagram must satisfy the condition:
\begin{equation}
    \frac{B}{P^2}> 0.17\times10^{12}\,\text{G\,s}^{-2}
    \label{eq: deathline92}
\end{equation}
in order to be active. If it is not, the pulsar is rejected. For each pulsar, the beaming fraction, $f_{b}$, is calculated using the empirical relation derived by \citet{Tauris1998}:
\begin{equation}
    f_{b} = 0.03 + 0.09\left[\text{log}_{10}\left(\frac{P}{1\,\text{s}} \right) -1 \right]^2.
    \label{eq: beaming}
\end{equation}
This is done by choosing a value, $x$, from a uniform distribution where $0<x<1$. The pulsar is beaming if $x<f_{b}$. We found an average beaming fraction of $\sim0.2$ for young (age $<100$\,kyr) pulsars, which fell to $\sim0.1$ for the rest of the population. The luminosity at 1400\,MHz, $L_{1400}$ is then calculated from $P$ and $\dot{P}$. It is a dithered log-normal distribution \citep[][equation 18]{FaucherGiguere2006}:
\begin{equation}
    \text{log}_{10}(L_{1400})=\text{log}_{10}\left[ 0.18 \left(\frac{P}{1\,\text{s}}\right)^{-1.5}\left(\frac{\dot{P}}{10^{-15}}\right)^{0.5}\right] + L_{\text{d}}
    \label{eq: fk06lum}
\end{equation}
where the dither, $L_{\text{d}}$, is itself normally distributed with a standard deviation of 0.8\,mJy\,kpc$^{2}$. A spectral index is assigned using a normal distribution with a mean of $-$1.6 and a standard deviation of 0.35.

The final step of \evolve{} is to check if the pulsar is detected by PMPS. The intrinsic pulse profile width, $W_0$, is calculated using an empirical log-normal distribution \citep{Bates2014}. The width observed by PMPS, $W$, is then calculated based on broadening by the predicted $\tau_{\text{sc}}$, dispersion smearing, $\tau_{\text{smear}}$, from the PMPS frequency channelisation, and the sampling time, $t_{\text{samp}}$, using:
\begin{equation}
    W = \sqrt{ W_0^2 + t_{\text{samp}}^2 + \tau_{\text{sc}}^2 + \tau_{\text{smear}}^2}.
    \label{eq: broadening}
\end{equation}
The S/N is calculated by obtaining the flux at 1400\,MHz, scaling to 1374\,MHz (the central frequency of PMPS) and using the definition of pseudo-luminosity as $L_{1400}\sim S_{1400}D^2$ and the modified radiometer equation \citep{Dewey1985}. A new pulsar is created and the process repeated until 110 pulsars are detected. $\sim2\times10^{5}$ active pulsars are typically generated. 50 of these sub-populations are produced, of which a random set of 10 are combined to produce a full population of $\sim2\times10^{6}$ pulsars. The risk of repeated sets is judged to be low, as there are $\frac{50!}{(50-10)!10!}\sim10^{10}$ unique combinations, from which we make 100 draws. 

\subsubsection{Accounting for known pulsars}
\label{subsubsec: pop-detected}
\begin{figure}
    \centering
    \includegraphics[width=0.98\columnwidth]{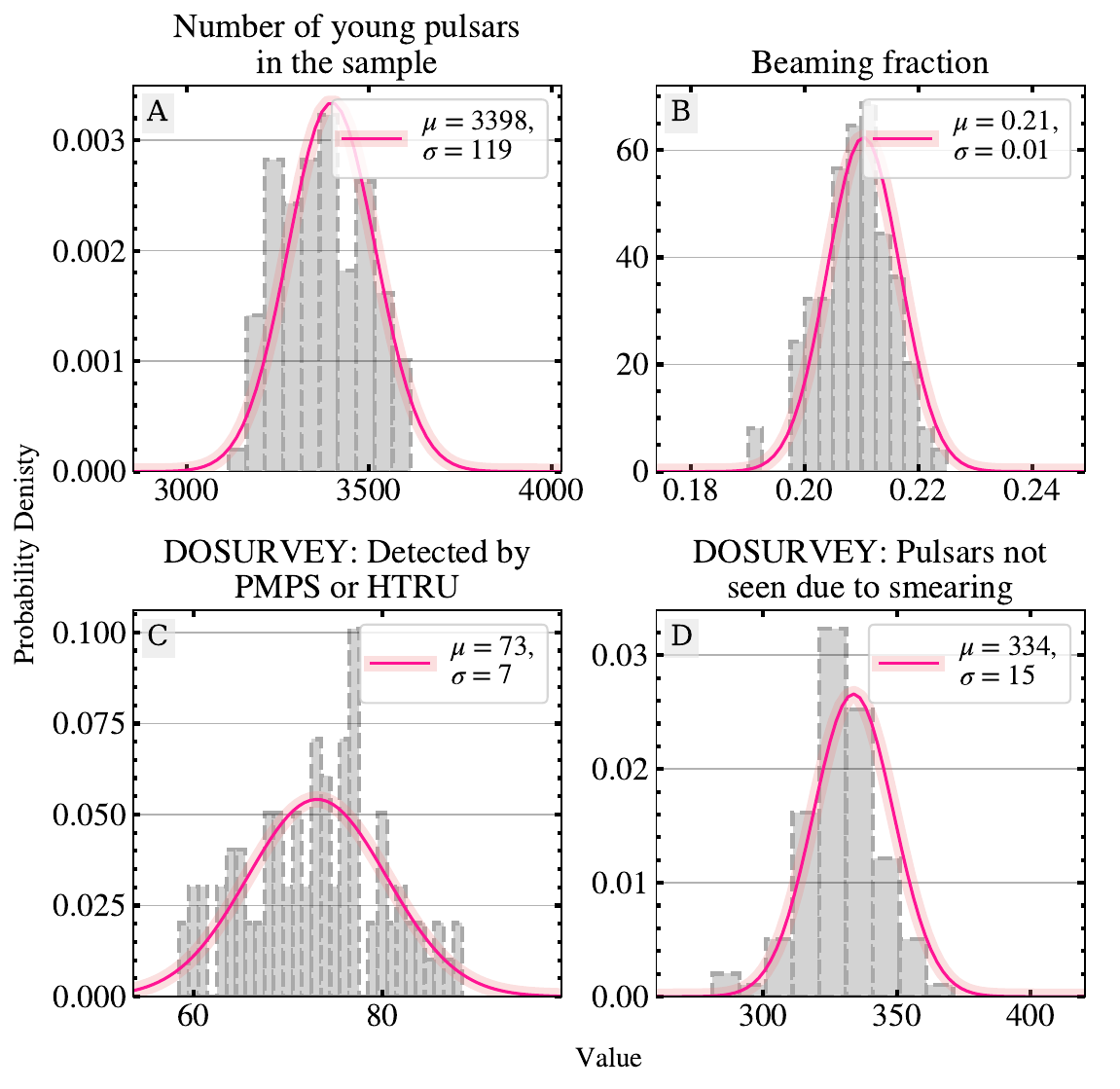}
    \caption[Properties of the synthesised population of young pulsars]{Histograms showing the properties of 100 synthesised populations after applying an age cut of 100\,kyr. On average, there are 3398 young pulsars, of which 73 are detected by PMPS and HTRU. 21 per cent of all young pulsars are beaming towards the Earth and of those, an average of 334 are too smeared to be seen by the two surveys. Each histogram has been overlaid with a normal distribution function described by the mean, $\mu$, and standard deviation, $\sigma$, of the results.}
    \label{fig:popprops}
\end{figure}

We must remove from our sample any pulsars that would otherwise have been detected by real surveys prior to TRAPUM. To do this, we pass the output of \evolve{} to the \pop2/{\tt dosurvey} routine that assigns flags to pulsars that are detected by any given survey. 74 (66 per cent) out of the 112 Galactic, radio-emitting pulsars in the ATNF pulsar catalogue \catversion{} (excluding our discovery, PSR \psrA) with a characteristic age, $\tau_{\text{c}}<100$\,kyr, were seen by both PMPS and the High Time Resolution Universe survey \citep[HTRU;][]{Keith2010}, another comprehensive pulsar survey with Murriyang. 
In the region of sky that corresponds to the majority of our observing time ($|b|\leq2$\degr, $-100$\degr$<l\leq30$\degr), this figure rises to 58 out of 71 pulsars, or 82 per cent. Recognising that a significant majority of young pulsars have been seen by PMPS and HTRU, we can reasonably assume using {\tt dosurvey} with these two surveys provides an appropriate flux density cut that accounts for known pulsars, so that those pulsars are not considered by our synthetic survey. The properties of the 100 synthesised populations and the results of \texttt{dosurvey} are shown in \autoref{fig:popprops}, for the young pulsars only. We find the populations contain an average of around 3400 young pulsars, which have an average beaming fraction of 21 per cent. We see that around $73\pm7$ pulsars are flagged as detected by the two surveys, which is consistent with the aforementioned 74 pulsars seen by PMPS and HTRU. Just under half of the pulsars beaming towards Earth are not seen due to being too smeared.\footnote{{\tt dosurvey} sets $\delta>1$ as the smearing threshold. In order to not interfere with the normalisation, we keep this value here. However, in \S\ref{subsubsec: pop-survey}, a more realistic cut of 0.75 is set when calculating TRAPUM S/N values.} Having established that our model populations can replicate the PMPS and HTRU populations, we now run the full populations through a routine that mimics the TRAPUM targeted survey.

\subsubsection{Replicating the TRAPUM survey}
\label{subsubsec: pop-survey}
\begin{figure}
    \centering
    \includegraphics[width=0.98\columnwidth]{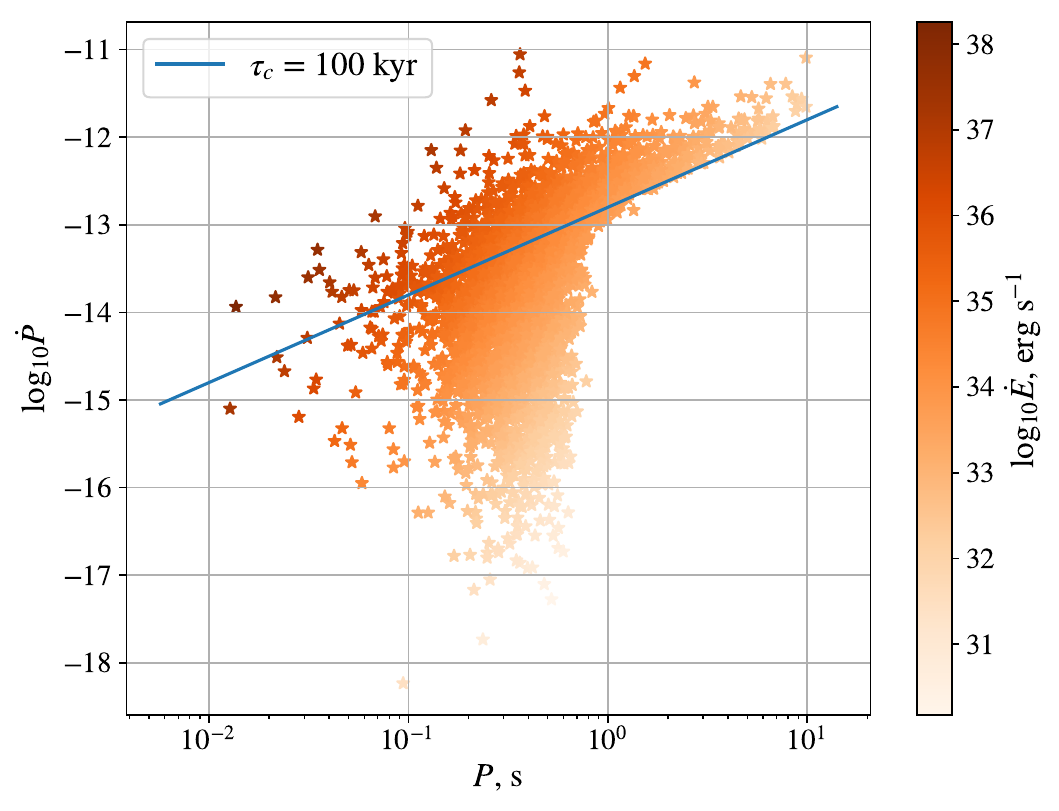}
    \caption[$P-\dot{P}$ diagram for a synthetic young pulsar population]{$P-\dot{P}$ diagram for approximately 3000 synthetic young (real age $<100$\,kyr) pulsars produced by the {\tt evolve} method under the chosen models and parameter distributions listed in \autoref{tab:popmodels}. The colours of the markers correspond to the spin-down power. The blue solid line shows the 100\,kyr characteristic age contour. The input models for $n$, $\chi$ and $B_0$ leads to a spread in $\dot{P}$ that for many pulsars results in characteristic ages much larger than their real ages.}
    \label{fig:fakeppfot}
\end{figure}
\begin{figure*}
    \centering
    \includegraphics[width=0.95\textwidth]{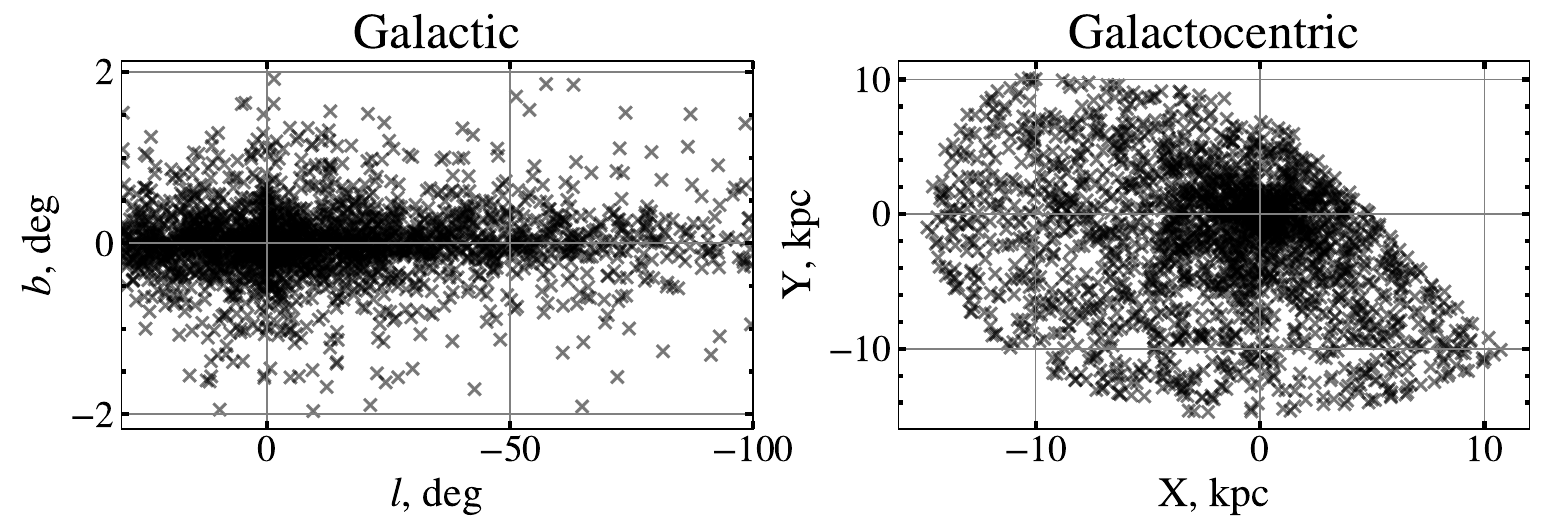}
    \caption[Distribution of synthetic pulsars within TRAPUM footprint]{Example of a distribution of the young pulsars in Galactic coordinates (\textit{left}) and looking down on the Galaxy in Galactocentric coordinates (\textit{right}) after applying the sky cut that accounts for the TRAPUM survey footprint.}
    \label{fig:pulsarlocs}
\end{figure*}

The first step of this routine\footnote{The routine is performed by \texttt{process\_population.py}, which is available in the repository \href{https://github.com/jamesdturner/PsrPopPy2\_trapumSNR}{github.com/jamesdturner/PsrPopPy2\_trapumSNR} by James Turner, a forked version of \pop2 by Devansh Agarwal.}, which is shown in the bottom strip of the flowchart in \autoref{fig:pop-flowchart}, is to cut all pulsars older than a SNR dissipation age of 100\,kyr from the population. As was seen in \autoref{fig:popprops}, this leaves $3398\pm119$ pulsars. This range overlaps with the expected number of 2000-3500 Galactic SNRs extrapolated from a SN rate of $2.8\pm0.6$ \citep[][page 36]{Vink2020}. However, it is above an estimate of between 1600-1900 from Galactic CCSN rates over $10^{5}\,\text{yr}$ ($1.63\pm0.46$\,century$^{-1}$, \citealt{Rozwadowska2021}; $1.9\pm1.1$\,century$^{-1}$, \citealt{Diehl2006}), though is within those propagated uncertainties. This difference could be due to the uniform age distribution or perhaps the choice of spin-down model (see \S\ref{subsubsec: disc-lum}) assumed for our population synthesis producing more young pulsars than exist. This could be due to e.g., a recent drop in the Galactic star formation rate, although there is strong evidence that the rate has not changed significantly over the lifetime of massive stars (10-100\,Myr) \citep[e.g.][]{Soler2023}. Ultimately, we are satisfied with these numbers and do not attempt to infer any astrophysics from them beyond stating that our full population appears to be appropriately normalised. The young pulsars from one of the populations are shown on the $P-\dot{P}$ diagram in \autoref{fig:fakeppfot}. The spread in $\dot{P}$ values is attributed to the spread in $B_0$ and $\chi$, especially for pulsars with braking indices near a value of 2.

We now introduce a different model of the birth velocities of the pulsars. At ages below 100\,kyr, the total displacement since birth is negligible compared to the size of the Galaxy, thus the effect of the prior velocity vector is not concerning. We did not change the default velocity model for \pop2/\evolve{} because it may have significantly changed the distribution of the older evolved pulsars. This could have affected the normalisation, and we wanted to be consistent with previous use of \pop. We deemed it was better to simply resample after the age cut instead. Each pulsar is reassigned a birth velocity value from the Maxwellian distribution derived by \citet{Hobbs2005}, where the 3D velocity vector has a mean of 400\,km\,s$^{-1}$ and root mean square (rms) of 265\,km\,s$^{-1}$. The pulsars are then assigned a SNR that is offset from the position of the pulsar by an amount $-\vec{v}_{\text{birth}}t$. The size of the remnant is determined by the pulsar's age, using the model described in \S\ref{subsubsec: pop-snr}. In the real survey, CBs were placed up to the edge of the shell, so if a pulsar has a sufficient transverse velocity component it will escape the shell and not be covered by the search area. If the tangential velocity is small, the pulsar may still be moving very quickly along the line of sight, but will still have been tiled by a coherent beam. At this stage, a flag is given if the pulsar is projected onto its SNR as seen from the Earth, which is taken to be at a position (0, 8.5, 0.006)\,kpc in Galactocentric coordinates. We found that approximately 9 per cent of pulsars that have left their remnants still appear to be projected onto them.

The penultimate step is to select the pulsars that will be searched by TRAPUM. A 2D sky cut selects the region $|b|\leq2$\degr, $-100$\degr$<l\leq30$\degr where $>$95 per cent of targets were concentrated. The distribution of the pulsars after this cut is shown in \autoref{fig:pulsarlocs}. Then, a radial sky cut is performed by excluding remnants with an angular diameter smaller than the smallest in the real sample (the candidate G28.56+0.00 from THOR, \citealt{Anderson2017} at 1.5\,arcmin across). This removes very distant SNRs that we would not realistically have searched in the real survey. We searched a total of 119 shell regions\footnote{A figure of 119 is found by subtracting 10 CCOs and 5 TeV sources where there was no SNR from the total of 134.}, so 119 pulsar-SNR pairs are sampled quasi-randomly; targets are twice as likely to be selected from $l>15$\degr to reflect our general preference against targets in the MMGPS footprints. In the real survey, we switched between two sampling times and either the core and full array configuration were used depending on the requirements of the search block. The search parameters assigned to the synthetic pulsars therefore match those ratios. Finally, the S/N values at 1284\,MHz are calculated. We obtain pulse widths using equation \ref{eq: broadening} and using the modified radiometer equation (\citealt{Dewey1985}; also Paper I, equation 1). We include a FFT sensitivity loss of 0.7, and also the degradation factor of 0.65 which corresponds to the average coherent sensitivity across the CB tiling. Pulsars are then flagged as detected by TRAPUM if they are beaming towards Earth, they are projected onto the SNR, the duty cycle is below 75 per cent and the S/N is above the spectral threshold of 9. The results are saved, and the process is repeated 100 times.

\subsubsection{Supernova remnant evolution model}
\label{subsubsec: pop-snr}
\begin{figure}
    \centering
    \includegraphics[width=0.98\columnwidth]{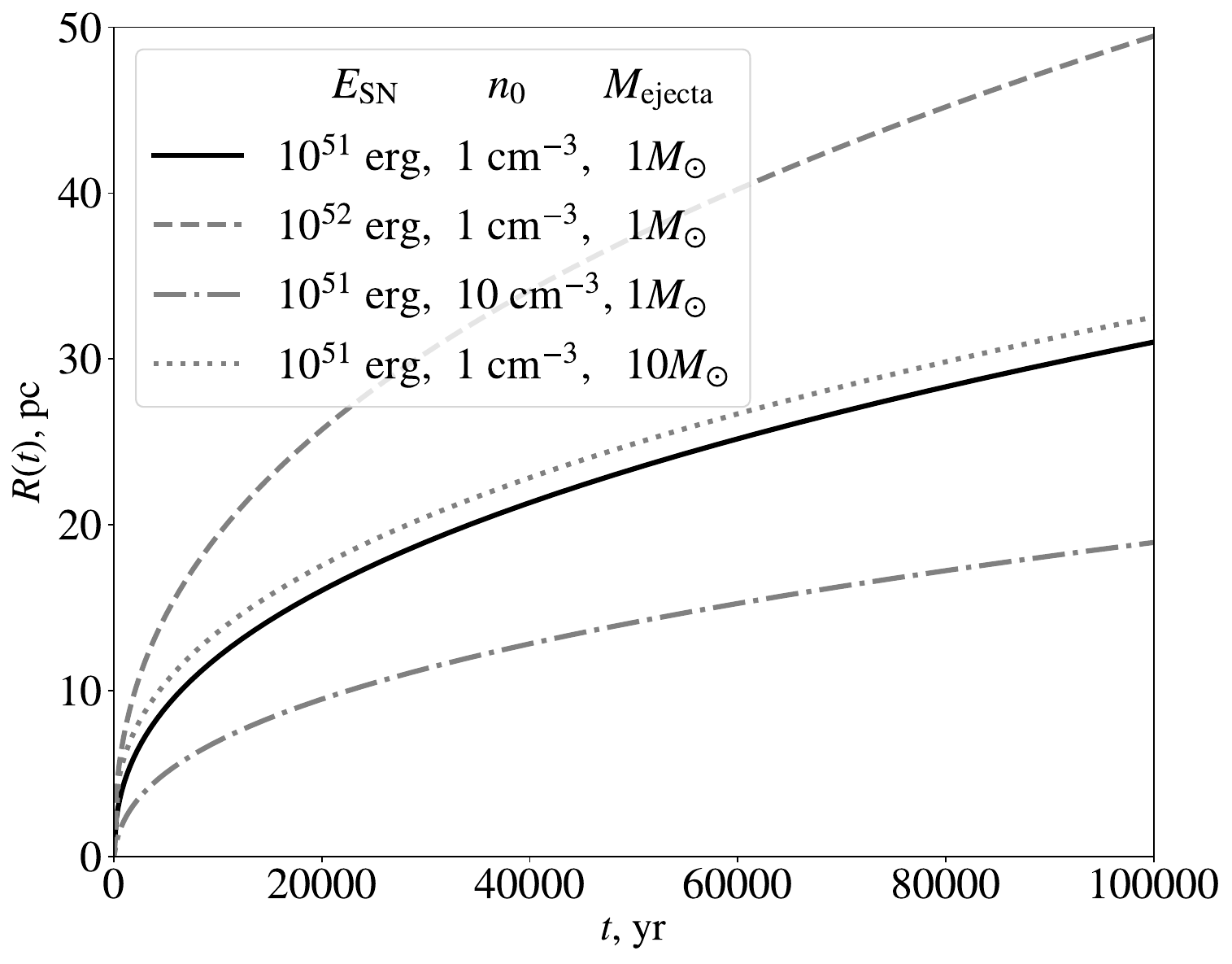}
    \caption[Models of SNR radius evolution]{Plot showing the growth of a SNR during phase 1 and phase 2 for different explosion energies, CSM/ISM number densities and ejecta masses. The chosen model is the solid line, which lies approximately between the different models.}
    \label{fig:radius}
\end{figure}

In this model, a supernova remnant is a spherical boundary that expands in time, $t$, according to a radius function $R(t)$. All pulsars are assigned a remnant retrospectively; the pulsar's real age and resampled 3D velocity is used to place a SNR at the birth site. As mentioned above, we choose the dissipation time for a SNR to be 100\,kyr, which is why we cut pulsars older than this from the analysis. Phase 1 (the ejecta-dominated phase) ends when the ejecta mass is approximately equal to the swept up mass. It can be shown that the radius at the end of phase 1 \citep[][equation 3]{Bamba2022} is
\begin{equation}
    R_{1} = 5.8\left(\frac{M_{\text{ejecta}}}{10\,\text{M}_{\odot}}\right)^{\frac{1}{3}} \left( \frac{n_{0}}{0.5\,\text{cm}^{-3}}\right)^{-\frac{1}{3}}  (\text{pc}),
    \label{eq:r1}
\end{equation}
where $n_{0}$ is the number density of the surrounding medium. During this time, we model the growth as entirely freely expanding, such that the radius expands at a constant rate:
\begin{equation}
    v_{\text{exp}} = \sqrt{\frac{2E_{SN}}{M_{\text{ejecta}}}},
    \label{eq:expvel}
\end{equation}
where $E_{\text{SN}}$ is the kinetic energy and $M_{\text{ejecta}}$ is the mass of the ejecta released by the explosion. Conventional values are adopted for the ejecta mass of 1\,M$_{\odot}$ and energy of 10$^{51}$\,erg. $n_{0}$ is set to be 1\,cm$^{-3}$ across the Galaxy. A pulsar's SNR size is therefore only dependent on its age. Phase 1 is followed by the Sedov-Taylor phase (phase 2). In our modelling, the SNR continues to expand with time as $t^{\frac{2}{5}}$ until dissipation. The chosen radius function is therefore:
\begin{equation}
  R(t)=\begin{cases}
    v_{\text{exp}}t, & \text{if}\,\,v_{\text{exp}}t\leq R_{1}. \\
    5.0\left(\frac{n_0}{1\,\text{cm}^{-3}}\right)^{-\frac{1}{5}}\left(\frac{E_{\text{SN}}}{10^{51}\,\text{erg}}\right)^{\frac{1}{5}}\left(\frac{t}{1\,\text{kyr}}\right)^{\frac{2}{5}} - \Delta R, & \text{otherwise}.
    \end{cases}
    \label{eq:snr-radius}
\end{equation}
where the second expression is equation 7 from \citet{Bamba2022} and $\Delta R \approx0.53$\,pc that accounts for the discontinuity between the two expressions. This radius function is shown in \autoref{fig:radius} where it is compared to different values of $E_{\text{SN}}$, $M_{\text{ejecta}}$ and $n_{0}$. Our choice of these values results in a sample of SNRs that represent the approximate average size. However, not including a transition to the Snowplough phase (phase 3), where the remnant expands as $R(t)\propto t^{\frac{1}{5}}$ results in a sample of larger remnants. This is amplified by setting $n_{0}$ to be a constant low value as \autoref{fig:radius} shows, as we know the density can be much higher in the Galactic plane. The effect that this bias has on our analysis is discussed in \S\ref{subsec: pop-disc}.

\subsection{Results}
\label{subsec: pop-results}
\begin{figure}
    \centering
    \includegraphics[width=0.95\columnwidth]{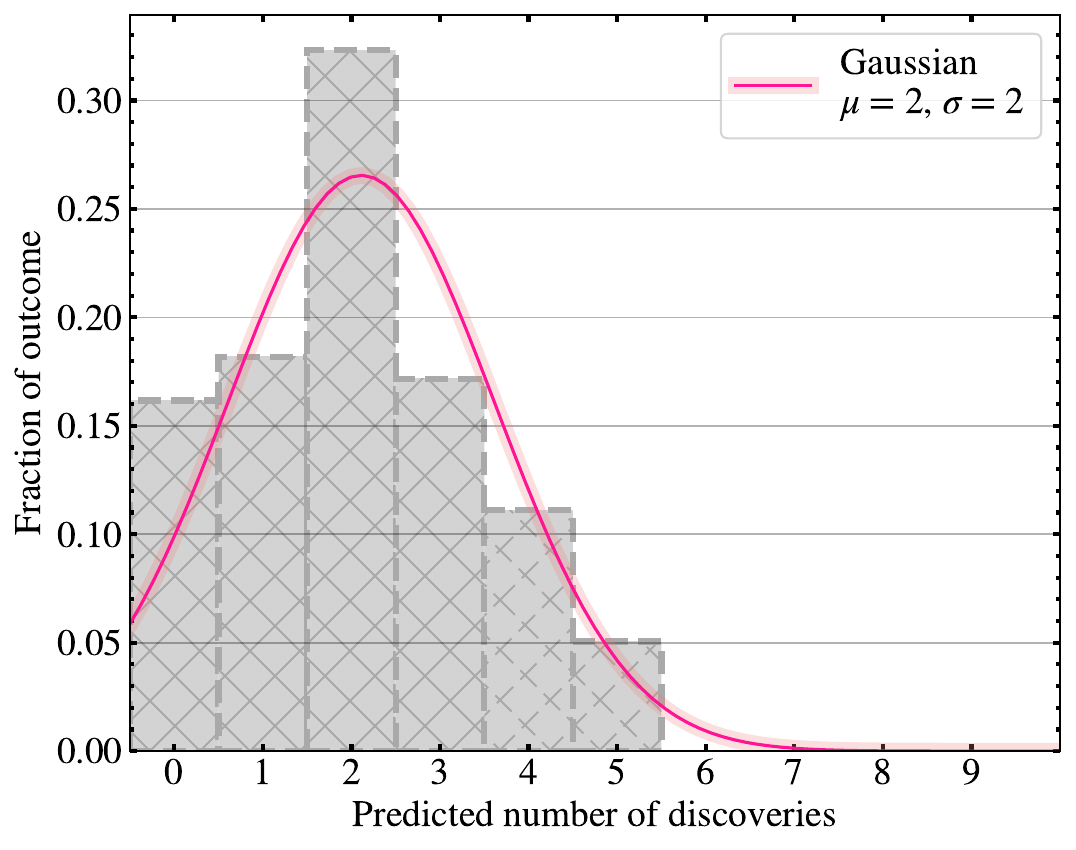}
    \includegraphics[width=0.98\columnwidth]{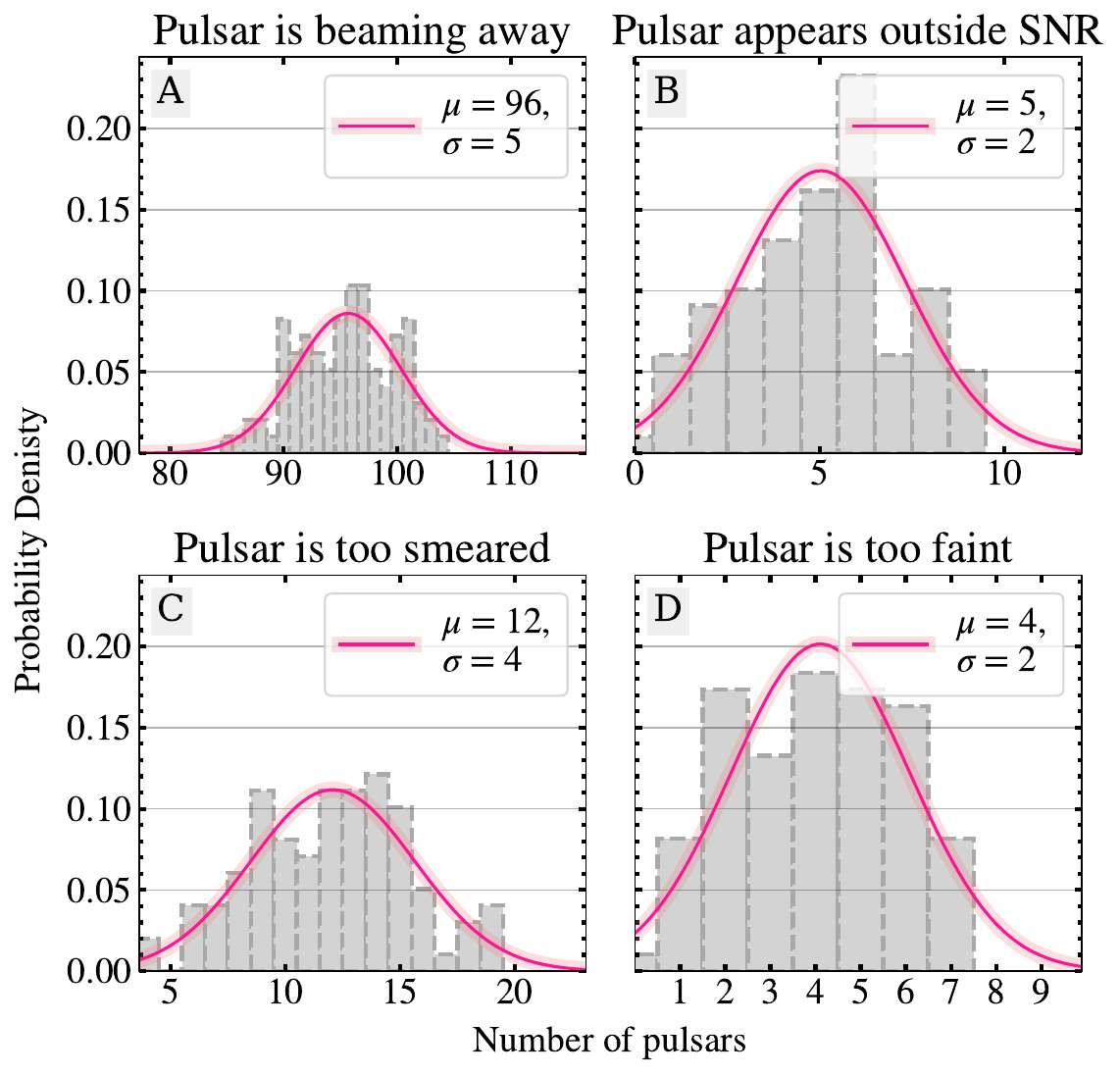}
    \caption[Predicted number of discoveries from 100 synthesised pulsar populations]{Histograms for the results of the TRAPUM SNR survey routine applied to 100 synthetic young pulsar populations. As with \autoref{fig:popprops}, the red solid lines are Gaussian functions plotted for the mean and standard deviation of each distribution.}
    \label{fig:popresultsAB}
\end{figure}
The results of the synthetic TRAPUM SNR survey routine iterated through 100 simulated populations are shown in \autoref{fig:popresultsAB}, and also listed in \autoref{tab: popresults}. The upper panel shows the prediction of $2\pm2$ new pulsars. Reminding ourselves that PSR \psrB{} is a serendipitously discovered old foreground pulsar, the real number of our discoveries is one. This scenario represents 18 of the 100 outcomes. The lower panels break down the non-detections. Panel A shows that $96\pm5$ of the pulsars would be beaming away (consistent with the beaming fraction in \autoref{fig:popprops}), panel B shows that on average $5\pm2$ pulsars are not projected onto their SNR. Then panel C shows that $12\pm4$ pulsars are smeared below the S/N threshold by propagation effects and finally panel D shows $4\pm2$ of the left over pulsars were too faint to be seen. As a quick sanity check we can sum the mean of all five panels, which returns 119 as it should. 

Of the smeared pulsars, we recalculated their S/N values again for zero smearing, and find that $6\pm2$ have sufficient flux densities to be detected. This demonstrates that, in these simulations, propagation effects actually explain 6 out of 17 pulsars that are beamed towards the Earth (henceforth `beaming pulsars') being missed. For the real survey, we chose to keep the number of frequency channels at the maximum of 4096. As was explained in \S\ref{subsec: datarate}, this required trading off either the number of coherent beams or the sampling time, but had the benefit of ensuring that DM smearing was not significant at periods above 30\,ms even at very high DMs of over 1000\dm. This, however, does not provide any defences against the smearing due to scattering. Scattering is therefore the dominant contribution to pulse broadening for all realistic DMs under the parameters of both the real and simulated surveys. Nevertheless, starting from a position of 23 or 24 potentially detectable pulsars (based on the $\sim21$ per cent beaming fraction), we find that the combined non-detections due to locations outside SNR shells, propagation effects and luminosity are able to explain our low discovery rate with a probability of 18 per cent.

In order to increase the number of discoveries, the observing frequency could be increased. The mean pulsar spectral index is negative, so the integration time would also need to be increased. To investigate this further, we reran the prescription outlined above but with survey parameters of an equivalent survey conducted with the MeerKAT S-band receivers (centred on 2406.25\,MHz). The CBs are smaller at S-band compared to L-band for an equivalent overlap level, so we reduced the number of frequency channels from 4096 to 1024. This would be a realistic requirement in order to decrease the data rate and allow for more CBs. The ISM, pulsar and receiver parameters that scale with frequency were converted using a spectral index of $-$1.6. The result was a decrease of 5 in the number of smeared pulsars. However, the number of discoveries only increased to $3\pm2$ as the number of faint pulsars increased by 4. These results are provided in \autoref{tab: popresults} against the results of the simulations at L-band. In order to match the flux density limits at L-band, the S-band integration times would need to increase by a factor of about 3.4. We recalculated the S/N values of the pulsars with $T_{\text{obs}}\times3.4$,  which resulted in a further 2 faint pulsars becoming detectable. This is also shown in \autoref{tab: popresults}.

In the next section, we compare the properties of the simulated and real samples, discuss the short-comings of the analysis in more detail, and explore what these results could be telling us about the properties of young pulsars in the Galaxy.

\begin{table}
\begin{center}
\caption{The results of the synthetic TRAPUM survey showing the magnitude of the selection effects at L$-$band, S$-$band and at S$-$band with S/N values recalculated for a factor 3.4 increase in the observing time.}\label{tab: popresults}
\begin{tabular}{lrrr}
\hline
Selection Effect & L$-$band & S$-$band & S$-$band, $T_{\rm{obs}}\times3.4$ \\
\hline
Beaming away & $96\pm5$ & $95\pm5$ & $95\pm5$ \\
Appears outside SNR & $5\pm2$ & $6\pm2$ & $6\pm2$ \\
Too smeared  & $12\pm4$ & $7\pm2$ & $7\pm2$ \\
Too faint & $4\pm2$ & $8\pm3$ & $6\pm3$ \\
\hline
Number of discoveries  & $2\pm2$ & $3\pm2$ & $5\pm2$ \\
\hline
\end{tabular}
\end{center}
\end{table}

\subsection{Discussion in the context of the survey}
\label{subsec: pop-disc}
The variance of the results is notable. In fact, the number of predicted discoveries is consistent with zero at $2\sigma$. While 100 iterations is relatively small for a Monte Carlo approach, further iterations may not change the overall picture given that we are interested in integer numbers of pulsars. The benefit of using \pop2/\evolve{} was that real pulsar ages are considered, which could then be used to model the SNRs. However, the extra physics that \pop2 requires for a cohesively modelled and normalised population of pulsars could have injected too much complexity and inflated the final uncertainty. Future modelling of pulsar-SNR populations might use a more selective approach to how pulsars are modelled. We now consider individually some specific results and possible biases and limitations of this work.
\subsubsection{Distances}
\label{subsubsec: disc-dist}
\begin{figure}
    \centering
    \includegraphics[width=1.0\columnwidth]{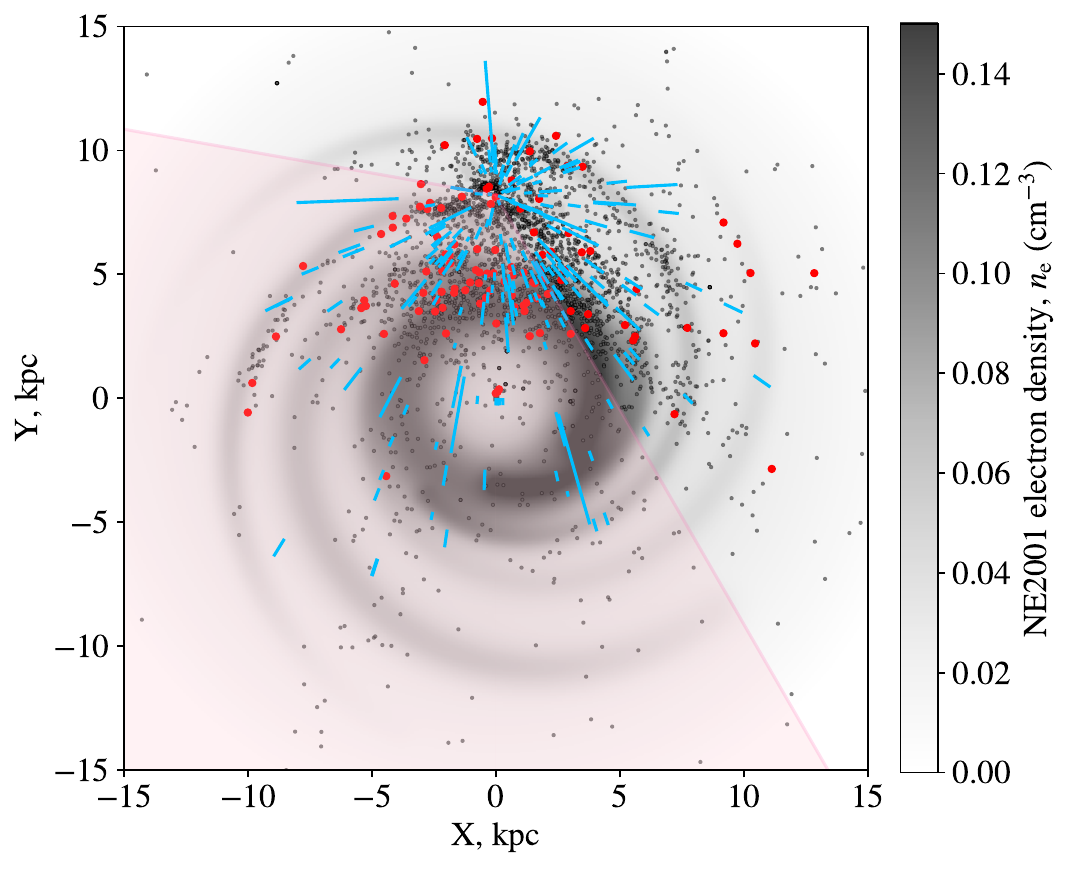}
    \caption[Galactic positions of pulsars and supernova remnants]{Positions of known pulsars (dots) in the Milky Way shown against the {\sc ne}2001 model. Young pulsars ($\tau_{\text{c}}<100$\,kyr) are highlighted as larger red dots. The positions of known supernova remnants that have distance measurements are shown as blue lines corresponding to the most up to date uncertainty or range to have been reported in the literature. The main survey region $|b|\leq2$\degr, $-100$\degr$<l\leq30$\degr{} has been shaded pink. Pulsar positions are acquired from \catversion{} of the ATNF pulsar catalogue \citep{Manchester2005}. This figure is based on Figure 4.2 of \citet{Lyne2022}.}
    \label{fig:psrsnrposn}
\end{figure}

At the start of this section, we singled out the distance as a limitation of population studies when translating between the properties of a real and synthetic pulsar population. For the majority of real pulsars, the most accurate estimate of their distance is derived from $n_{\text{e}}$ models, with realistic uncertainties of 30 percent, maybe even 50 per cent on the Galactic plane \citep{Deller2009, Yao2017}. This is a similar, if not worse, issue for SNRs as distances are arduous to measure with high precision. In \autoref{fig:psrsnrposn}, we show the best positions for the young pulsar and SNR populations in the Galaxy. They can both be seen to cluster along the spiral arms and in the central bulge. The vast majority of these known sources are located on the near side of the Galaxy, as is expected given that the detectability of both objects degrades significantly with distance.

In \pop2/\evolve{}, we populated the Galactic disk with young pulsars and SNRs uniformly up to a radius of 15\,kpc. Our angular diameter cut at 1.5\,arcmin filtered out $\lesssim1$\,per cent of supernova remnants located on the far side of the Galaxy. This resulted in an average distance of 11.5\,kpc to the pulsar-SNR pairs. The real average for all known SNRs in \autoref{fig:psrsnrposn} is 5.2\,kpc. But for those that we targeted, we find the average is 7.5\,kpc for the 40\,per cent that have measured distances. For known young ($\tau_{\text{c}}<100$\,kyr) pulsars, the average distance is 8.3\,kpc assuming the {\sc ne}2001 model. Both are smaller than for what we simulated, so therein lies a bias towards apparently faint and highly scattered pulsars in the simulation. Our real sample of SNRs was chosen to be as diverse as possible, and as a result contains a large number of candidate supernova remnants which have no distance measurements. Can we infer anything about the distances of the candidate supernova remnants in our sample? It is non-trivial to infer the distance to a SNR based on its flux density or surface brightness, the latter of which is independent of the distance. Instead, the brightness can be loosely connected to the size of the remnant by using the surface brightness-diameter relation \citep{Clark1976}. These variables do show evidence of a correlation \citep[e.g.][]{Green2025} as would be expected if the radio brightness of the shell fades over time. The scatter around the correlation however makes it an extremely unreliable tool for estimating distances. With this being said, applying the basic relation to the fluxes and angular sizes of all the candidates discovered by the GLOSTAR \citep{Dokara2021} and THOR \citep{Anderson2017} surveys, we find those targets are on average located at larger distances compared to those in \autoref{fig:psrsnrposn}. On the other hand, the candidates from GLEAM \citep{Hurley-Walker2019b} do not appear to be further away. It could be that we have sampled pulsar locations on the other side of the Galaxy by choosing to observe many GLOSTAR and THOR candidates, but it is highly unlikely that we sampled the disk up to 15\,kpc as uniformly as in the simulation. In ensuing sections, we discuss how the discrepancy between the distances of the real and synthetic pulsar-SNR populations has affected other modelled properties of our pulsar-SNR populations.

\begin{figure*}
    \centering
    \includegraphics[width=0.75\textwidth]{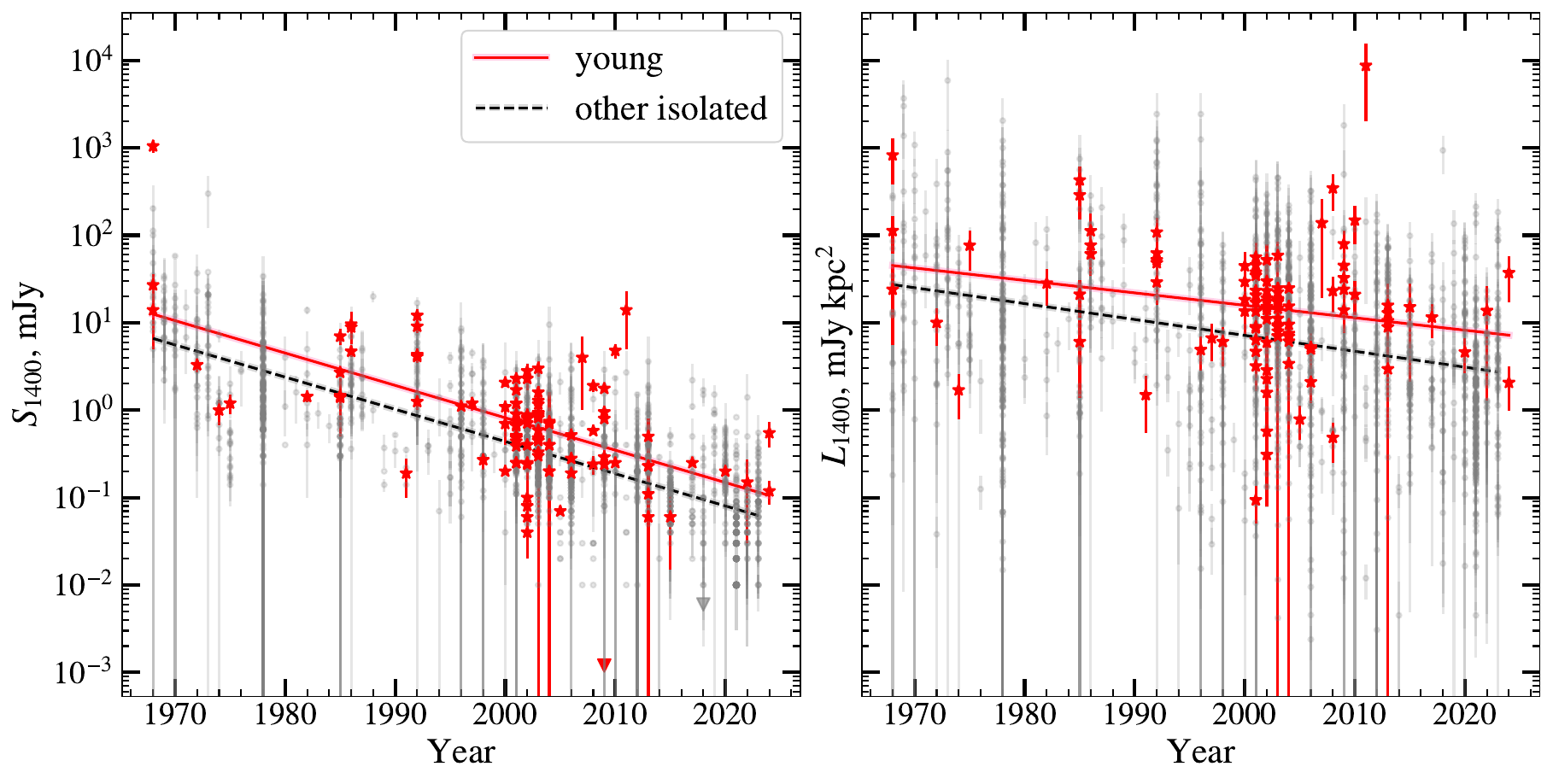}
    \caption[Flux density and pseudo-luminosity of pulsar discoveries over time]{Flux densities ({\it left}) and pseudo-luminosities ({\it right}) of pulsars at 1400\,MHz against the year they were reported in the literature. The young pulsars (red stars) are differentiated from the other isolated pulsars (grey dots). Each of these are fitted with a straight line using the \python{} module {\tt numpy.polyfit} for the purpose of showing the trend. Upper limits are represented as downward triangles. $S_{1400}$ and the `best' distances used to calculate $L_{1400}$ were obtained from the ATNF pulsar catalogue \catversion{} \citep{Manchester2005}. Uncertainties are shown at their $3\sigma$ value.}
    \label{fig:lumtime}
\end{figure*}

\subsubsection{Misidentification of Type II SNRs}
\label{subsubsec: disc-misid}
Our analysis has so far made no consideration of remnants that do not have an associated pulsar. Examples of this include those produced by Type Ia SNe, pair-instability SNe and, even more exotically, CCSNe that produce black holes. Indeed, black holes can be associated with SNRs, as was observationally proven to near certainty by the stellar black hole binary system \mbox{SS 433} located inside SNR W50 \citep{Ryle1978}. Although we excluded from our sample any SNRs with evidence for a Type Ia origin, we cannot rule out that some were observed. Type Ia SNe that have produced understudied, poorly visible or candidate SNRs may especially be vulnerable to being included in our sample. If we take the estimate of 18 per cent of SNRs to be Type Ia \citep[][for spiral galaxies]{Li2011}, and 1 per cent to be pair-instability \citep{Heger2002, Heger2003}, the worst case scenario is that a fraction of $0.18+0.01$ of our sample have no pulsar. We also do not account for sources such as HII regions, bubbles or overlapping filaments that are misidentified as SNRs. Powerful radio imaging surveys are probing remnants located along ever more confusing and densely populated lines of sight. Misidentification is therefore even more likely to be important for the candidate SNRs, which constitute a significant portion of the TRAPUM survey. We can attempt to quantify this by looking at successive catalogue publications up to and including G25 \citep{Green2009,Green2014,Green2019,Green2022}. We find that 14 out of 310 sources were mistakenly categorised as SNRs. If we believe all confirmed SNRs but bluntly apply this ratio to the sample of candidate SNRs, then the fraction of targets without a pulsar becomes $0.18+0.01+0.02=0.21$. If we retroactively apply this fraction to the result of each population, this raises the probability of finding one new pulsar in the simulated survey from 0.18 to 0.28.

\subsubsection{CCOs and magnetars in the real sample}
\label{subsubsec: disc-ccos}
In \S\ref{subsubsec: ccos-upperlims}, we outlined the deep upper limits we set on the flux density and luminosity on pulsed radio emission on six CCOs and four CCO candidates, adding to the evidence that these objects are radio-quiet in nature. 
Magnetars, like CCOs, are also located in supernova remnants. They emit hard X-rays and \grays{}, and are not always radio-quiet \citep[e.g.][]{Levin2010}. However their radio activity is highly variable and it is unlikely that we would detect one in a 40 minute observation. Given that our analysis does not account for radio-quiet NSs, it is worth examining the level of bias they may introduce. This is particularly important for the sources deep in the Galactic plane located on the far side of the Milky Way. The absorption of soft X-rays from CCOs by the gaseous, molecular and dust phases of the ISM \citep[][and references therein]{Wilms2000} could be attenuating them below detection thresholds. Those SNRs might then be included in our sample. CCOs and magnetars constitute a significant portion of the NS-SNR associations that have been identified. 23 non-radio Galactic neutron stars listed in the ATNF pulsar catalogue \catversion{} \citep{Manchester2005} have SNR associations. Including the SNRs of the 13 CCOs without a pulsed detection, 36/310 SNRs have radio-quiet NSs. \citet{Lorimer1998}, \citet{Straal2019} and \citet{Sett2021} invoke radio-quiet NSs as partly responsible for the non-detections of their searches. We consider this a possibility here due to searching far-away targets, but it is not required to explain the low discovery rate according to the simulation.

\subsubsection{Luminosity}\label{subsubsec: disc-lum}
Normalising to the results of PMPS required using \pop2 with input models that describe the canonical population well, as these constitute the bulk of active pulsars that are simulated by \evolve{}. A possible source of uncertainty in our luminosity modelling may come from the spin-down model chosen. As mentioned in \S\ref{subsubsec: pop-synth}, \citet{Spitkovsky2006} suggest that a spin-down model for a plasma-filled magnetosphere results in $\dot{P}$ up to 2 times higher for oblique rotators ($\chi\rightarrow90$\degr) compared to the pulsars we have simulated. This has two consequences for our chosen prescription. Firstly, our simulated oblique pulsars evolve more slowly in $P-\dot{P}$ space and, secondly, their luminosities will be lower (see equation \ref{eq: fk06lum}). Given that we normalise each population using a pulsar survey, this may result in more pulsars than necessary being simulated to recreate the PMPS results, as they will be fainter on average. A lower average luminosity could thus be inflating the importance of luminosity as a selection effect. However, we do not expect this to be introducing a more significant uncertainty than the pulsars distances, for which we now provide a discussion.

The observed luminosities of real young pulsars are similar to the rest of the canonical population \citep[e.g.][]{Kramer2003}, so on that basis we would expect the luminosities of our synthetic young sample to be well modelled. The flux cut applied by \texttt{dosurvey} will remove the brighter or nearby beaming pulsars. Also, given that the distances to the simulated pulsars are likely larger than for our real sample, we are biasing towards pulsars with a low flux density in the analysis. If the simulated pulsars were closer to the measured distances of our real sample, the average flux would increase by a factor of $(\frac{11.5\,\text{kpc}}{7.5\,\text{kpc}})^{2}\approx2$ and the predicted number of discoveries would increase. However we note that the 7.5\,kpc value is poorly constrained as only 40\,per cent of our observed sample of SNRs have a distance estimate as outlined above.

It is important to note that a faint pulsar could be just bright enough to surpass our threshold, but then be scattered by a small amount and drop below the detection threshold. In our population analysis, the duty cycle cut for smearing is applied before the S/N is calculated, so we took the pulsars that were too smeared and recalculated their non-smeared S/N values. We find that $6\pm2$ out of the 12 smeared pulsars would be bright enough to have been seen at the TRAPUM sensitivity. Therefore, the number of beaming pulsars that are projected onto their SNR but that are unable to be detected due to having a low luminosity is $10\pm3$.

After beaming, intrinsic faintness was suggested as the dominant selection effect for the searches by \citet{Kaspi1996}, \citet{Gorham1996} and \citet{Sett2021}. Our findings therefore align with these statements as, despite smearing accounting for a plurality of the non-detections, a total of $10\pm3$, or 30-56 per cent of beaming pulsars are too faint in a scenario with no smearing. Of course, reducing the effects of scattering would require observing at a higher frequency, for which it would be harder to achieve the same sensitivity limits. The average luminosity upper limit for the targets with distances in Table 2 of Paper I and \autoref{tab:targets2} is $\sim0.8$\,mJy\,kpc$^2$. This is an improvement over past searches. For example, \citet{Lorimer1998} achieved a mean limit scaled to L-band of $L_{1284}=22$\,mJy\,kpc$^{2}\times(\frac{1284}{400})^{-1.6}\approx3.4$\,mJy\,kpc$^2$. Those authors state that a large population of very faint pulsars is not needed to explain their non-detections. We find this to generally be true for our survey too, recognising the problem as it relates to the volume of the Galaxy that is sampled. As sensitivity improves, the distances being probed increase, whereas the volume scales as the square of the distance if we assume the Milky Way to be a flat disk. Thus if the TRAPUM survey probes the far side of the Galaxy, we do not need there to exist a large population of very faint pulsars to explain the low detection rate, because this survey will realistically only be sensitive to the brightest pulsars in that region. A corollary of this is that the luminosity distribution of young pulsars must already be well-sampled. To check this, the flux densities of young and normal pulsars, and their inferred pseudo-luminosities, are shown against the year they were discovered in \autoref{fig:lumtime}. Trend lines are also fitted to show how the brightness of discoveries evolve over time. We see that the flux continually drops as surveys become more sensitive. However the luminosity trend is much shallower which demonstrates the volume problem; probing further across the Galaxy increases the average luminosity of the pulsars that will be found.

\subsubsection{Birth velocities}
\label{subsubsec: disc-vel}
Under the models and assumptions of the simulation, 5 beaming pulsars, approximately a quarter, are not projected onto their SNRs due to their sufficiently high transverse birth velocity. This is a very interesting result as we now have an idea of the level of importance escaped pulsars have when counting non-detections. There are two important caveats to note. Firstly, the SNR expansion model results in wider SNRs than would realistically exist on average. The real shapes of SNRs can significantly diverge from a spherical shell, as many are elliptical or have uncertain shapes or poorly defined edges \citep{Green2025}. Assuming a constant $n_0$ across the entire Galaxy is also unrealistic. We could have dithered $n_0$ to reflect the variation in the ISM density, though this would likely have just inflated the final uncertainty. It may instead have been useful to tie $n_0$ to the {\sc ne}2001 model or at least to first order apply a scale height dependence. Furthermore, the choice to hold $n_0$ constant does not take into account issues regarding the visibility and expansion rate of SNRs depending on the CSM/ISM density. The detectability of remnants in dense environments could be enhanced by an increased surface brightness. However the visibility timescale is expected to be shortened in turn \citep[e.g.][]{Sarbadhicary2017}. 

The second caveat is that the birth velocity distributions can be biased towards high velocities due to the difficulty of measuring a smaller proper motion. The chosen birth velocity distribution of \citet{Hobbs2005} has been suggested by \citet{Igoshev2021} and more recently \citet{Disberg2025} to overestimate birth velocities. \citet{Igoshev2021} considered binary formation channels and, while we are only interested in synthesising isolated pulsars, we recognise that some real SNRs in our sample may be from the CCSN of a star that was or remains in a binary. In any case, if the velocities of the simulated pulsars are larger than in the real sample, this is offset by our remnants being larger than is realistic. Motivated by the recent discoveries of high-velocity pulsars located far outside their SNR edges \citep{Motta2023, Ahmad2025}, we state that it is possible that 5 non-detections due to escaped pulsars could actually be an underestimate.

\subsubsection{Scattering}
\label{subsubsec: disc-scatt}
\begin{figure}
    \centering
    \includegraphics[width=0.98\columnwidth]{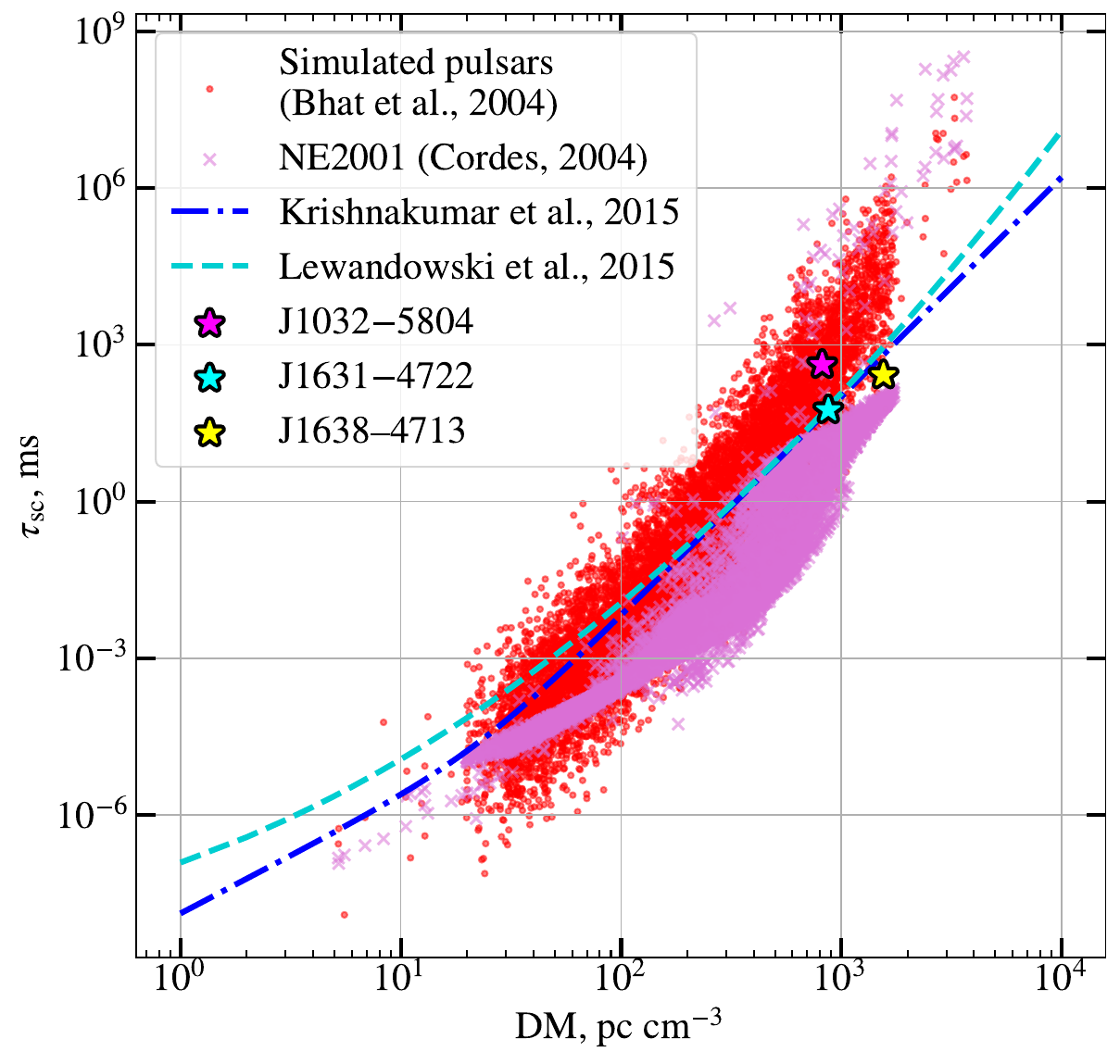}
    \caption[Comparison of models for the $\tau_{\text{sc}}$-DM relation]{Scattering timescale at 1400\,MHz against DM for 10,000 pulsars selected at random from one of the simulated populations (red dots). The scattering timescales were recalculated using the {\sc ne}2001 model. These are shown as magenta crosses. Two empirical relationships are also shown. The dark blue dot-dashed line from \citet[][Figure 3]{Krishnakumar2015} was fitted against measurements of $\tau_{\text{sc}}$ at 327\,MHz. The logarithmic parabolic relation from \citet[][equation 8]{Lewandowski2015} is calculated for $\tau_{\text{sc}}$ at 1\,GHz is shown as the light blue shaded dashed line. This expression appears to be printed erroneously by the authors, and we ascertain that the correct expression is: $\text{log}\,W_{\text{scatt}} \text{(ms)}\,=\,-6.344\,+\,1.467\,\text{log DM}\,+\,0.509\,(\text{log DM})^{2}$. Due to an oversight, we used the erroneous expression for Figure 2 of Paper I, which consequently underestimates scattering. These values are all scaled to 1400\,MHz using the scattering index of $-3.86$ from \citet{Bhat2004}.}
    \label{fig:scatt_fits}
\end{figure}

The importance of scattering as a selection effect for the simulated beaming pulsars is notable, though not completely unexpected given how we have populated the simulated pulsars in the Galaxy. $\tau_{\text{sc}}$ is calculated from the DM, which itself is assigned using the {\sc ne}2001 model of $n_{\text{e}}$. The scattering timescale is therefore very sensitive to the distances of the pulsars, which as explained in \S\ref{subsubsec: disc-dist}, are larger on average than the 40\,per cent of targets in the real sample that have distances. \autoref{fig:scatt_fits} shows the $\tau_{\text{sc}}$-DM relation for 10,000 pulsars from one of the simulated populations. Two values of $\tau_{\text{sc}}$ are shown for each pulsar: the value from \citet{Bhat2004} and the prediction from the {\sc ne}2001 model. We also show two empirical parabolic relationships and the three highly scattered young pulsars that were discovered while the TRAPUM survey was being undertaken: PSR J1032$-$5804 \citep{Wang2024a}, PSR J1631$-$4722 \citep{Ahmad2025} and PSR J1638$-$4713 \citep{Lazarevic2024}. The figure also shows that at DMs above 1000\dm, the \citet{Bhat2004} relation sits approximately one or two orders of magnitude above the empirical and {\sc ne}2001 predictions, which could be further increasing the effect of scattering for the simulated sample.

If we recalculate S/N values by reducing the scattering timescale of all the pulsars by an order of magnitude, the number of predicted discoveries increases to $4^{+2}_{-1}$, with 16 per cent of outcomes resulting in 7 new pulsars.  
As can be seen in \autoref{fig:psrsnrposn}, the region $15^{\circ}<l<30^{\circ}$ which was the focus of the survey, is directed along an entire edge of the central Galactic density torus. A significant number of our targets are thus likely to be located along lines of sight with a considerable scattering timescale. The $\tau_{\text{sc}}$-DM relation in this region is explored by \citet{He2024}. They find evidence that the relation can be decomposed into a local and an inner disk contribution to $\tau_{\text{sc}}$. The centre of the correlation for the inner disk sits around three orders of magnitude higher than the correlation for local pulsars. If the scattering on these sight lines is very strong, then it may explain why we only saw one pulsar in a sample of targets with a smaller average distance than these synthetic pulsars. This would reduce the aforementioned bias to high scattering timescales from using the \citet{Bhat2004} relation and the high average distance to the synthetic pulsars.

\begin{figure}
    \centering
    \includegraphics[width=0.98\columnwidth]{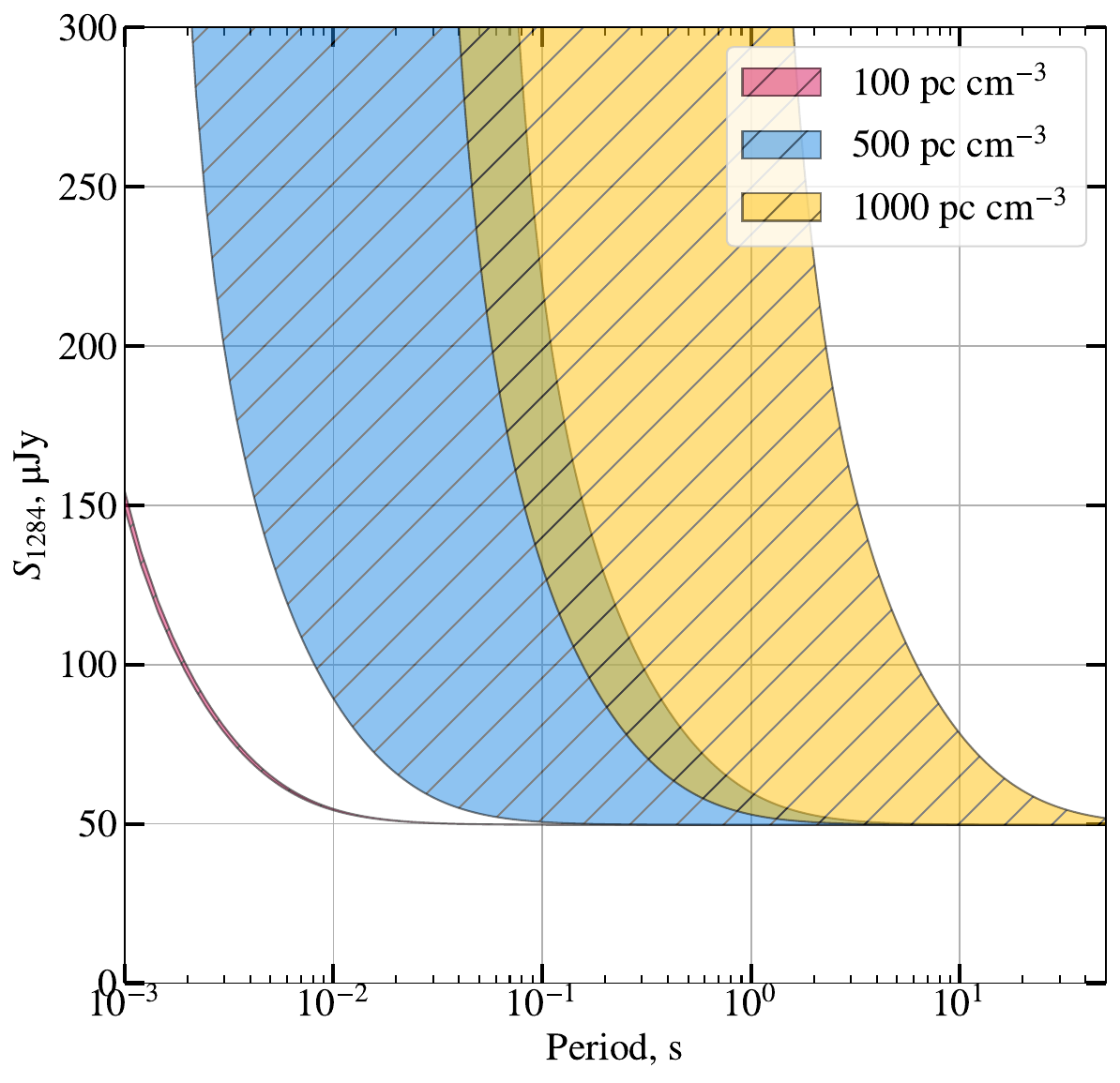}
    \caption[Effect of scattering on the sensitivity of the TRAPUM survey]{Flux density limits of the TRAPUM survey at the central frequency of the L-band receiver at DMs of 100\dm, 500\dm{} and 1000\dm. The shaded regions correspond to the limit calculated for log$_{10}(\tau_{\text{sc}})\pm0.8$.}
    \label{fig:smin-dither}
\end{figure}

The dither of 0.8 applied to log$_{10}(\tau_{\text{sc}})$ in the simulations is intended to capture the spread around the correlation between $\tau_{\text{sc}}$ and DM. This spread can be extremely important for our results. In \autoref{fig:smin-dither}, we show a plot of the survey's sensitivity but between a range covering the dithered timescale. We see that at 1284\,MHz, at DMs of 1000 or above (which are predicted on the Galactic plane by \citealt{Cordes2001}, \citealt{Yao2017}), a pulsar with $P=100$\,ms could either be negligibly smeared by scattering, or be scattered to be broader than the pulse period. Given the dominance of scattering among these selection effects, and the extreme unpredictability on the plane where these sources tend to lie, we would expect that targeted observations at higher radio frequencies would recover a large number of the pulsars that have been missed by this and other surveys. This is supported by the increased discoveries at S-band in \autoref{tab: popresults}, but also by the discovery of the three scattered pulsars discussed in \S\ref{subsubsec: general-disc}, including one in a target that we searched ourselves. Future searches in SNRs should therefore aim to search above 2\,GHz. More recent surveys have delivered hundreds of new candidate SNRs \citep{Ball2023, Anderson2025} many of which are located in confused and complex regions deep in the Galactic plane. We further emphasise the importance of observing at higher frequencies for future surveys that would sample these targets, rather than simply going deeper at L-band or lower frequencies.

\section{Updated information on the two discoveries}
\label{sec: psr-update}
\subsection{Timing}\label{subsec: timing}

\begin{table}
\centering
\caption{Timing solutions for PSR \psrB{} and PSR \psrA{}. The parentheses provided at the end of the fitted parameters are the 1-$\upsigma$ uncertainty on the final digit. Uncertainties are provided only for parameters that were included in the \tempo{} fit.}\label{tab: timingsolns}
{\scriptsize
\begin{tabular}{lll}
\hline
\multicolumn{3}{c}{Fit and Data-set} \\
\hline
Pulsar name\dotfill & J1818$-$1502 & J1831$-$0941 \\ 
MJD range\dotfill & 60088.1---60457.9 & 59733.8---60807.8 \\ 
Data span (yr)\dotfill & 1.01 & 2.94 \\ 
Number of TOAs\dotfill & 43 & 143 \\
RMS timing residual ($\mu s$)\dotfill & 3635 & 28036 \\
Reduced $\chi^2$ value \dotfill & 1.2 & 202 \\
\hline
\multicolumn{3}{c}{Measured and Set Quantities} \\ 
\hline
Spin frequency, $f$ (s)\dotfill & 1.74657512915(20) & 3.3236763242(12) \\
First derivative of $f$, $\dot{f}$\dotfill &  $-$1.5(3)$\times 10^{-16}$ & $-$9.4452(3)$\times 10^{-13}$ \\
Right ascension (hh:mm:ss)\dotfill & 18:18:54.31 &  18:31:24.85(12) \\
Declination (dd:mm:ss)\dotfill & $-$15:02:05.4 & $-$09:41:51(8) \\  
Dispersion measure, DM (cm$^{-3}$pc)\dotfill & 435 & 370.1 \\
Epoch of $f$ determination (MJD)\dotfill & 60207.6 & 59733.8 \\  
Epoch of position determination (MJD)\dotfill & 60207.6 & 59733.8 \\ 
Epoch of DM determination (MJD)\dotfill & 60207.6 & 59733.8 \\ 
\hline
\multicolumn{3}{c}{Derived Quantities} \\
\hline
$\tau_{\text{c}}$ (Myr) \dotfill & 178 & 0.056 \\
$\log_{10}$($B$, G) \dotfill & 11.2 & 12.7 \\
$\log_{10}$($\dot{E}$, ergs/s) \dotfill & 31.0 & 35.1 \\
\hline
\multicolumn{3}{c}{Assumptions} \\
\hline
Clock correction procedure\dotfill & TT(TAI) & TT(TAI) \\
Solar system ephemeris model and units\dotfill & DE405, TCB & DE405, TCB \\
Binary model\dotfill & NONE & NONE \\
\hline
\end{tabular}
}
\end{table}

\begin{figure}
    \centering
    \includegraphics[width=0.98\columnwidth]{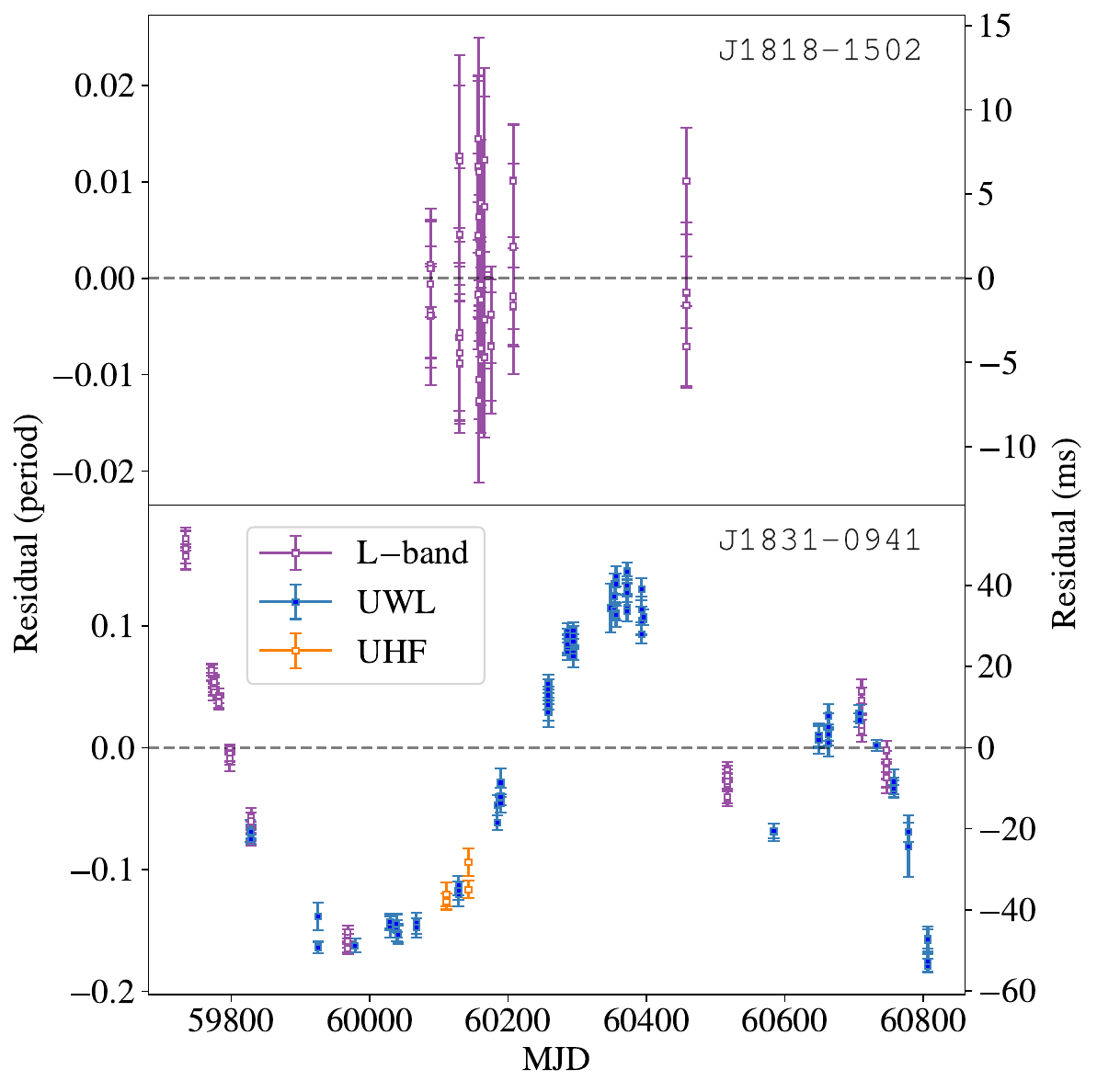}
    \caption[Residuals of PSR \psrB{} and PSR \psrA{}]{Timing residuals of PSR \psrB{} ({\it upper}) and PSR \psrA{} ({\it lower}), obtained with the updated timing solutions using the latest observations using the MeerKAT L$-$band (856-1712\,MHz) and UHF-band (544-1088\,MHz) receivers and Murriyang's UWL (704-4032\,MHz) receiver.}
    \label{fig:residuals}
\end{figure}

The up-to-date timing models obtained for both pulsars discovered by this survey are provided in \autoref{tab: timingsolns}. The residuals of the TOAs against arrival times predicted by these timing models are shown in \autoref{fig:residuals}. In the case of PSR \psrB{}, we have measured a period derivative of $(5.0\pm0.8)\times 10^{-17}$ from which we can now infer with confidence the large $\tau_{\mathrm{c}}$ and weak $B$-field strength of this pulsar, as was posited in Paper I. 

In the case of PSR \psrA{}, \autoref{fig:residuals} shows a significant amount of residual noise left over after fitting for $f$, $\dot{f}$, and the pulsar's position in RA and DEC. An erroneous value for $\dot{f}$ results in a parabolic evolution in residual space. In Paper I, which presented the data up to the UHF-band observations, fitting for RA, DEC and $\dot{f}$ was enough to model the variability of that data. This has now changed and so we attempted to model the timing noise using the method described in \S\ref{subsubsec: timingfits}. We compared the evidence of a model that included only the prescribed red noise and white noise parameters (henceforth RN) against one that also included $\ddot{f}$ (RN+F2). The RN model finds a power spectrum with log$_{10}(A)=-8.45\pm0.08$ and $\gamma=-5.5\pm0.6$, whereas for RN+F2 it is log$_{10}(A)=-8.46\pm0.08$, $\gamma=-5.4\pm0.7$ and $\ddot{f}=(4\pm7)\times10^{-23}$\,s$^{-3}$. The negligible change in red noise terms itself suggests little correlation between these and $\ddot{f}$, which is confirmed in the corner plot of the RN+F2 fit in \autoref{fig:cornerrn+f2}. The RN model has a higher evidence, though the difference is a factor of $8\pm5$. Thus is not preferred over RN+F2 to a statistically significant degree.

In Paper I we made the presumption that PSR \psrA{} must be a glitching pulsar because of its age, and thus the effect of glitch recoveries could induce, or perhaps dominate, the $\ddot{f}$ component of the pulsar's rotation. $\ddot{f}$ is consistent with zero, but if we use our value, we can set an approximate upper limit on the braking index of 270, which is large even for interglitch values \citep{Liu2024b}. We conclude that in order to reliably constrain the long-term timing noise and real $\ddot{f}$, we would need to observe this pulsar for several more years. For this reason, we do not include the results of either model in the final timing solutions presented in \autoref{tab: timingsolns}.

\subsection{Polarisation of PSR J1831$-$0941}\label{subsec: pol}
\begin{figure}
    \centering
    \includegraphics[width=0.98\columnwidth]{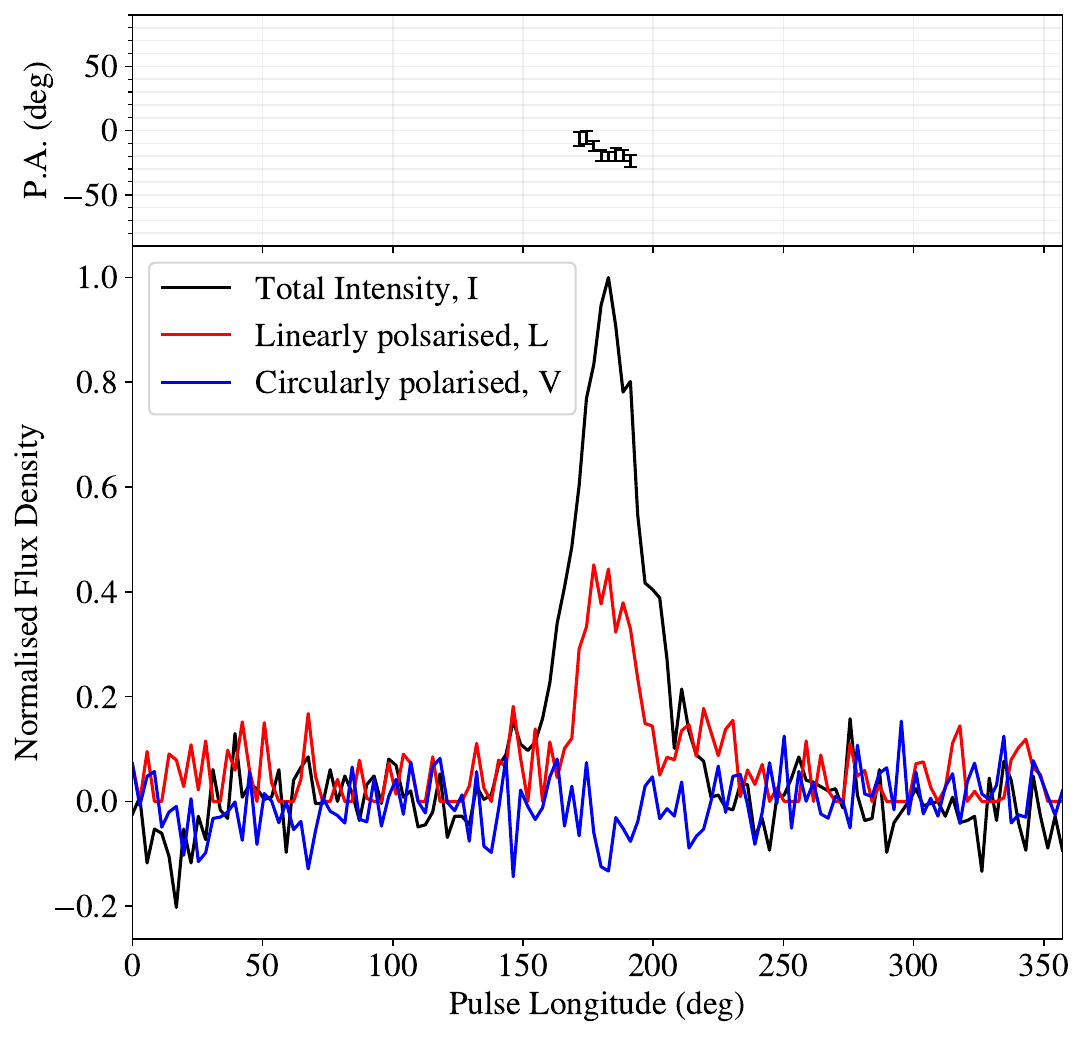}
    \caption[Polarisation of PSR \psrA{}]{RM-corrected pulse profile of PSR \psrA{} showing the total intensity (black), linearly polarised (red) and circularly polarised (blue) components measured from 704-4032\,MHz. The total intensity profile has a S/N of 45.1. In the upper panel, the angle of polarisation is shown for bins where $L$ has a significance of $\geq3\sigma$.}
    \label{fig:polprof}
\end{figure}

Following the method of \S\ref{subsubsec: pol}, the fraction of linear polarisation, $L/I$ and circular polarisation, $|V|/I$ and PA were computed for PSR \psrA{}. Their flux across the integrated pulse profile and the PA are shown in \autoref{fig:polprof}. The PA is seen to decrease steadily across the pulse peak from by about 18\degr. A smooth S-shaped PA swing is predicted by the Rotating Vector Model \citep[RVM;][]{Radhakrishnan1969}, and the polarisation properties of many pulsars conform to this model \citep[e.g.][]{Johnston2023}. A shallow PA swing suggests either that the magnetic and rotational axes of the pulsar are more closely aligned, or that the swing has been smeared and flattened due to scattering \citep{Karastergiou2009, Noutsos2009}. We find the latter to be a more likely explanation, as in Paper I we measured a scattering timescale of approximately half the pulse width within the bottom portion of the L-band (856-1070\,MHz) data. We are not able to confirm the scattering explanation by measuring the PA in the higher frequency portion of the UWL band, as the S/N of the linearly polarised component is insufficient. We find the average value for $L/I$ across the profile to be $0.37\pm0.23$. High $\dot{E}$ pulsars such as PSR \psrA{} tend to show linear polarisation fractions of above 40\,per cent \citep{Weltevrede2008, Mitra2025}. We see no significant circular polarisation component. 

\section{Conclusions}
\label{sec: pop-conc}
In this work, we presented the second half of the TRAPUM targeted survey for pulsars in supernova remnants and pulsar wind nebulae and justified the selection of the targets. In general, we avoided regions of the Galactic plane that are covered by ongoing competitive pulsar surveys, and chose to observe targets that would maximise the number of discoveries. No new pulsars were discovered since the last reported observations. 10 CCOs and 13 TeV sources from the H.E.S.S. survey, most of which have unidentified emission sources, were observed. The survey has uncovered two new pulsars, but only one, PSR \psrA{}, is young and associated with its target. We have undertaken a study of both the polarimetry and timing noise of this pulsar, and also provided updated timing solutions for the two discoveries. We set a mean best upper limit of 34\uJy{} for the 134 targets of the survey, though we estimate a more realistic average upper limit of 52\uJy{} that accounts for the average degradation in sensitivity of 0.65 across the coherent beam tilings. The serendipitous detection of six known pulsars in the field by the search pipeline was analysed, for which we find the pulsars are redetected with S/N values fully consistent with their predicted significance.

The low discovery rate of the survey follows the trend in other targeted surveys for SNRs. We have simulated 100 populations of isolated Galactic young pulsars and applied a prescription of assigning supernova remnants to them retroactively to create a synthetic version of the TRAPUM survey of SNRs. The routine has some limitations surrounding some simplifications and the pulsars' distances compared to the real sample. Nevertheless, the simulations reproduce the actual discovery rate of the survey to within $1\sigma$. We find that 80 per cent of the simulated young pulsars are beaming away from the Earth. Of the remaining non-detections, around half are due to scattering-dominated smearing at L-band. Of the overly smeared pulsars, approximately half would be too faint to be detected anyway. Thus the intrinsic luminosity is the dominant selection effect, precluding about 45 per cent of the pulsars beaming towards Earth from being discovered. The simulations predict that around 25 per cent of pulsars have a sufficient birth velocity to no longer be projected onto their supernova remnants, so would not be searched under our strategy. The importance of scattering is a key interpretation of these results. We strongly recommend that, in order to make the highest number of new detections, future targeted surveys be operated at frequencies above 2\,GHz whilst also improving the flux density limits of below $\sim$100\uJy{} that were achieved by this survey. This recommendation is supported by the prediction of our simulations for an equivalent survey conducted with the MeerKAT S-band receivers, where the number of discoveries increases between 50 and 150 per cent. To navigate the reduced FoV of observing at higher frequencies, such searches could take advantage of targeting regions of interest, for example, steep spectrum and potentially polarised radio sources in and near the SNR identified by deeper Galactic plane imaging surveys.

\section*{Acknowledgements}
The MeerKAT telescope is operated by the South African Radio Astronomy Observatory (SARAO), which is a facility of the National Research Foundation, itself an agency of the Department of Science and Innovation. All the authors thank the staff at SARAO for scheduling the MeerKAT observations presented here. TRAPUM observations used the FBFUSE and APSUSE computing clusters for data acquisition, storage and analysis. These instruments were designed, funded and installed by the Max-Planck Institut f\"{u}r Radioastronomie (MPIfR) and the Max-Planck-Gesellschaft. 
The Parkes Radio Observatory is a part of the Australia Telescope National Facility (ATNF), which funded by the Government of Australia and administered by the Commonwealth Scientific and Industrial Research Organisation (CSIRO) national science agency. We acknowledge the Wiradjuri people as the traditional owners of the Parkes Observatory site.
JDT acknowledges funding from the United Kingdom’s Research and Innovation Science and Technology Facilities Council (STFC) Doctoral Training Partnership, project code 2659479.
JDT thanks Michelle Tsirou for her helpful conversations about unidentified H.E.S.S. TeV sources, and Bhavnesh Bhat for his help understanding timing noise modelling. He also thanks Emma Carli, Rene Breton and Colin Clark for their general guidance. The authors would like to thank the reviewer for their useful comments that improved this manuscript.
This work used version \catversion{} of the ATNF Pulsar Catalogue and the October 2024 version of the Galactic SNR Catalogue by Green D. A, Cavendish Laboratory, Cambridge, United Kingdom. SAOImage DS9 was used for image analysis. We also made use of APLpy, an open-source plotting package for Python \citep{aplpy2012, aplpy2019} and Astropy \citep{astropy2013}, a community-developed core Python package and an ecosystem of tools and resources for astronomy.

\section*{Data Availability}
Data that are not available through the public archive of the South African Radio Astronomy Observatory, and all source code, will be shared on reasonable request to the corresponding author. The project code for the TRAPUM Science Working Group is SCI-20180923-MK-03. This paper includes archived data obtained through the Parkes Pulsar Data archive on the CSIRO Data Access Portal (\url{http://data.csiro.au}). Data from the Parkes Observatory for PSR \psrA{} can be retrieved using project code P1054.

\bibliographystyle{mnras}
\bibliography{main} 




\appendix
\section{Empirical sensitivity using detections of known young pulsars}
\label{sec: disc-known}

\begin{figure}
    \centering
    \includegraphics[width=0.98\columnwidth]{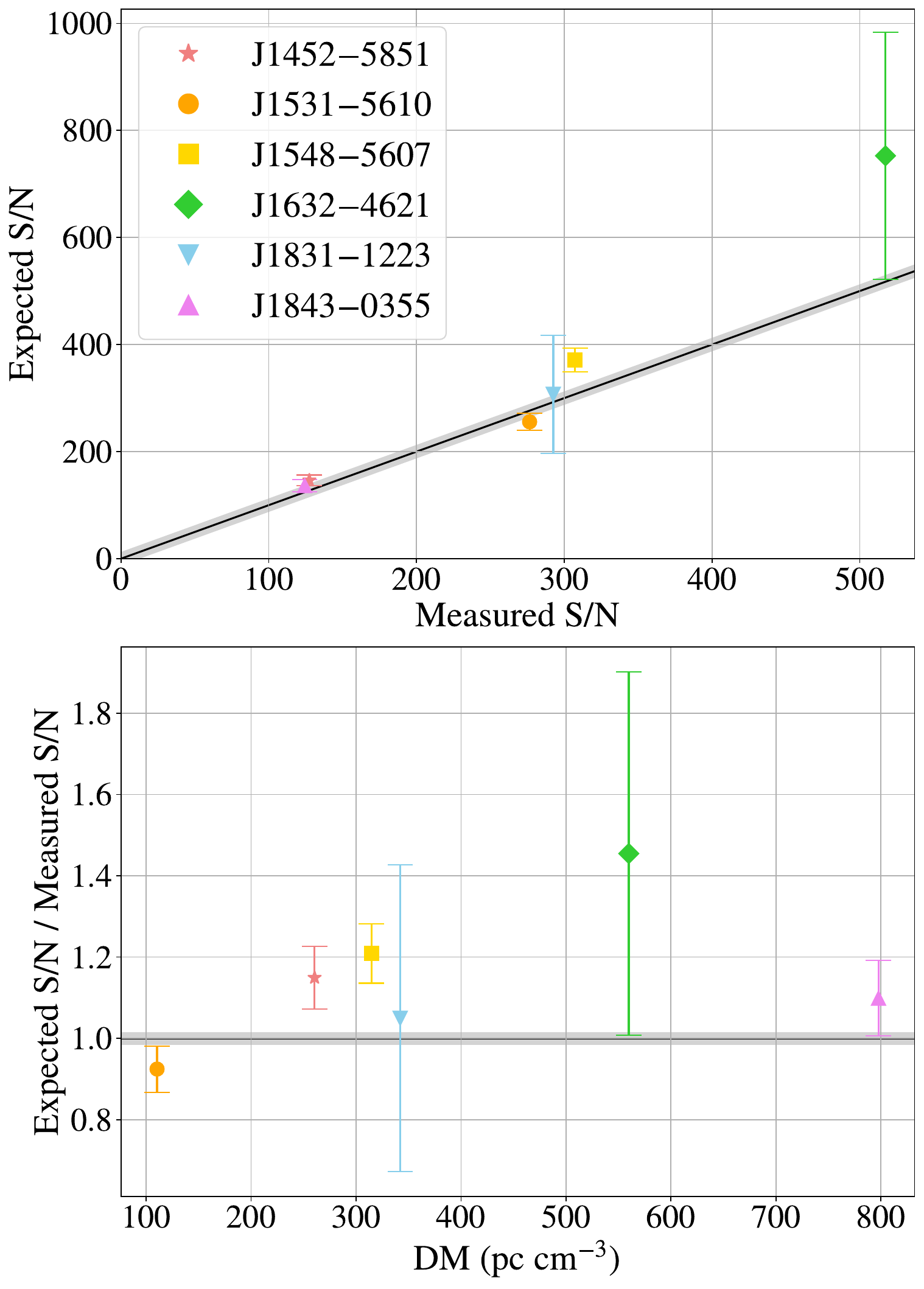}
    \caption[Predicted and measured S/N values of known pulsars]{\textit{Upper:} Expected S/N values against the S/N in the survey data for 6 redetected pulsars. The line of unity is shown, and all pulsars are consistent within their $3\sigma$ uncertainty. 5 out of 6 of the pulsars lie above this line, suggesting they are slightly fainter in the TRAPUM survey data than would be expected. \textit{Lower:} Ratio of the two S/N values against the pulsar's DM. A ratio of 1 is highlighted to help identify any trend, but the points appear to be randomly scattered.}
    \label{fig:knownpulsars}
\end{figure}

\citet{Padmanabh2023} compared the measured and expected S/N values of pulsars, mostly MSPs, that they had serendipitously seen in MMGPS-L. We judged it important to carry out an equivalent analysis with a sample of canonical pulsars rather than MSPs. Very few pulsars would be automatically covered by our tessellations of supernova remnant shells, so we targeted the positions of pulsars within the primary beam with a CB. There were no pulsars in our sample that have both flux density, $S_{\nu}$, and spectral index, $\alpha$, measurements from MeerKAT data, so we cross-matched against the data from \citet{Jankowski2018} instead. That study used Murriyang at bands centred on 728, 1382 and 3100\,MHz. There were 6 pulsar matches in our set, with periods from 84\,ms to 2.9\,s. All of the pulsars were seen in the filtered and sifted candidate lists, albeit occasionally only as harmonics presumably due to the removal of the primary period candidate by the filter.
To obtain their measured S/N and width values, the data were folded with \dspsr{}\footnote{\href{https://dspsr.sourceforge.net/}{https://dspsr.sourceforge.net/}} using ephemerides from the ATNF pulsar catalogue \catversion{} \citep{Manchester2005}, then cleaned with \clfd{}\footnote{\href{https://github.com/v-morello/clfd}{https://github.com/v-morello/clfd} by Vincent Morello} before S/N values and widths were found using the \pdmp{} command of {\sc psrchive}\footnote{\href{https://psrchive.sourceforge.net/}{https://psrchive.sourceforge.net/}}. 

Estimating the expected S/N requires a careful consideration of the pulsar's location within a multibeam tiling. The S/N at a position inside a CB will degrade further away from the CB centre by a factor, $f_{\text{CB}}$. Across the tiling, this is modulated by the degradation of the IB, $f_{\text{IB}}$, due to the offset from the boresight. The combined degradation is therefore \cite[][equation A2]{Padmanabh2023}:
\begin{equation}
    d = f_{\text{CB}} f_{\text{IB}}.
    \label{eq: IBCBdegredation}
\end{equation}
As we placed a CB at the best position, we can therefore assume that $f_{\text{CB}}=1$. We calculate $f_{\text{IB}}$ using the \textsc{beam-corrections}\footnote{\href{https://github.com/BezuidenhoutMC/beam-corrections}{github.com/BezuidenhoutMC/beam-corrections} by M. C. Bezuidenhout} tool, which uses the primary beam model of \textsc{katbeam}\footnote{\href{https://github.com/ska-sa/katbeam}{github.com/ska-sa/katbeam} by Ludwig Schwardt and  Mattieu de Villiers}. This model approximates the IB at an elevation of 60\degr as a half-cosine tapered function \citep[][equation 12]{DeVilliers2022}. The weighted flux of each pulsar at L-band is calculated by scaling the fluxes from \citet{Jankowski2018} to 1\,MHz-wide subbands across the 856\,MHz bandwidth. The frequency coverage of the mask applied by \clfd{} is used to calculate the effective bandwidth. The sky temperature at 1284\,MHz is estimated at each pulsar's position using GlobalSkyModel2016 \citep{Zheng2017}. Finally, the radiometer equation is used for the measured width to get the best S/N, then multiplied by $d$ to obtain the expected S/N. The results are shown in \autoref{fig:knownpulsars}, where the expected versus measured signal to noise is shown, as is the fraction of the two values plotted against the pulsars' DMs. The expected and measured S/N estimates are in good agreement. The young pulsar in the set, PSR \mbox{J1531$-$5610} is redetected with the smallest deviation from the expected S/N. We see no trend in the DM against the ratio of the two values in the lower panel. The reduced-$\chi^{2}$ around unity is 0.59. However the scatter slightly prefers the pulsars to be dimmer than expected in our survey. There are some factors that this analysis does not account for. For example, the measured S/N can be affected by scintillation changing the pulse brightness in both frequency and time, or even intrinsic temporal variability in the pulse flux, though we do not see very much of either phenomenon in this set of pulsars. The S/N may also be affected by inaccuracies of the IB model, though we saw no trend against the offset from the boresight, confirming that simplifications of the IB model are insignificant compared to the other uncertainties present. Under our considerations, we expect the S/N uncertainty to be dominated by the errors on $S_{\nu}$ and $\alpha$. We find our results to be in close agreement with the data for MSPs seen by MMGPS-L as analysed by \citet{Padmanabh2023}. 

We also targeted a further four young pulsars with CBs during the survey: PSR \mbox{B1758$-$23}, PSR \mbox{J1821$-$1419}, PSR \mbox{J1833$-$1034} and PSR \mbox{J1841$-$0345}. Detecting PSR \mbox{B1758$-$23} was particularly interesting to us, due to its high DM of 1068\dm{} and highly scattered profile. All pulsars were redetected by the pipeline except for PSR J1821$-$1419. PSR \mbox{J1821$-$1419} is a 1.6-s pulsar with a high DM of 1123\dm{} and is relatively understudied in the literature. The discovery pulse profile at 1374\,MHz measured by \citet{Hobbs2004} does not show particularly strong scattering. It is very close to the magnetars on the $P-\dot{P}$ diagram, and is considered a magnetar candidate \citep{Winters2020}. It was not seen in optimised folds with \pdmp{}. We are therefore left to speculate that it was undetected due to RFI losses that affected over 25\,per cent of the band or perhaps due to magnetar-like radio quiescence or intermittency. In addition, we regularly detected pulsars serendipitously in the IB data. One of those detected was the 182-ms period young pulsar PSR \mbox{J1702$-$4128}. We did not place a CB on it due to its high flux density of above 1\,mJy at L-band \citep{Jankowski2018}. We usually avoid targeting pulsars brighter than 1\,mJy, as they can cause issues not only for the TRAPUM searches, but also for any single pulse searches that may piggyback on our observations. The pulsar was approximately 26\,arcmin from the boresight and the folded S/N returned by \pulsarx/\psrfoldfil{} was 19.

\section{Sky and beam maps for H.E.S.S. sources}
\begin{figure*}
    \centering
    \includegraphics[width=0.245\textwidth]{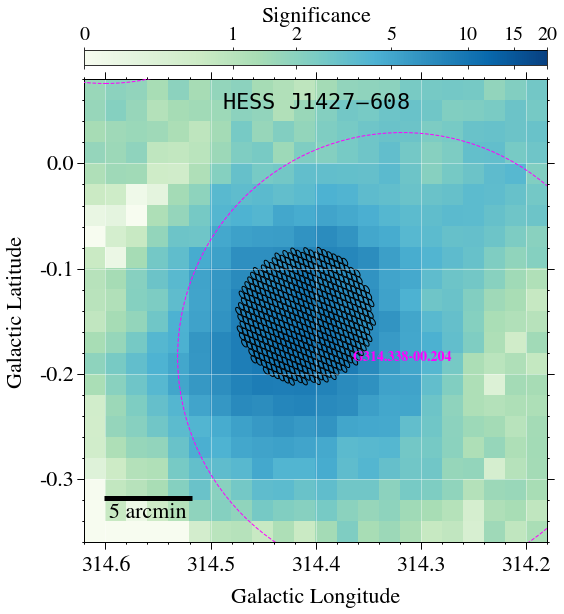}
    \includegraphics[width=0.245\textwidth]{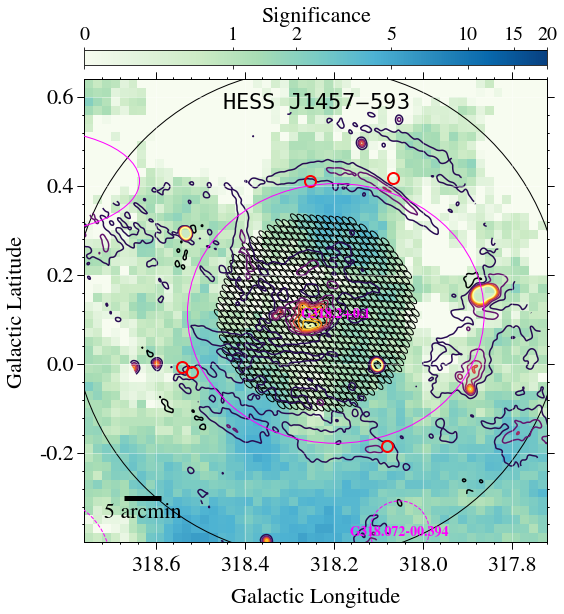}
    \includegraphics[width=0.245\textwidth]{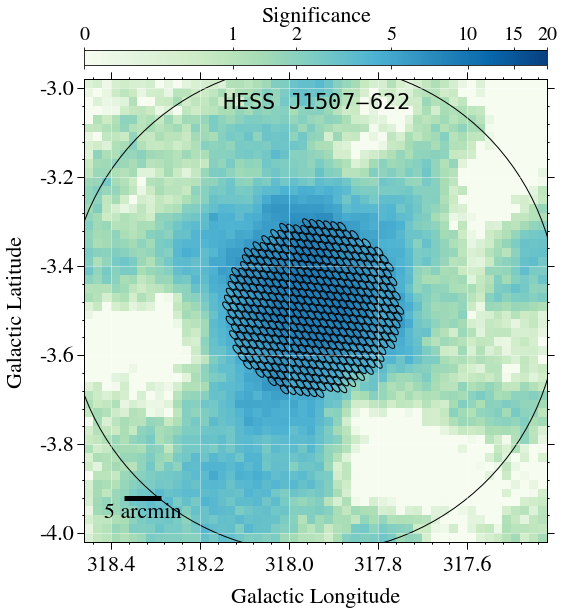}
    \includegraphics[width=0.245\textwidth]{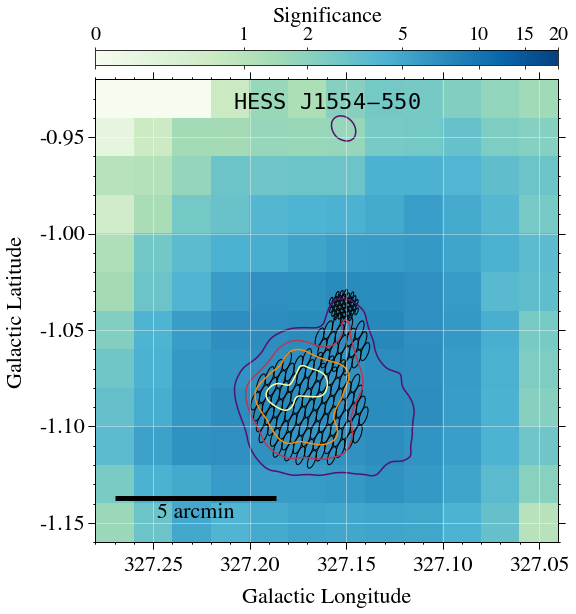}
    \includegraphics[width=0.245\textwidth]{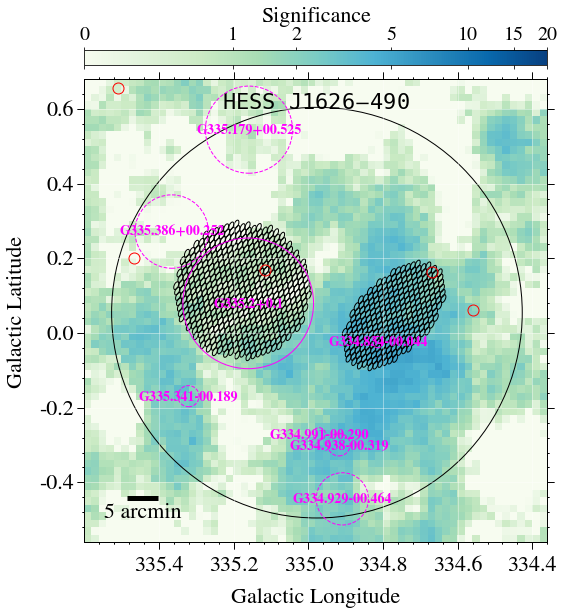}
    \includegraphics[width=0.245\textwidth]{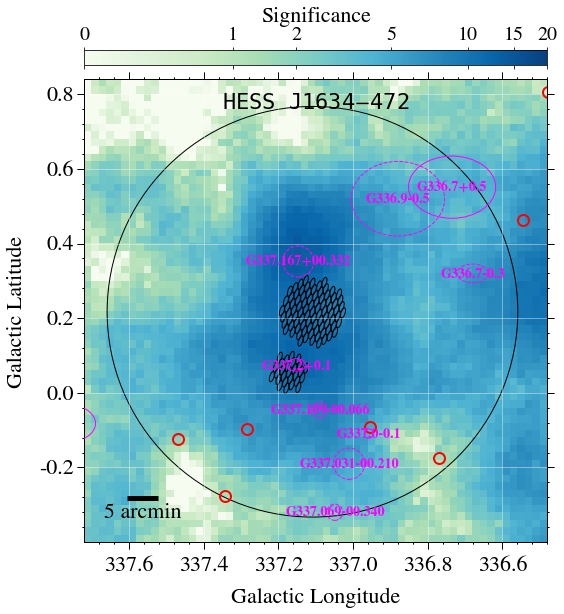}
    \includegraphics[width=0.245\textwidth]{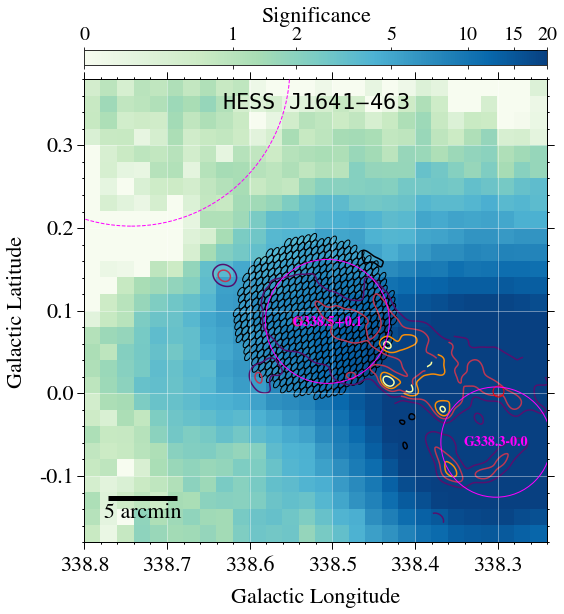}
    \includegraphics[width=0.245\textwidth]{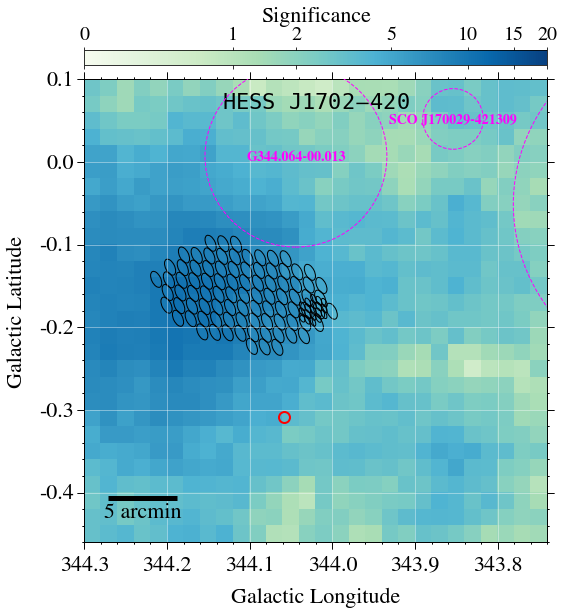}
    \includegraphics[width=0.245\textwidth]{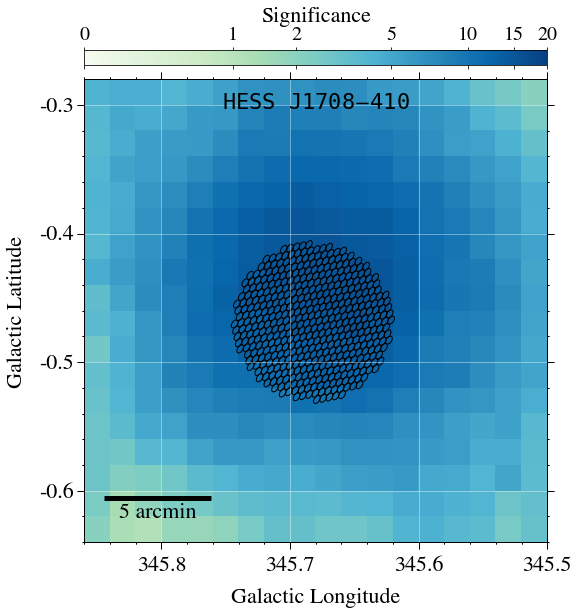}
    \includegraphics[width=0.245\textwidth]{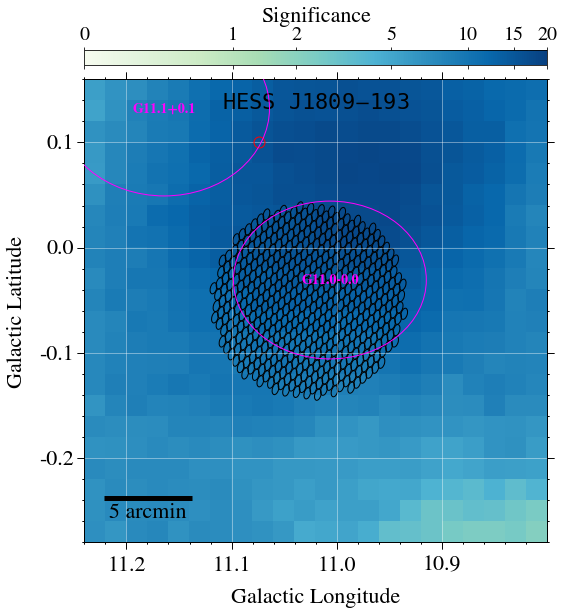}
    \includegraphics[width=0.245\textwidth]{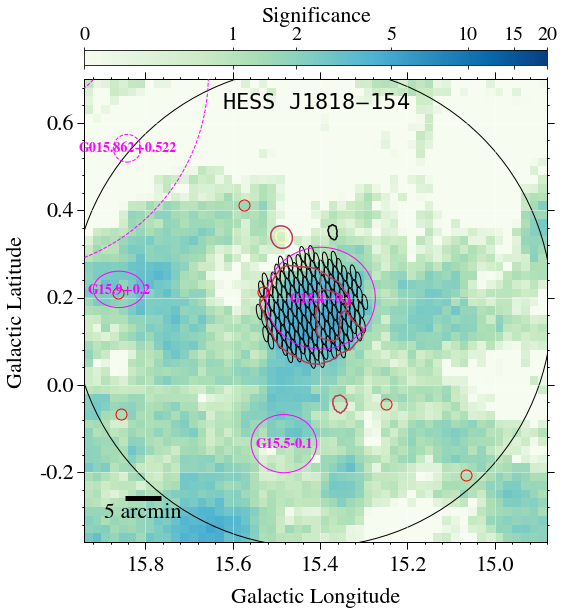}
    \includegraphics[width=0.245\textwidth]{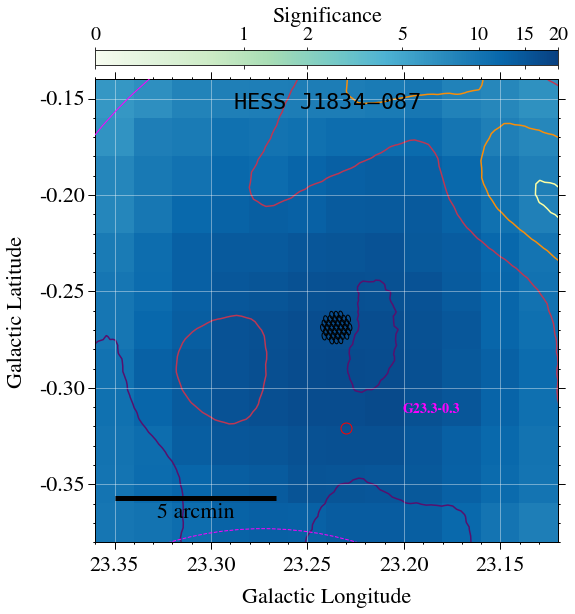}
    \includegraphics[width=0.245\textwidth]{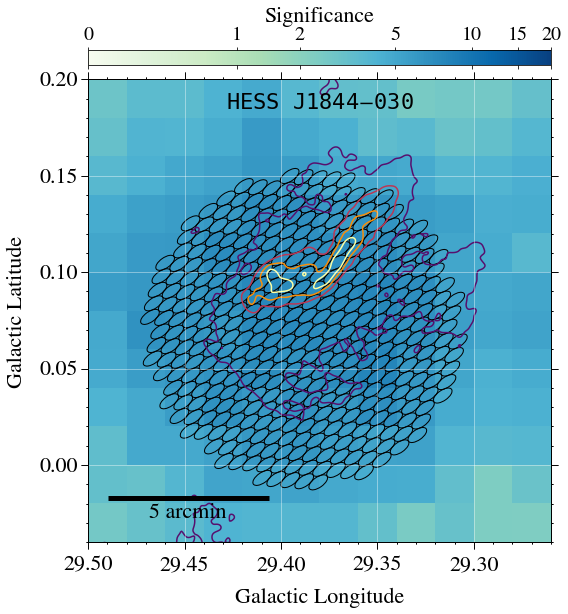}
    \caption[Beam maps for 8 H.E.S.S. source targets]{Sky maps of the fields for 13 H.E.S.S. sources that were searched, showing the locations of the CBs (small black ellipses) and, where appropriate, the IB (large black circle). The maps show the significance levels of the VHE \grays{} $>1$\,TeV detected by H.E.S.S. at a resolution of 0.1\degr (\citealt{Abdalla2018a}; available to download at \href{https://www.mpi-hd.mpg.de/hfm/HESS/hgps/}{https://www.mpi-hd.mpg.de/hfm/HESS/hgps/}). Magenta circles or ellipses mark the boundaries of supernova remnants; solid lines are G25 SNRs and dashed lines are candidate SNRs from the literature. Beams may appear offset from SNR boundaries due to the centroid position precision provided by G25; beam tilings have been centred on more precise SNR positions. Red circles are the locations of pulsars from the ATNF pulsar catalogue \catversion{} \citep{Manchester2005}. Coloured contours trace the significance levels of radio components in the field: for HESS J1457$-$593, HESS J1554$-$550 and HESS J1641$-$463 data are from the Molonglo Observatory Synthesis Telescope (MOST) Supernova Remnant Catalogue at 843\,MHz \citep{Whiteoak1996}, for HESS J1818$-$154 and HESS J1834$-$087 data are from the GaLactic and Extragalactic All-sky MWA survey \citep[GLEAM;][]{Hurley-Walker2019c} at 170-231\,MHz, and finally HESS J1844$-$030 is overlaid with data from the 1.4\,GHz NRAO VLA Sky Survey \citep[NVSS;][]{Condon1998}.}
    \label{fig:hessset}
\end{figure*}

In \autoref{fig:hessset}, beam maps for the thirteen H.E.S.S. TeV sources that we searched are shown.

\section{RN+F2 fit with {\sc run\_enterprise}}
\begin{figure*}
    \centering
    \includegraphics[width=0.98\textwidth]{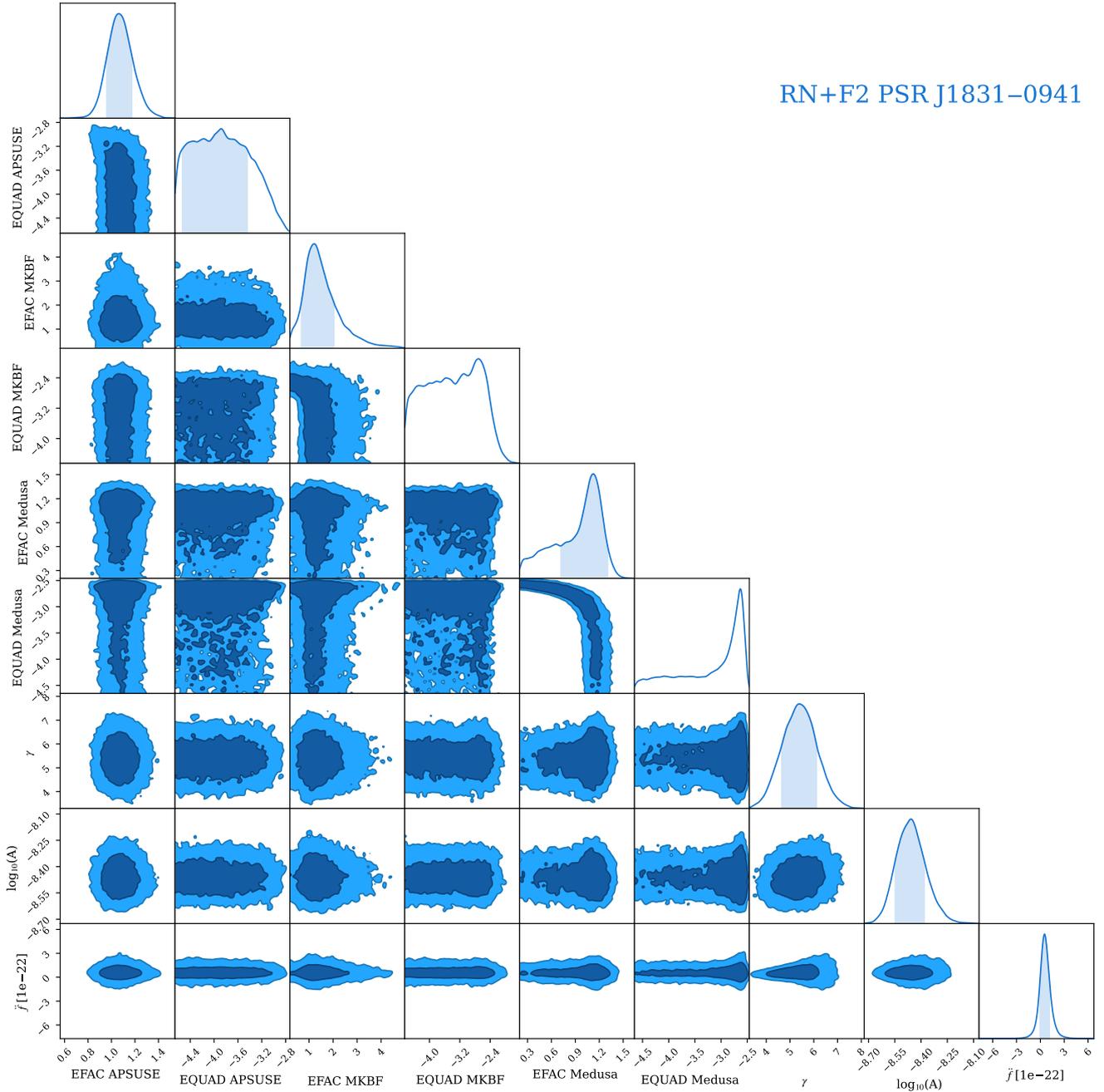}
    \caption[Corner plot of the 10,000 sample MCMC fit for PSR \psrA{}]{Corner plot for the fit of white and red noise parameters and F2 to the PSR \psrA{} TOAs and ephemeris in \autoref{tab: timingsolns}. Labels from left to right along x-axis: {\sc efac} and {\sc equad} for L-band data captured with the APSUSE back end, {\sc efac} and {\sc equad} for UHF-band data captured with the PTUSE back end, {\sc efac} and {\sc equad} for UWL data captured with the Medusa back end, red noise amplitude, red noise spectral index and $\ddot{f}$. Plot has been made using the {\sc chainconsumer} package \citep{Hinton2016}.}
    \label{fig:cornerrn+f2}
\end{figure*}

The corner plot produced by {\sc enterprise} for the white noise, red noise and $\ddot{f}$ parameters of PSR \psrA{} is provided in \autoref{fig:cornerrn+f2}.

\bsp	
\label{lastpage}
\end{document}